%% file: ml_geometries.tex
\documentclass[journal=jacsat,manuscript=article]{achemso}

\usepackage[version=3]{mhchem} 
\usepackage{amsmath}
\usepackage{amssymb}
\usepackage{subfig}
\usepackage{multicol}
\usepackage{chemfig}
\usepackage{tikz}
\usepackage{pgfplots}
\usepackage{algorithm}
\usepackage{algorithmic}
\usepackage{mathptmx}
\usetikzlibrary {graphs}
\usepackage{color}
\usepackage{amsfonts}
\DeclareSymbolFont{symbolsb}{OMS}{cmsy}{m}{n}
\SetSymbolFont{symbolsb}{bold}{OMS}{cmsy}{b}{n}
\usetikzlibrary{positioning, patterns,shapes,shadows,arrows}
\usepgfplotslibrary{groupplots}
\pgfplotsset{width=7cm,compat=1.8}
\usetikzlibrary{shapes.geometric, arrows}
\usepackage[useregional]{datetime2}
\usepackage{graphicx}
\usepackage{lscape}
\usepackage[export]{adjustbox}
\usepackage[compact]{titlesec}
\titlespacing{\section}{0pt}{*0}{*0}
\usepackage[normalem]{ulem}

\newcommand{\tset}{\mathbb{T}}
\newcommand{\cset}{\mathbb{C}}

\newcommand{\ve}[1]{\boldsymbol{\mathrm{#1}}}
\newcommand{\mx}[1]{\boldsymbol{\mathrm{#1}}}
\author{Carlos Manuel de Armas-Morej\'on}
\email{carlosdearmasm@gmail.com}
\affiliation[Nanobio]
{Nano-Bio Spectroscopy Group and ETSF Scientific Development Centre, Department
of Materials Physics, University of the Basque Country, CFM CSIC-UPV/EHU-MPC
and DIPC, Tolosa Hiribidea 72, E-20018 Donostia-San Sebasti\'an}
\alsoaffiliation[mpsd]
{Theory Department, Max Planck Institute for the Structure and Dynamics of
Matter and Center for Free-Electron Laser Science, Luruper Chaussee 149,
22761 Hamburg, Germany}
\alsoaffiliation[UHavana]
{Laboratorio de Química Computacional y Teórica,
          Facultad de Química, Universidad de La Habana,
          10400. La Habana, Cuba.}
\author{Ask Hjorth Larsen}
\email{asklarsen@gmail.com}
\affiliation[Nanobio]
{Nano-Bio Spectroscopy Group and ETSF Scientific Development Centre, Department
of Materials Physics, University of the Basque Country, CFM CSIC-UPV/EHU-MPC
and DIPC, Tolosa Hiribidea 72, E-20018 Donostia-San Sebasti\'an}
\author{Luis A. Montero-Cabrera}
\email{lmc@fq.uh.cu}
\affiliation[UHavana]
{Laboratorio de Química Computacional y Teórica,
          Facultad de Química, Universidad de La Habana,
          10400. La Habana, Cuba.}
\author{Angel Rubio}
\email{angel.rubio@mpsd.mpg.de}
\affiliation[mpsd]
{Theory Department, Max Planck Institute for the Structure and Dynamics
of Matter and Center for Free-Electron Laser Science, Luruper Chaussee
149, 22761 Hamburg, Germany}
\author{Joaquim Jornet-Somoza}
\email{j.jornet.somoza@gmail.com}
\affiliation[Nanobio]
{Nano-Bio Spectroscopy Group and ETSF Scientific Development Centre, Department
of Materials Physics, University of the Basque Country, CFM CSIC-UPV/EHU-MPC
and DIPC, Tolosa Hiribidea 72, E-20018 Donostia-San Sebasti\'an}
\alsoaffiliation[mpsd]
{Theory Department, Max Planck Institute for the Structure and Dynamics of
Matter and Center for Free-Electron Laser Science, Luruper Chaussee 149,
22761 Hamburg, Germany}
\title[An \textsf{achemso} demo]
  {A basic electro-topological descriptor for the prediction of organic molecule geometries by simple machine learning}

\keywords{DFT/TDDFT, Geometry Optimization, machine learning}

\begin{document}



\begin{abstract}
This paper proposes a machine learning (ML) method to predict stable molecular geometries from their chemical composition.
The method is useful for generating molecular conformations which may serve as initial geometries for saving time during expensive structure optimizations by quantum mechanical calculations of large molecules.
Conformations are found by predicting the local arrangement around each atom in the molecule after trained from a database of previously optimized small molecules.
It works by dividing each molecule in the database into minimal building blocks of different type.
The algorithm is then trained to predict bond lengths and angles for each type of building block using an electro-topological fingerprint as descriptor. A conformation is then generated by joining the predicted blocks.
Our model is able to give promising results for optimized molecular geometries from the basic knowledge of the chemical formula and connectivity. The method trends to reproduce interatomic distances within test blocks with RMSD under $0.05$ \r{A}.

\end{abstract}

\section{Introduction}
\label{sec:int}

This work assesses the problem of generating reliable conformers of molecules from proposed chemical compositions. Realistic initial bond lengths and angles are essential for efficient geometry optimizations. They are normally the first step of the usual computational workflow of systematic variations of the atomic coordinates inside a molecule and the calculation of the potential energy and forces of the system in order to find a minimum value for the potential energy, which indicates a theoretically optimized conformation. Moreover, approximate and reliable molecular geometries serve for many modelling and process simulation purposes from docking to Molecular Dynamics. 

The function combining all possible variations of the geometry and the potential energy forms a high-dimensional surface and it is the well known \emph{potential energy surface} (\textbf{PES}). All possible conformers for a given compound are comprehended as minima in the appropriate PES.
Two processes are required in order to obtain the model of the conformer with the lowest potential energy of a given molecule: (1) a procedure to attain its corresponding and plausible PES and (2) a method to navigate it to search for minima. Procedures to find PES's and then computing potential energies can vary in efficiency depending on several factors, and mostly the number of atoms of the compound. Options exist from the easily computed empirical force fields based on classical considerations of bodies in a molecule \cite{ml_85, ml_86}, passing fast quantum mechanical semi-empirical calculations being parameterised for specific scenarios \cite{ml_127}, to more general and reliable but computationally expensive \emph{ab-initio} and DFT calculations \cite{ml_102}. The method to navigate the PES can also be computationally expensive depending on how fast the global minimum can be found \cite{ml_59}. Obviously, if the initial choice of a guess conformation is near to the final optimized structure, the correspnding PES's minimum will be faster reached. This is crucial for accurate geometry optimizations of large molecules. 

Machine learning (\emph{ML}) has been recently used in a variety of topics inside the field of quantum mechanics. \cite{ml_3, ml_11, ml_12, ml_13, ml_10, ml_34, ml_56, ml_79, ml_92, ml_93, ml_94, ml_95, ml_9, ml_97, ml_98, ml_99, ml_130, ml_131, ml_132, ml_133, ml_134, ml_135, ml_136, ml_137, ml_138, ml_139, ml_140, ml_141, ml_142, ml_143, ml_144, ml_145} For molecular geometry optimizations, progress has been made mostly by using neural networks to parameterise classical force fields. \cite{ml_72, ml_81}

The proposal made in this work is to use ML, and specifically the Kernel Ridge Regression algorithm, for predicting molecular conformations by producing the local arrangement of each atom belonging to a molecule.  It can be achieved by using a large and confident database of optimised small molecules \cite{ml_6} as a source for both the training and testing sets. For this purpose, certain molecular structural blocks are characterised and defined by an electro-topological descriptor \cite{ml_8, ml_9} from the structures of the previously optimised molecules. Blocks are then reconstructed by applying ML tools. Using the ETKDG\cite{ml_30, ml_31, ml_84} (Experimental Torsion Knowledge Distance Geometry) present in RDKit\cite{ml_146} with a new \emph{ab-initio} torsion angle database, we join all predicted blocks to produce the desired molecular structures with reasonable reliability. Results promise fast and confident predictions of molecular geometries and conformations from their formulas taken as structural graphs.

\section{Learning data}

The database used in this work is part of a larger one\cite{ml_125, ml_126} of quantum PES's minima geometries of small organic molecules containing up to 8 C, O, N  and/or F atoms. The optimised geometries of this database are reported to be found using DFT/B3LYP\cite{ml_43} with the 6-31G(2df,p) basis set as a commonly accepted reliable PES. We will refer to this database as \emph{8CONF}. The size of this resulting database subset is $~21$k molecules.\cite{ml_6}

To facilitate predictions based on this data, we seek a representation which minimizes the amount of redundant information in the learning set, and also could group together similar kinds of data.

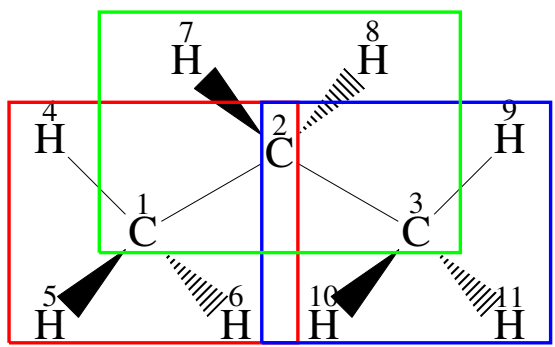
\begin{figure}[ht!]
\centering
{\scalebox{0.8}{%
    \input{./images/mol_block_example}}}
\caption{Block decomposition for the molecule C$_3$H$_8$. Each C atom is bonded to multiple atoms
and hence defines a block. The molecule can therefore be divided in $3$ blocks, two of which (for atoms 1 and 3) belong to the same \emph{block-class}. In each block we have redundant atoms to indicate where blocks join together.}
\label{fig:block_exampl}
\end{figure}

First, each molecule is split into \emph{blocks}. A block is the main building part of our model, and it is characterised by: (1) a central atom with more than one bond and (2) the first neighbors of such central atom.  Figure \ref{fig:block_exampl} shows a block decomposition for an example molecule from the 8CONF database.  Note that each atom can normally be included in multiple blocks: the block centered around itself as well as each of the blocks surrounding its neighboring atoms.  Atoms with only one neighbor are not considered to define a block. Blocks therefore have from two to four neighboring atoms in the selected molecular sets where all atoms belong to the first and second rows of the periodic table.

A unique Cartesian representation is not well suited for predictions because coordinate values depend on the chosen reference center. Instead, we represent local coordinates within a block by 1) subtracting the molecular Cartesian coordinates of the central atom position in that block to define it as the local coordinate origin, and 2) computing the matrix $\mx B$ of scalar products $\ve a_i \cdot \ve a_j$ between each pair $(i, j)$ of coordinates of local position vectors corresponding to the non-central atoms in the block. This matrix is the feature we use for training and predictions.

The matrix $\mx B$ of scalar products is symmetric and at most $4 \times 4$ in size, and therefore has up to ten unique degrees of freedom.  $\mx B$ contains enough information to rebuild the set of molecular Cartesian coordinates (see Appendix~\ref{app:sp}) for a block except for translations, rotations, and chirality, with which the matrix is invariant.

The non-central atoms in a block have no natural ordering. Hence a way must be chosen to assign an index $i$ to each of them with a minimum of ambiguity for ML training and testing. To this end we define an \emph{equivalence relation} for the set of all blocks, i.e., each block belongs to a single, specific equivalence class or \emph{block-class}. The ML algorithm is then independently trained for each block-class.

Two blocks belong to the same class if 1) the species of the central atom in each block is the same, 2) the species of each neighboring atom is the same (some ambiguity is solved by sorting by atomic numbers) and 3) the arrangement of the atoms in space is the same, i.e. either tetrahedral (\emph{TH}), triangular (\emph{TR}) or linear (\emph{L}) .

The definition of block-classes can be applied with different levels of restrictions. As a result, the number of different block-classes can vary, as blocks with the same atoms can appear in very different environments.

\begin{figure}
    \centering
    {\scalebox{0.8}{%
    \input{./images/block_dist_tikz.tex}}}
    \caption{Distribution of block-classes within all molecules in the database. A few common and uncommon block-classes as O-H-H and C-O-H-H are referred.}
    \label{fig:block_dist}
\end{figure}
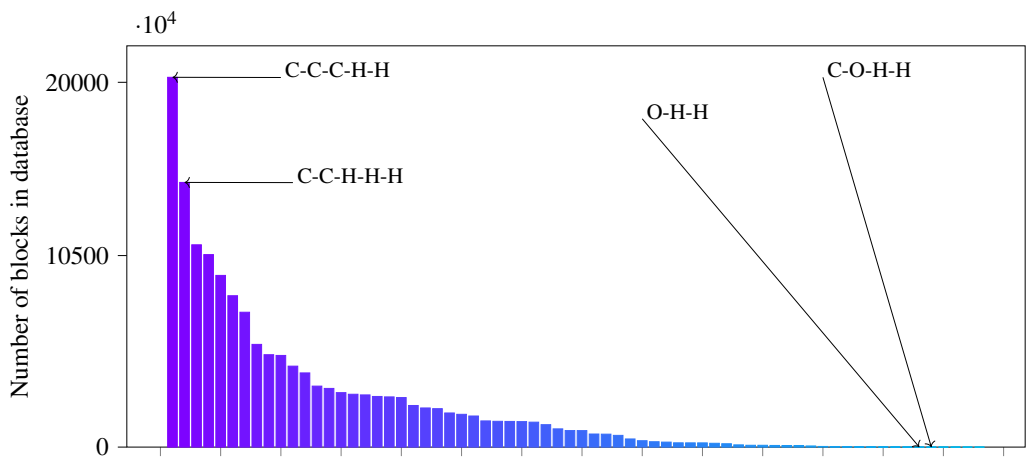

We denote a block-class by a series of chemical symbols followed by certain indicators of spatial configurations when necessary. The first symbol is that of the central atom, followed by the symbols of the neighboring atoms ordered with higher atomic numbers first. Figure \ref{fig:block_dist} shows the distribution of blocks inside the database.


When choosing the definition of block-classes there is a tradeoff: We can make predictions easier by maximizing the amount of chemical knowledge which defines a block. It can be achieved by dividing the blocks into a large number of classes each of which contains very similar blocks. However doing so also decreases the size of the learning sets, resulting in block-classes with few or no members. In general block-classes must be defined in such a way to get neither too many nor too few.


\section{Descriptors}

To complete our learning data, a property is required that can correlate with the desired feature. This property is named as a \emph{descriptor} or \emph{fingerprint}. In this case, desired features are the scalar products of a block being directly related to the pursued molecular geometry predictions. \emph{Descriptors} needs to fulfill a number of required characteristics such as: (1) easiness to establish or compute, (2) good representability and (3) low dimensionality\cite{ml_4}. We took the so-called electro-topological\cite{ml_8, ml_9} state index (\emph{e-state}), which is a combination of both electronic and topological characteristics of atoms in a molecule.

If $Z^v$ is the number of valence electrons of a certain element and $h$ the number of bonded hydrogen atoms according to the position in a molecule, then $\delta^v$ could be defined as $(Z^v - h)$, the count of valence electrons of a certain atom left for being engaged in the skeleton of a molecule. Similarly, $\delta$ could be defined as the count of engaged $\sigma$ electrons $(\sigma - h)$. Then, $I$ can be defined as the \emph{intrinsic state value} of an atom in a molecule, given by $I = \frac{\delta^v + 1}{\delta}$. It gets related to the backbone valence of an atom other than Hydrogen.

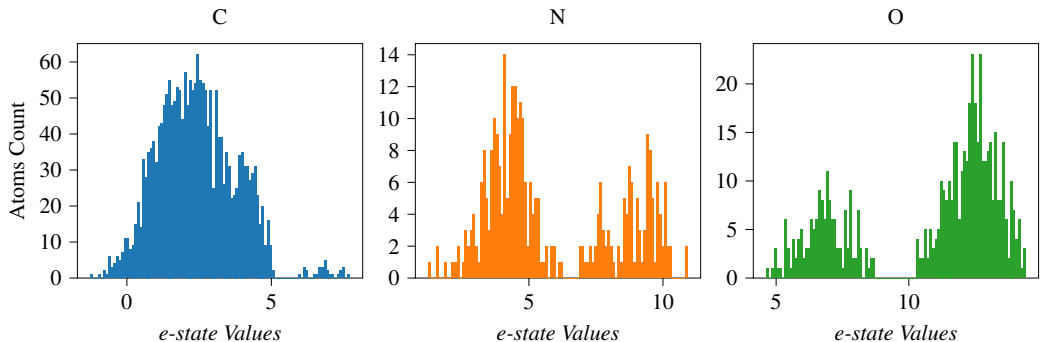
\begin{figure}[ht!]
\centering
{\scalebox{0.7}{\input{./images/env_stats_e_state_tikz}}}
\caption{Distribution of e-state values for some atoms species present in our database.}
\label{fig:env_stats_e_state}
\end{figure}

Then, the \emph{e-state} to relate bonded atoms with their positions in a molecule can be taken as $S = I + \varDelta I$. It combines the previously defined intrinsic state value $I$ and a certain $\varDelta I = \sum_{j=1}^N \frac{I_i + I_j}{r_{i, j}^2}$ where $r_{i, j}$ is a rough expression of distances between atoms $i$ and $j$ given by the count of atoms in the shortest path between them, including themselves. $\varDelta I$ relates the intrinsic state value with the bonded environment. For each atom we use a vector of e-state composed by $[I, \varDelta I_0, \varDelta I_n]$, where $\varDelta I_0$ are the contributions of the first neighbours and $\varDelta I_n$ are the contributions of more distant neighbours. The choice of this descriptor fulfill to a fine degree all our desired characteristics. Figure \ref{fig:env_stats_e_state} shows the distribution of e-state values of some elements in our database.

\section{Learning method}

The kernel ridge regression (\textbf{KRR}) formulation proposed by \emph{Ramakrishnan, R. et al.}\cite{ml_128} is our selected learning method. It have been tested with success in other \emph{machine learning/Quantum Mechanical} applications. \cite{ml_6, ml_128} It serves to provide a given property of any query molecule as a linear combination of similarity measures (as ”distances”) between the query's property and those of a finite set of training points. For the \emph{tests} we randomly selected $450$ molecules from 8CONF ($\cset$ from now on). The rest of the molecules were chosen as the training set ($\tset$ form now on).

Our query's property is the scalar product matrix obtained as the result of applying the transformation $\mx B = \mx X \mx X^T$ to each block, where $\mx X's$ are the nonzero position vectors resulting after a translation of the central atom of a block to the origin of coordinates. The KRR method needs training to grasp correlation between the descriptor and the property, and $\tset$ was used for that. Let $\mx R$ be the scalar product tensor of all blocks inside a block-class in the $\tset$. In matrix notation the training process can be defined as:
\begin{align}\label{eq:krr}
  \mx R = (\mx K + \lambda \mx I) \mx C
\end{align}
\noindent where $\mx C$ represents the coefficients to be computed, $K_{ij} = K(|d_i - d_j|) = \exp(-\frac{|d_i - d_j|}{\alpha})$ are the kernel terms as calculated by using $L_{1}$ norm distances $|\cdot|$ between the descriptors $d$ and $\alpha$ is a regularization parameter to be commented below. From solving eq. \ref{eq:krr}, we obtain the coefficients $\mx C$, used to get the prediction $\mx P$ by:

\begin{align}
  \mx P = \mx D \mx C
\end{align}

\noindent where $\mx D$ is the matrix of distances between descriptors $d$ of known $j$ learning points and the new $i$ points computed using $K(-\frac{|d_{i \in \mathrm{new}} - d_{j \in \mathrm{known}}|}{\alpha})$.

\begin{figure}[ht!]
\centering
{\scalebox{0.8}{\includegraphics[]{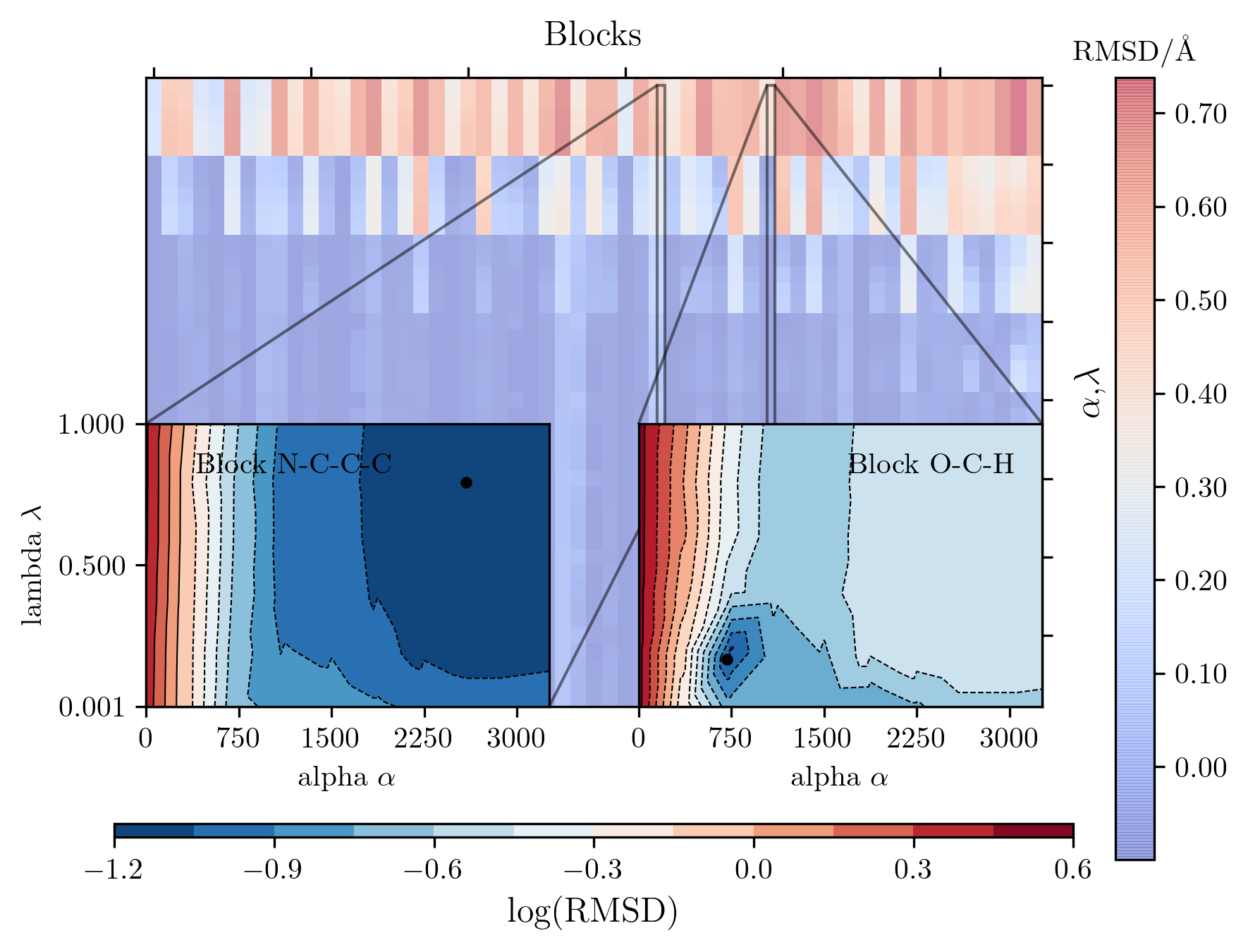}}}
\caption{Summary of the KRR's $\alpha$ and $\lambda$ hyper-parameter
optimizations. The background image shows the RMSD value for all blocks-class by each combination of the hyper-parameter $\alpha, \lambda$. The color scale shows (on the right) in dark blue the found RMSD minimum. In the bottom two examples of hyper-surface formed with the hyper-parameters and $log(RMSD)$ for visibility purposes. A black dot (.) marks the minimum.}
\label{fig:hype_param_env}
\end{figure}

The KRR method depends on two so-called hyper-parameters: $\lambda$ to control the regularization factor of the kernel $\mx K$, and $\alpha$ which controls the radius of inclusion/similarity for the kernel function. Both have a tremendous impact on the process behaviour and the best combination for each application needs to be previously found. For this purpose we conduct a grid optimization for each block-class: using a linear distribution of $85$ values of $\alpha$ in the range $[10^{-3}, 3000]$ and a linear distribution of $25$ values of $\lambda$ in the range $[10^{-3}, 1]$. Some blocks-class found their minimum at $\alpha = 2500$. This behaviour came from the fact that KRR gives the same weight to all blocks in a certain block-class as it needs all possible information to produce the best results. Most blocks require less data because they have a small deviation in their scalar product values. The values of $\lambda$ behave as a compensation for how far from the optimum value the $\alpha$ hyper-parameter is. But it can not correct by itself the deviation in the Root Mean Squared Distance (RMSD) produced by wrong $\alpha$ values. As soon $\alpha$ move closer to the optimum value, $\lambda$ lose his influence as long as the values stays in the range $(0, 1)$. However, the values of $\alpha = 10$ and $\lambda = 0.9$ resulted from $\tset$ provided an appropriate trend of improvement of the RMSD  for all block-classes. Figure \ref{fig:hype_param_env} shows some results of this optimisation.

\begin{scheme}
\centering
{\scalebox{0.8}{%
\input{images/learn_diag1}}}
\caption{Flow diagram for the prediction of a new coming molecule representation, from which the algorithm can extracts connectivity (e.g. codified by SMILES), selects the necessary data from the training set $\tset$ to train the KRR method and produces blocks Cartesian coordinate.}
\label{fig:flow_dig}
\end{scheme}

A new set of Cartesian coordinates can be predicted with our algorithm as outlined below. Scheme \ref{fig:flow_dig} shows the flow diagram. A SMILES \cite{ml_129} representation of the molecule is taken as input. It is then analyzed to extract connectivity, the blocks and their associated classes. The corresponding e-states are then computed. The learning data-set is built by using the block-classes to compute the scalar product and the e-states in order to train the KRR algorithm. The newly parameterized KRR is then used to predict each scalar product. Finally, the blocks are reconstructed again from the scalar product predictions to a Cartesian coordinate output.\cite{ml_121, ml_122} (The formulation used to reconstruct from scalar products to Cartesian coordinates is presented as an Appendix)

\section{Joining blocks.}

The RDKit \cite{ml_146} code was modified for the purpose of joining blocks. Geometry reconstruction by this kit implements the ETKDG\cite{ml_30, ml_31, ml_84} (Experimental Torsion Knowledge Distance Geometry) algorithm, which depends on a data base of torsion angles extracted from other sources, mostly experimental data. Using the same approach proposed by \emph{Scharfer, C.; Schulz-Gasch, T., et al.}\cite{ml_30} we extracted a new torsion angle data base from the ground 8CONF.
This newly \emph{modified ETKDG} (Theoretical-Torsion Knowledge Distance Geometry and Machine Learning, TTKDG-ML from now on) was used to produce reasonable conformer geometries with the input of predicted ML blocks.

\section{Results and discussions}\label{sec:rslt}
Two experiments were performed with the $\cset$ testing set. 
The first consists on block's prediction. Such control group was divided in block-classes to predict all scalar products followed by the reconstruction of their coordinates. Values of terms in the scalar product matrices $\mx B$ are obtained individually as each one gives a specific information. The diagonal describes bond lengths and off-diagonal terms provide information on angles. The first attempt to predict vectors, such as the upper triangular part of a matrix, resulted problematic, mostly because KRR does not particularize each unique component of the matrix and distribute the errors among all values.

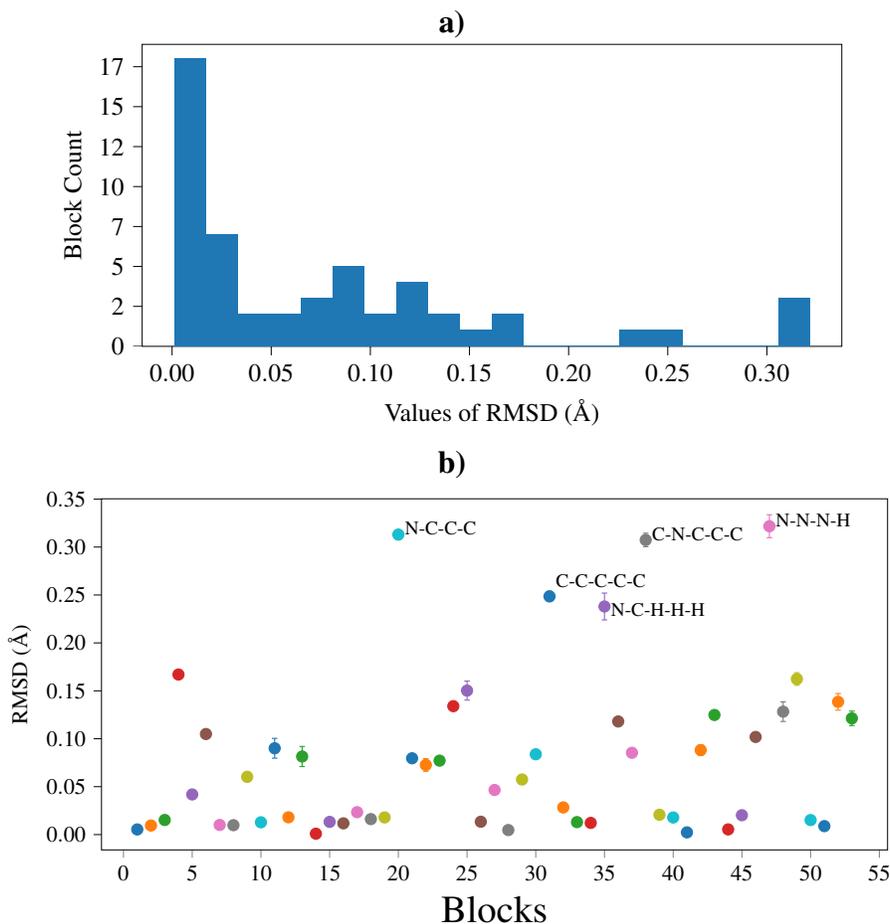
\begin{figure}[ht!]
\centering
\begin{tabular}[c]{ c }
\textbf{a)}\\
{\scalebox{0.8}{%
    \input{./images/error_block_hist_tikz}}}\\
\textbf{b)}\\
{\scalebox{0.7}{%
    \input{./images/error_block_tikz}}}
\end{tabular}
\caption{a) Histograms with the mean RMSD distribution for the block-classes present in the $450$ molecules of $\cset$. b) Details of block-classes where those with RMSD errors above $0.3$ \r A are explicit. Each color identify a block-class present in the test group $\cset$.}
\label{fig:env_error_block_hist}
\end{figure}

Results of this computing experiment appear in Figure \ref{fig:env_error_block_hist}. RMSD histograms show how good the prediction/reconstruction of blocks really is, with most block-classes showing errors under $0.05$ \r{A}. However, methods must be also evaluated by their worst cases as some block-classes rose them to around $0.3$ \r{A}. Among the highest RMSD's values, two different kinds of problematic blocks are found: those belonging to a group on which there are not enough data to properly train the KRR algorithm (e.g. \emph{N-N-N-H} with only 7 occurrences) and those belonging to molecules where the blocks are located inside rings that tend to distort the bond lengths and angles (e.g. \emph{C-C-C-C-C} which is a very abundant block-class with multiple combinations).

The second experiment with the $\cset$ testing set involved the complete rebuilding of the molecules. After predicting the blocks and reconstructing the coordinates, TTKDG-ML procedure was fed with the distance matrices of each block. Some restrictions were put in place in the form of an arbitrary high number (the decimal value $10$ was used) to be interpreted as a weight by the algorithm. So, the better conformers were left within those non modifying the predicted distances by the KRR method. The best is selected by the minimun RMSD value.

\begin{figure}[ht!]
\centering
\begin{tabular}[c]{ c }
{\scalebox{0.8}{%
    \input{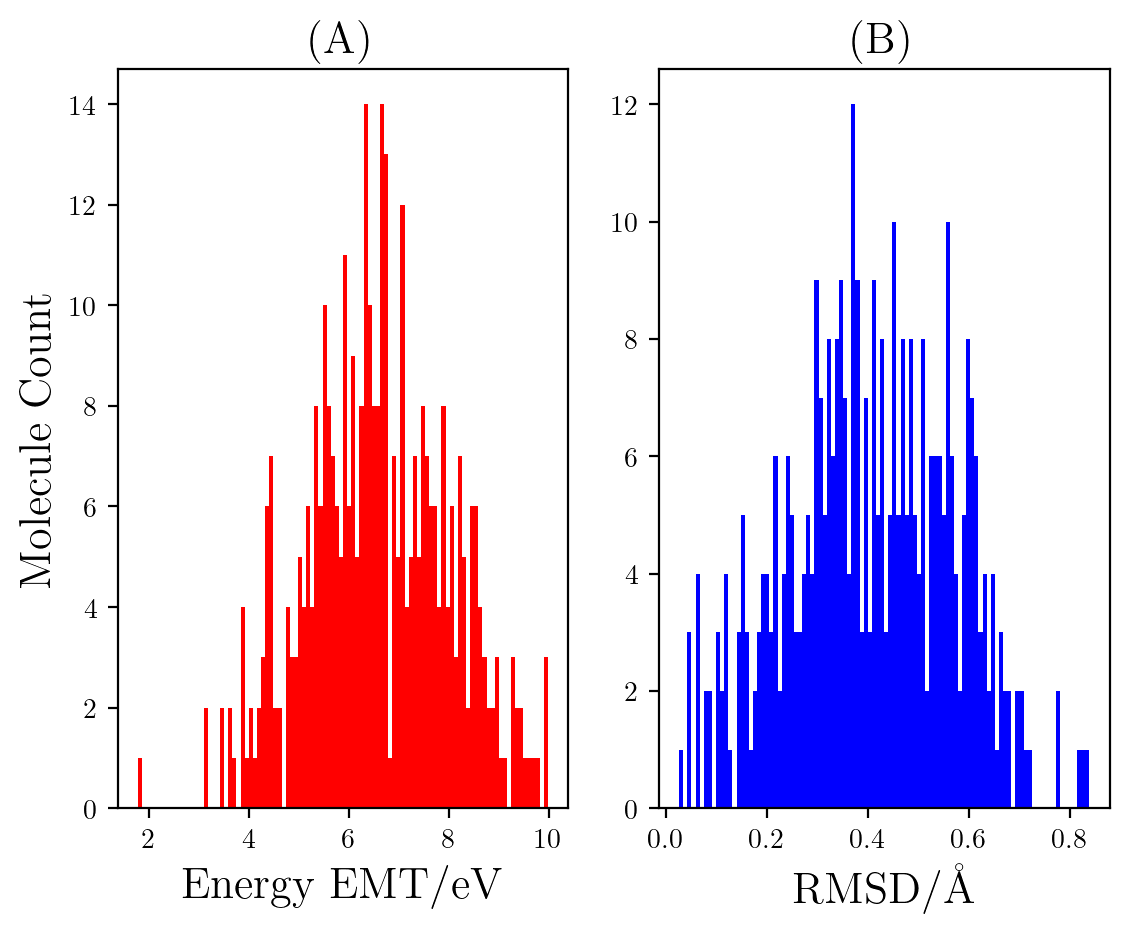}}}\\
{\scalebox{0.8}{%
    \input{./images/rmsd_error_mol_set}}}
\end{tabular}

\caption{The first row (from the top) shows two histograms with the EMT (plot \emph{A}) energies values and RMSD's (plot \emph{B}) obtained after the complete reconstruction of molecules contained in the test-set $\cset$. The similarities of the histogram shapes are noticeable as the EMT express bigger energy values when molecular assembling is more complex. As points of reference, the second row shows the EMT energy of the original molecules as present in the test-set $\cset$ (plot \emph{C}) to be compared with those in (plot \emph{A}). The RMSD's (plot \emph{D}) between original molecules present in test-set $\cset$ and their conformers obtained by the ETKDG in RDkit show very encouraging results.}
\label{fig:rslt_conf_compl}
\end{figure}

Histograms comparing both RMSD results and potential energy calculations using a simple \emph{Effective Medium Theory} (EMT)\cite {ml_100, ase_paper} for the $450$ molecules of the test-set are shown in Figure \ref{fig:rslt_conf_compl}. This method consists of an expression derived for obtaining a value related with the total energy of a system of interacting atoms. It is based on an ansatz for the total electron density of the system as a superposition of atom densities taken from calculations for the atoms embedded in a homogeneous electron gas. We use these energy values here to illustrate molecular complexity compared with total RMSD predictions. As EMT was developed only for solids and not for molecules, the resulting values of energy must be only taken as just a reference. It can be realized that there are eight molecules from the test-set on which the TTKDG does not perform well. It must be originated in the fact that they contain one or more blocks with poor prediction results (See Figure \ref{fig:env_error_block_hist}) such as \emph{N-C-C-C.Tr} and \emph{C-C-C-C-C.Td}. The results show that the proposed TTKDG-ML  performs better than the original ETKDG implemeted in RDkit (See Figure \ref{fig:rslt_conf_compl} \emph{B} and \emph{D}). 

\section{Conclusions}

To predict accurate geometries of molecules is a complex problem able to be multiple approached. A possible solution is described here by using a machine learning tool based on a Kernel Ridge Regression routine. The described procedure departing from bonded atom blocks results in a very flexible and easy-to-expand way to describe the canonical geometry of an atom in a molecule and its environment. The descriptor used showed a very good correlation with bond distances and angles when the coordinates were transformed into scalar products. RMSD values obtained from the experiments performed on blocks support our decision on this kind of descriptors and validate the block structure. To solve most of the \emph{problematic block} predictions will be required to add more related training data because the involved atoms in such blocks are rare in common molecules and therefore scarce in the training set. For those atoms belonging to rings, a better data treatment could be a solution. But this needs to be exercised with caution, because too much pruning of the data base could lead to \emph{over-fitting} the KRR method results.

For approaching the final purpose of getting a complete machine learning method, the selected descriptor was tested for torsion angles. It completely failed. The symmetries involved when joining two blocks were very difficult to grasp by the e-state with the only use of the KRR algorithm. Even the selection of the correct torsion angle for the learning process proved to be non-trivial. Several difficulties were faced in this area especially with linear carbon chains where the definition rules resulted to be ambiguous. Nevertheless, the TTKDG-ML from an \emph{ab-initio} database proved to provide very consistent conformers, i.e. those related to extreme torsion angles. The used weights helped to maintain the obtained results by the KRR method.

All optimized geometries used can be downloaded from \url{http://www.quantum-machine.org/datasets/},

\begin{acknowledgement}
This work was supported by the European Research Council (ERC-2015-AdG694097), the Cluster of Excellence 'CUI: Advanced Imaging of Matter' of the Deutsche Forschungsgemeinschaft (DFG) - EXC 2056 - project ID 390715994, Grupos Consolidados (IT1249-19) and  the SFB925 "Light induced dynamics and control of correlated quantum systems”. We kindly recognize the partial support of the project ID PN223LH010-002 "Inteligencia Artificial Aplicada. Espectroscopía y Bioactividad" of the Cuban Ministry of Science, Technology and Environment as well as the overall support given to LAMC by the Universidad de La Habana and the Donostia International Physics Center.

\end{acknowledgement}

\section{Appendix}

\subsection{Scalar product formulations and reconstruction}\label{app:sp}

For the sake of argument clarity, tetravalent blocks are taken here as an example for the scalar product formulation. Each block has five atoms, or a total of 15 coordinates. Rotations and translations account for six degrees of freedom (DOF's), leaving nine of them to consider.

Translations were automatically eliminated by measuring neighbouring atom coordinates as displacements from the central atom. The elimination of rotational dependencies were achieved by forming the matrix of all scalar products between the nonzero position vectors $X_{ac}$, where $a=1\ldots4$ specifies direct neighbouring atoms and $c$ is one of x, y, z:

\begin{align}
\mx B = \mx X \mx X^T = \left[
\begin{matrix}
  a^2 & \ve a \cdot \ve b & \ve a \cdot \ve c & \ve a \cdot \ve d \\
  \ve b \cdot \ve a & b^2 & \ve b \cdot \ve c & \ve b \cdot \ve d\\
  \ve c \cdot \ve a & \ve c \cdot \ve b & c^2 & \ve c \cdot \ve d\\
  \ve d \cdot \ve a & \ve d \cdot \ve b & \ve d \cdot \ve c & d^2\\
\end{matrix}\right]
\end{align}

The bond lengths can be obtained directly from the four diagonal elements. The information about the angle is provided by the six distinct scalar products. The system is over determined by one DOF. However the matrix has only rank 3 because it was formed from a $4\times3$ matrix. Therefore one eigenvalue is zero, leaving a three-dimensional eigenspace which spans the nine DOF's remaining as well determined.

The matrix can be rewritten by using its eigendecomposition as:
\begin{align}\label{eq:brecostruct}
  \mx B = \mx Q \mx \Lambda \mx Q^T
  = (\mx Q \mx \Lambda^{1/2})((\mx \Lambda^{1/2})^T \mx Q^T)
  = (\mx Q \mx \Lambda^{1/2})(\mx Q \mx \Lambda^{1/2})^T
  = \bar{\mx X}' (\bar{\mx X}')^T
\end{align}

This defines a matrix $\bar{\mx X}' = \mx Q \mx \Lambda^{1/2}$ which produces the same scalar products.  The matrix $\bar{\mx X}'$ will contain one row of zeros, corresponding to the eigenvalue 0.

This row/column is discarded, and the remaining matrix $\mx R'$ will be an eligible set of reconstructed positions.  In weeding out numerical garbage, the lowest eigenvalue is always discarded even if it is not exactly zero. The remaining eigenvalues must be positive since the scalar product is positive definite.

\bibliography{ml_geometries}
\end{document}

%% file: images/mol_block_example.tex
\begin{tikzpicture}[]

\node(mol)[]{\LARGE 
        \chemfig{\chemabove{C}{\scriptstyle 1}(-[3]\chemabove{H}{\scriptstyle 4})(<[5]\chemabove{H}{\scriptstyle 5})(<:[7]\chemabove{H}{\scriptstyle 6})
                 -[:30,1.2]\chemabove{C}{\scriptstyle 2}(<[3]\chemabove{H}{\scriptstyle 7})(<:[1]\chemabove{H}{\scriptstyle 8})
                 -[:-30, 1.2]\chemabove{C}{\scriptstyle 3}(-[1]\chemabove{H}{\scriptstyle 9})(<[5]\chemabove{H}{\scriptstyle 10})(<:[7]\chemabove{H}{\scriptstyle 11})}};

\draw[red, ultra thick] (-4.5, -2.5) -- (0.3, -2.5) -- (0.3, 1.5) -- (-4.5, 1.5) -- (-4.5, -2.5);
\draw [blue, ultra thick](-0.3, -2.5) -- (4.5, -2.5) -- (4.5, 1.5) -- (-0.3, 1.5) -- (-0.3, -2.5);
\draw [green, ultra thick](-3, -1) -- (3, -1) -- (3, 3) -- (-3, 3) -- (-3, -1);



\end{tikzpicture}

%% file: images/block_dist_tikz.tex
\begin{tikzpicture}

\definecolor{color0}{rgb}{0.5,0,1}
\definecolor{color1}{rgb}{0.492156862745098,0.0123196595352384,0.999981027348727}
\definecolor{color2}{rgb}{0.484313725490196,0.024637449195382,0.999924110114831}
\definecolor{color3}{rgb}{0.476470588235294,0.0369514993891449,0.999829250458053}
\definecolor{color4}{rgb}{0.468627450980392,0.0492599410928169,0.999696451977872}
\definecolor{color5}{rgb}{0.46078431372549,0.0615609061339428,0.999525719713366}
\definecolor{color6}{rgb}{0.452941176470588,0.073852527474874,0.999317060143023}
\definecolor{color7}{rgb}{0.445098039215686,0.086132939496146,0.999070481184493}
\definecolor{color8}{rgb}{0.437254901960784,0.0984002782796427,0.99878599219429}
\definecolor{color9}{rgb}{0.429411764705882,0.110652681891501,0.998463603967434}
\definecolor{color10}{rgb}{0.42156862745098,0.122888290664714,0.998103328737044}
\definecolor{color11}{rgb}{0.413725490196078,0.135105247481393,0.997705180173873}
\definecolor{color12}{rgb}{0.405882352941176,0.147301698054637,0.997269173385788}
\definecolor{color13}{rgb}{0.398039215686275,0.159475791209981,0.996795324917199}
\definecolor{color14}{rgb}{0.390196078431373,0.17162567916636,0.996283652748429}
\definecolor{color15}{rgb}{0.382352941176471,0.18374951781657,0.995734176295034}
\definecolor{color16}{rgb}{0.374509803921569,0.195845467007167,0.995146916407064}
\definecolor{color17}{rgb}{0.366666666666667,0.207911690817759,0.994521895368273}
\definecolor{color18}{rgb}{0.358823529411765,0.219946357839669,0.993859136895274}
\definecolor{color19}{rgb}{0.350980392156863,0.231947641453898,0.993158666136636}
\definecolor{color20}{rgb}{0.343137254901961,0.243913720108377,0.992420509671936}
\definecolor{color21}{rgb}{0.335294117647059,0.255842777594436,0.991644695510743}
\definecolor{color22}{rgb}{0.327450980392157,0.267733003322468,0.99083125309156}
\definecolor{color23}{rgb}{0.319607843137255,0.279582592596744,0.989980213280707}
\definecolor{color24}{rgb}{0.311764705882353,0.291389746889325,0.989091608371146}
\definecolor{color25}{rgb}{0.303921568627451,0.303152674113044,0.988165472081259}
\definecolor{color26}{rgb}{0.296078431372549,0.314869588893508,0.987201839553569}
\definecolor{color27}{rgb}{0.288235294117647,0.326538712840083,0.986200747353403}
\definecolor{color28}{rgb}{0.280392156862745,0.338158274815817,0.985162233467507}
\definecolor{color29}{rgb}{0.272549019607843,0.349726511206261,0.984086337302604}
\definecolor{color30}{rgb}{0.264705882352941,0.361241666187153,0.982973099683902}
\definecolor{color31}{rgb}{0.256862745098039,0.372701991990914,0.981822562853537}
\definecolor{color32}{rgb}{0.249019607843137,0.384105749171926,0.980634770468978}
\definecolor{color33}{rgb}{0.241176470588235,0.395451206870543,0.979409767601366}
\definecolor{color34}{rgb}{0.233333333333333,0.4067366430758,0.978147600733806}
\definecolor{color35}{rgb}{0.225490196078431,0.417960344886783,0.976848317759601}
\definecolor{color36}{rgb}{0.217647058823529,0.429120608772609,0.975511967980437}
\definecolor{color37}{rgb}{0.209803921568627,0.440215740830987,0.97413860210451}
\definecolor{color38}{rgb}{0.201960784313725,0.451244057045323,0.972728272244605}
\definecolor{color39}{rgb}{0.194117647058824,0.462203883540313,0.971281031916114}
\definecolor{color40}{rgb}{0.186274509803922,0.47309355683601,0.969796936035009}
\definecolor{color41}{rgb}{0.17843137254902,0.483911424100302,0.968276040915759}
\definecolor{color42}{rgb}{0.170588235294118,0.494655843399779,0.966718404269187}
\definecolor{color43}{rgb}{0.162745098039216,0.505325183948948,0.965124085200289}
\definecolor{color44}{rgb}{0.154901960784314,0.515917826357751,0.963493144205983}
\definecolor{color45}{rgb}{0.147058823529412,0.526432162877356,0.961825643172819}
\definecolor{color46}{rgb}{0.13921568627451,0.53686659764418,0.960121645374628}
\definecolor{color47}{rgb}{0.131372549019608,0.547219546922111,0.958381215470122}
\definecolor{color48}{rgb}{0.123529411764706,0.557489439342886,0.956604419500441}
\definecolor{color49}{rgb}{0.115686274509804,0.56767471614459,0.954791324886644}
\definecolor{color50}{rgb}{0.107843137254902,0.577773831408251,0.952942000427157}
\definecolor{color51}{rgb}{0.1,0.587785252292473,0.951056516295154}
\definecolor{color52}{rgb}{0.092156862745098,0.597707459266094,0.949134944035901}
\definecolor{color53}{rgb}{0.0843137254901961,0.607538946338817,0.94717735656404}
\definecolor{color54}{rgb}{0.0764705882352941,0.617278221289793,0.94518382816082}
\definecolor{color55}{rgb}{0.0686274509803921,0.626923805894106,0.943154434471277}
\definecolor{color56}{rgb}{0.0607843137254902,0.636474236147141,0.941089252501372}
\definecolor{color57}{rgb}{0.0529411764705883,0.645928062486787,0.938988360615057}
\definecolor{color58}{rgb}{0.0450980392156863,0.655283850013454,0.936851838531311}
\definecolor{color59}{rgb}{0.0372549019607843,0.664540178707858,0.934679767321111}
\definecolor{color60}{rgb}{0.0294117647058824,0.673695643646557,0.932472229404356}
\definecolor{color61}{rgb}{0.0215686274509804,0.682748855215185,0.93022930854674}
\definecolor{color62}{rgb}{0.0137254901960784,0.69169843931937,0.927951089856575}
\definecolor{color63}{rgb}{0.00588235294117645,0.700543037593291,0.925637659781556}
\definecolor{color64}{rgb}{0.00196078431372548,0.709281307605853,0.923289106105489}
\definecolor{color65}{rgb}{0.00980392156862742,0.717911923064442,0.920905517944954}
\definecolor{color66}{rgb}{0.0176470588235293,0.726433574016224,0.918486985745923}
\definecolor{color67}{rgb}{0.0254901960784314,0.734844967046976,0.916033601280334}

\begin{axis}[
tick align=outside,
tick pos=left,
width=1\textwidth,
height=0.5\textwidth,
xticklabel=\empty,
xmin=-2.79, xmax=71.79,
y grid style={white!69.0196078431373!black},
ylabel={Number of blocks in database},
ymin=0, ymax=22000,
ytick style={color=black},
ytick={0,10500,20000,23000},
yticklabels={\(\displaystyle 0\),
            \(\displaystyle 10500\),
             \(\displaystyle 20000\),
             \(\displaystyle 23000\)}
]
\draw[draw=color0,fill=color0] (axis cs:0.6,0) rectangle (axis cs:1.4,20274);
\draw[draw=color1,fill=color1] (axis cs:1.6,0) rectangle (axis cs:2.4,14511);
\draw[draw=color2,fill=color2] (axis cs:2.6,0) rectangle (axis cs:3.4,11091);
\draw[draw=color3,fill=color3] (axis cs:3.6,0) rectangle (axis cs:4.4,10553);
\draw[draw=color4,fill=color4] (axis cs:4.6,0) rectangle (axis cs:5.4,9413);
\draw[draw=color5,fill=color5] (axis cs:5.6,0) rectangle (axis cs:6.4,8299);
\draw[draw=color6,fill=color6] (axis cs:6.6,0) rectangle (axis cs:7.4,7394);
\draw[draw=color7,fill=color7] (axis cs:7.6,0) rectangle (axis cs:8.4,5631);
\draw[draw=color8,fill=color8] (axis cs:8.6,0) rectangle (axis cs:9.4,5065);
\draw[draw=color9,fill=color9] (axis cs:9.6,0) rectangle (axis cs:10.4,5025);
\draw[draw=color10,fill=color10] (axis cs:10.6,0) rectangle (axis cs:11.4,4438);
\draw[draw=color11,fill=color11] (axis cs:11.6,0) rectangle (axis cs:12.4,4067);
\draw[draw=color12,fill=color12] (axis cs:12.6,0) rectangle (axis cs:13.4,3343);
\draw[draw=color13,fill=color13] (axis cs:13.6,0) rectangle (axis cs:14.4,3217);
\draw[draw=color14,fill=color14] (axis cs:14.6,0) rectangle (axis cs:15.4,2985);
\draw[draw=color15,fill=color15] (axis cs:15.6,0) rectangle (axis cs:16.4,2897);
\draw[draw=color16,fill=color16] (axis cs:16.6,0) rectangle (axis cs:17.4,2861);
\draw[draw=color17,fill=color17] (axis cs:17.6,0) rectangle (axis cs:18.4,2776);
\draw[draw=color18,fill=color18] (axis cs:18.6,0) rectangle (axis cs:19.4,2752);
\draw[draw=color19,fill=color19] (axis cs:19.6,0) rectangle (axis cs:20.4,2716);
\draw[draw=color20,fill=color20] (axis cs:20.6,0) rectangle (axis cs:21.4,2281);
\draw[draw=color21,fill=color21] (axis cs:21.6,0) rectangle (axis cs:22.4,2146);
\draw[draw=color22,fill=color22] (axis cs:22.6,0) rectangle (axis cs:23.4,2107);
\draw[draw=color23,fill=color23] (axis cs:23.6,0) rectangle (axis cs:24.4,1871);
\draw[draw=color24,fill=color24] (axis cs:24.6,0) rectangle (axis cs:25.4,1797);
\draw[draw=color25,fill=color25] (axis cs:25.6,0) rectangle (axis cs:26.4,1705);
\draw[draw=color26,fill=color26] (axis cs:26.6,0) rectangle (axis cs:27.4,1435);
\draw[draw=color27,fill=color27] (axis cs:27.6,0) rectangle (axis cs:28.4,1414);
\draw[draw=color28,fill=color28] (axis cs:28.6,0) rectangle (axis cs:29.4,1407);
\draw[draw=color29,fill=color29] (axis cs:29.6,0) rectangle (axis cs:30.4,1397);
\draw[draw=color30,fill=color30] (axis cs:30.6,0) rectangle (axis cs:31.4,1367);
\draw[draw=color31,fill=color31] (axis cs:31.6,0) rectangle (axis cs:32.4,1229);
\draw[draw=color32,fill=color32] (axis cs:32.6,0) rectangle (axis cs:33.4,999);
\draw[draw=color33,fill=color33] (axis cs:33.6,0) rectangle (axis cs:34.4,905);
\draw[draw=color34,fill=color34] (axis cs:34.6,0) rectangle (axis cs:35.4,903);
\draw[draw=color35,fill=color35] (axis cs:35.6,0) rectangle (axis cs:36.4,717);
\draw[draw=color36,fill=color36] (axis cs:36.6,0) rectangle (axis cs:37.4,709);
\draw[draw=color37,fill=color37] (axis cs:37.6,0) rectangle (axis cs:38.4,643);
\draw[draw=color38,fill=color38] (axis cs:38.6,0) rectangle (axis cs:39.4,446);
\draw[draw=color39,fill=color39] (axis cs:39.6,0) rectangle (axis cs:40.4,350);
\draw[draw=color40,fill=color40] (axis cs:40.6,0) rectangle (axis cs:41.4,294);
\draw[draw=color41,fill=color41] (axis cs:41.6,0) rectangle (axis cs:42.4,273);
\draw[draw=color42,fill=color42] (axis cs:42.6,0) rectangle (axis cs:43.4,234);
\draw[draw=color43,fill=color43] (axis cs:43.6,0) rectangle (axis cs:44.4,232);
\draw[draw=color44,fill=color44] (axis cs:44.6,0) rectangle (axis cs:45.4,231);
\draw[draw=color45,fill=color45] (axis cs:45.6,0) rectangle (axis cs:46.4,209);
\draw[draw=color46,fill=color46] (axis cs:46.6,0) rectangle (axis cs:47.4,184);
\draw[draw=color47,fill=color47] (axis cs:47.6,0) rectangle (axis cs:48.4,122);
\draw[draw=color48,fill=color48] (axis cs:48.6,0) rectangle (axis cs:49.4,100);
\draw[draw=color49,fill=color49] (axis cs:49.6,0) rectangle (axis cs:50.4,99);
\draw[draw=color50,fill=color50] (axis cs:50.6,0) rectangle (axis cs:51.4,90);
\draw[draw=color51,fill=color51] (axis cs:51.6,0) rectangle (axis cs:52.4,87);
\draw[draw=color52,fill=color52] (axis cs:52.6,0) rectangle (axis cs:53.4,86);
\draw[draw=color53,fill=color53] (axis cs:53.6,0) rectangle (axis cs:54.4,62);
\draw[draw=color54,fill=color54] (axis cs:54.6,0) rectangle (axis cs:55.4,35);
\draw[draw=color55,fill=color55] (axis cs:55.6,0) rectangle (axis cs:56.4,21);
\draw[draw=color56,fill=color56] (axis cs:56.6,0) rectangle (axis cs:57.4,18);
\draw[draw=color57,fill=color57] (axis cs:57.6,0) rectangle (axis cs:58.4,10);
\draw[draw=color58,fill=color58] (axis cs:58.6,0) rectangle (axis cs:59.4,8);
\draw[draw=color59,fill=color59] (axis cs:59.6,0) rectangle (axis cs:60.4,8);
\draw[draw=color60,fill=color60] (axis cs:60.6,0) rectangle (axis cs:61.4,6);
\draw[draw=color61,fill=color61] (axis cs:61.6,0) rectangle (axis cs:62.4,4);
\draw[draw=color62,fill=color62] (axis cs:62.6,0) rectangle (axis cs:63.4,1);
\draw[draw=color63,fill=color63] (axis cs:63.6,0) rectangle (axis cs:64.4,1);
\draw[draw=color64,fill=color64] (axis cs:64.6,0) rectangle (axis cs:65.4,1);
\draw[draw=color65,fill=color65] (axis cs:65.6,0) rectangle (axis cs:66.4,1);
\draw[draw=color66,fill=color66] (axis cs:66.6,0) rectangle (axis cs:67.4,1);
\draw[draw=color67,fill=color67] (axis cs:67.6,0) rectangle (axis cs:68.4,1);
\draw[->,draw=black] (axis cs:10,20264) -- (axis cs:1,20274);
\draw (axis cs:10,20264) node[
  scale=0.5,
  anchor=base west,
  text=black,
  rotate=0.0
]{\LARGE C-C-C-H-H};
\draw[->,draw=black] (axis cs:11,14501) -- (axis cs:2,14511);
\draw (axis cs:11,14501) node[
  scale=0.5,
  anchor=base west,
  text=black,
  rotate=0.0
]{\LARGE C-C-H-H-H};
\draw[->,draw=black] (axis cs:40,18000) -- (axis cs:63,1);
\draw (axis cs:40,18000) node[
  scale=0.5,
  anchor=base west,
  text=black,
  rotate=0.0
]{\LARGE O-H-H};
\draw[->,draw=black] (axis cs:55,20275) -- (axis cs:64,1);
\draw (axis cs:55,20275) node[
  scale=0.5,
  anchor=base west,
  text=black,
  rotate=0.0
]{\LARGE C-O-H-H};
\end{axis}

\end{tikzpicture}

%% file: images/env_stats_e_state_tikz.tex
\begin{tikzpicture}

\definecolor{color0}{rgb}{0.12156862745098,0.466666666666667,0.705882352941177}
\definecolor{color1}{rgb}{1,0.498039215686275,0.0549019607843137}
\definecolor{color2}{rgb}{0.172549019607843,0.627450980392157,0.172549019607843}

\begin{groupplot}[group style={group size=3 by 1}]
\nextgroupplot[
tick align=outside,
tick pos=left,
title={C},
x grid style={white!69.0196078431373!black},
xmin=-1.71264073129252, xmax=8.1598998015873,
xtick style={color=black},
xlabel={\emph{e-state Values}},
ylabel={Atoms Count},
xtick={-5,0,5,10},
xticklabels={
  \(\displaystyle {−5}\),
  \(\displaystyle {0}\),
  \(\displaystyle {5}\),
  \(\displaystyle {10}\)
},
y grid style={white!69.0196078431373!black},
ymin=0, ymax=65.1,
ytick style={color=black},
xlabel={\emph{e-state Values}},
ytick={0,10,20,30,40,50,60,70},
yticklabels={
  \(\displaystyle {0}\),
  \(\displaystyle {10}\),
  \(\displaystyle {20}\),
  \(\displaystyle {30}\),
  \(\displaystyle {40}\),
  \(\displaystyle {50}\),
  \(\displaystyle {60}\),
  \(\displaystyle {70}\)
}
]
\draw[draw=none,fill=color0] (axis cs:-1.26388888888889,0) rectangle (axis cs:-1.17413852040816,1);
\draw[draw=none,fill=color0] (axis cs:-1.17413852040816,0) rectangle (axis cs:-1.08438815192744,0);
\draw[draw=none,fill=color0] (axis cs:-1.08438815192744,0) rectangle (axis cs:-0.994637783446712,0);
\draw[draw=none,fill=color0] (axis cs:-0.994637783446712,0) rectangle (axis cs:-0.904887414965987,1);
\draw[draw=none,fill=color0] (axis cs:-0.904887414965986,0) rectangle (axis cs:-0.815137046485261,0);
\draw[draw=none,fill=color0] (axis cs:-0.815137046485261,0) rectangle (axis cs:-0.725386678004535,2);
\draw[draw=none,fill=color0] (axis cs:-0.725386678004535,0) rectangle (axis cs:-0.63563630952381,1);
\draw[draw=none,fill=color0] (axis cs:-0.63563630952381,0) rectangle (axis cs:-0.545885941043084,6);
\draw[draw=none,fill=color0] (axis cs:-0.545885941043084,0) rectangle (axis cs:-0.456135572562359,3);
\draw[draw=none,fill=color0] (axis cs:-0.456135572562358,0) rectangle (axis cs:-0.366385204081633,4);
\draw[draw=none,fill=color0] (axis cs:-0.366385204081633,0) rectangle (axis cs:-0.276634835600907,6);
\draw[draw=none,fill=color0] (axis cs:-0.276634835600907,0) rectangle (axis cs:-0.186884467120181,5);
\draw[draw=none,fill=color0] (axis cs:-0.186884467120181,0) rectangle (axis cs:-0.097134098639456,7);
\draw[draw=none,fill=color0] (axis cs:-0.097134098639456,0) rectangle (axis cs:-0.00738373015873028,11);
\draw[draw=none,fill=color0] (axis cs:-0.00738373015873028,0) rectangle (axis cs:0.0823666383219952,11);
\draw[draw=none,fill=color0] (axis cs:0.0823666383219952,0) rectangle (axis cs:0.172117006802721,8);
\draw[draw=none,fill=color0] (axis cs:0.172117006802721,0) rectangle (axis cs:0.261867375283447,9);
\draw[draw=none,fill=color0] (axis cs:0.261867375283447,0) rectangle (axis cs:0.351617743764172,15);
\draw[draw=none,fill=color0] (axis cs:0.351617743764172,0) rectangle (axis cs:0.441368112244898,21);
\draw[draw=none,fill=color0] (axis cs:0.441368112244898,0) rectangle (axis cs:0.531118480725623,14);
\draw[draw=none,fill=color0] (axis cs:0.531118480725623,0) rectangle (axis cs:0.620868849206349,33);
\draw[draw=none,fill=color0] (axis cs:0.620868849206349,0) rectangle (axis cs:0.710619217687075,28);
\draw[draw=none,fill=color0] (axis cs:0.710619217687075,0) rectangle (axis cs:0.8003695861678,35);
\draw[draw=none,fill=color0] (axis cs:0.8003695861678,0) rectangle (axis cs:0.890119954648526,36);
\draw[draw=none,fill=color0] (axis cs:0.890119954648526,0) rectangle (axis cs:0.979870323129252,38);
\draw[draw=none,fill=color0] (axis cs:0.979870323129252,0) rectangle (axis cs:1.06962069160998,32);
\draw[draw=none,fill=color0] (axis cs:1.06962069160998,0) rectangle (axis cs:1.1593710600907,42);
\draw[draw=none,fill=color0] (axis cs:1.1593710600907,0) rectangle (axis cs:1.24912142857143,43);
\draw[draw=none,fill=color0] (axis cs:1.24912142857143,0) rectangle (axis cs:1.33887179705215,48);
\draw[draw=none,fill=color0] (axis cs:1.33887179705215,0) rectangle (axis cs:1.42862216553288,51);
\draw[draw=none,fill=color0] (axis cs:1.42862216553288,0) rectangle (axis cs:1.51837253401361,55);
\draw[draw=none,fill=color0] (axis cs:1.51837253401361,0) rectangle (axis cs:1.60812290249433,48);
\draw[draw=none,fill=color0] (axis cs:1.60812290249433,0) rectangle (axis cs:1.69787327097506,49);
\draw[draw=none,fill=color0] (axis cs:1.69787327097506,0) rectangle (axis cs:1.78762363945578,53);
\draw[draw=none,fill=color0] (axis cs:1.78762363945578,0) rectangle (axis cs:1.87737400793651,52);
\draw[draw=none,fill=color0] (axis cs:1.87737400793651,0) rectangle (axis cs:1.96712437641723,44);
\draw[draw=none,fill=color0] (axis cs:1.96712437641723,0) rectangle (axis cs:2.05687474489796,57);
\draw[draw=none,fill=color0] (axis cs:2.05687474489796,0) rectangle (axis cs:2.14662511337868,48);
\draw[draw=none,fill=color0] (axis cs:2.14662511337868,0) rectangle (axis cs:2.23637548185941,55);
\draw[draw=none,fill=color0] (axis cs:2.23637548185941,0) rectangle (axis cs:2.32612585034014,52);
\draw[draw=none,fill=color0] (axis cs:2.32612585034014,0) rectangle (axis cs:2.41587621882086,54);
\draw[draw=none,fill=color0] (axis cs:2.41587621882086,0) rectangle (axis cs:2.50562658730159,62);
\draw[draw=none,fill=color0] (axis cs:2.50562658730159,0) rectangle (axis cs:2.59537695578231,55);
\draw[draw=none,fill=color0] (axis cs:2.59537695578231,0) rectangle (axis cs:2.68512732426304,54);
\draw[draw=none,fill=color0] (axis cs:2.68512732426304,0) rectangle (axis cs:2.77487769274376,52);
\draw[draw=none,fill=color0] (axis cs:2.77487769274376,0) rectangle (axis cs:2.86462806122449,42);
\draw[draw=none,fill=color0] (axis cs:2.86462806122449,0) rectangle (axis cs:2.95437842970521,52);
\draw[draw=none,fill=color0] (axis cs:2.95437842970521,0) rectangle (axis cs:3.04412879818594,25);
\draw[draw=none,fill=color0] (axis cs:3.04412879818594,0) rectangle (axis cs:3.13387916666667,52);
\draw[draw=none,fill=color0] (axis cs:3.13387916666667,0) rectangle (axis cs:3.22362953514739,39);
\draw[draw=none,fill=color0] (axis cs:3.22362953514739,0) rectangle (axis cs:3.31337990362812,39);
\draw[draw=none,fill=color0] (axis cs:3.31337990362812,0) rectangle (axis cs:3.40313027210884,26);
\draw[draw=none,fill=color0] (axis cs:3.40313027210884,0) rectangle (axis cs:3.49288064058957,35);
\draw[draw=none,fill=color0] (axis cs:3.49288064058957,0) rectangle (axis cs:3.58263100907029,31);
\draw[draw=none,fill=color0] (axis cs:3.58263100907029,0) rectangle (axis cs:3.67238137755102,22);
\draw[draw=none,fill=color0] (axis cs:3.67238137755102,0) rectangle (axis cs:3.76213174603175,23);
\draw[draw=none,fill=color0] (axis cs:3.76213174603175,0) rectangle (axis cs:3.85188211451247,25);
\draw[draw=none,fill=color0] (axis cs:3.85188211451247,0) rectangle (axis cs:3.9416324829932,34);
\draw[draw=none,fill=color0] (axis cs:3.9416324829932,0) rectangle (axis cs:4.03138285147392,35);
\draw[draw=none,fill=color0] (axis cs:4.03138285147392,0) rectangle (axis cs:4.12113321995465,31);
\draw[draw=none,fill=color0] (axis cs:4.12113321995465,0) rectangle (axis cs:4.21088358843538,31);
\draw[draw=none,fill=color0] (axis cs:4.21088358843537,0) rectangle (axis cs:4.3006339569161,27);
\draw[draw=none,fill=color0] (axis cs:4.3006339569161,0) rectangle (axis cs:4.39038432539683,29);
\draw[draw=none,fill=color0] (axis cs:4.39038432539683,0) rectangle (axis cs:4.48013469387755,31);
\draw[draw=none,fill=color0] (axis cs:4.48013469387755,0) rectangle (axis cs:4.56988506235828,23);
\draw[draw=none,fill=color0] (axis cs:4.56988506235828,0) rectangle (axis cs:4.659635430839,15);
\draw[draw=none,fill=color0] (axis cs:4.659635430839,0) rectangle (axis cs:4.74938579931973,20);
\draw[draw=none,fill=color0] (axis cs:4.74938579931973,0) rectangle (axis cs:4.83913616780045,9);
\draw[draw=none,fill=color0] (axis cs:4.83913616780045,0) rectangle (axis cs:4.92888653628118,16);
\draw[draw=none,fill=color0] (axis cs:4.92888653628118,0) rectangle (axis cs:5.0186369047619,9);
\draw[draw=none,fill=color0] (axis cs:5.0186369047619,0) rectangle (axis cs:5.10838727324263,2);
\draw[draw=none,fill=color0] (axis cs:5.10838727324263,0) rectangle (axis cs:5.19813764172336,0);
\draw[draw=none,fill=color0] (axis cs:5.19813764172336,0) rectangle (axis cs:5.28788801020408,0);
\draw[draw=none,fill=color0] (axis cs:5.28788801020408,0) rectangle (axis cs:5.37763837868481,0);
\draw[draw=none,fill=color0] (axis cs:5.37763837868481,0) rectangle (axis cs:5.46738874716553,0);
\draw[draw=none,fill=color0] (axis cs:5.46738874716553,0) rectangle (axis cs:5.55713911564626,0);
\draw[draw=none,fill=color0] (axis cs:5.55713911564626,0) rectangle (axis cs:5.64688948412698,0);
\draw[draw=none,fill=color0] (axis cs:5.64688948412698,0) rectangle (axis cs:5.73663985260771,0);
\draw[draw=none,fill=color0] (axis cs:5.73663985260771,0) rectangle (axis cs:5.82639022108843,0);
\draw[draw=none,fill=color0] (axis cs:5.82639022108843,0) rectangle (axis cs:5.91614058956916,0);
\draw[draw=none,fill=color0] (axis cs:5.91614058956916,0) rectangle (axis cs:6.00589095804989,1);
\draw[draw=none,fill=color0] (axis cs:6.00589095804989,0) rectangle (axis cs:6.09564132653061,0);
\draw[draw=none,fill=color0] (axis cs:6.09564132653061,0) rectangle (axis cs:6.18539169501134,3);
\draw[draw=none,fill=color0] (axis cs:6.18539169501134,0) rectangle (axis cs:6.27514206349206,2);
\draw[draw=none,fill=color0] (axis cs:6.27514206349206,0) rectangle (axis cs:6.36489243197279,0);
\draw[draw=none,fill=color0] (axis cs:6.36489243197279,0) rectangle (axis cs:6.45464280045351,0);
\draw[draw=none,fill=color0] (axis cs:6.45464280045351,0) rectangle (axis cs:6.54439316893424,1);
\draw[draw=none,fill=color0] (axis cs:6.54439316893424,0) rectangle (axis cs:6.63414353741497,1);
\draw[draw=none,fill=color0] (axis cs:6.63414353741497,0) rectangle (axis cs:6.72389390589569,3);
\draw[draw=none,fill=color0] (axis cs:6.72389390589569,0) rectangle (axis cs:6.81364427437642,3);
\draw[draw=none,fill=color0] (axis cs:6.81364427437642,0) rectangle (axis cs:6.90339464285714,5);
\draw[draw=none,fill=color0] (axis cs:6.90339464285714,0) rectangle (axis cs:6.99314501133787,2);
\draw[draw=none,fill=color0] (axis cs:6.99314501133787,0) rectangle (axis cs:7.08289537981859,1);
\draw[draw=none,fill=color0] (axis cs:7.08289537981859,0) rectangle (axis cs:7.17264574829932,1);
\draw[draw=none,fill=color0] (axis cs:7.17264574829932,0) rectangle (axis cs:7.26239611678004,0);
\draw[draw=none,fill=color0] (axis cs:7.26239611678004,0) rectangle (axis cs:7.35214648526077,1);
\draw[draw=none,fill=color0] (axis cs:7.35214648526077,0) rectangle (axis cs:7.4418968537415,2);
\draw[draw=none,fill=color0] (axis cs:7.4418968537415,0) rectangle (axis cs:7.53164722222222,3);
\draw[draw=none,fill=color0] (axis cs:7.53164722222222,0) rectangle (axis cs:7.62139759070295,0);
\draw[draw=none,fill=color0] (axis cs:7.62139759070295,0) rectangle (axis cs:7.71114795918367,1);

\nextgroupplot[
tick align=outside,
tick pos=left,
title={N},
x grid style={white!69.0196078431373!black},
xmin=0.763972222222222, xmax=11.4056574074074,
xtick style={color=black},
xtick={0,5,10,15},
xlabel={\emph{e-state Values}},
xticklabels={
  \(\displaystyle {0}\),
  \(\displaystyle {5}\),
  \(\displaystyle {10}\),
  \(\displaystyle {15}\)
},
y grid style={white!69.0196078431373!black},
ymin=0, ymax=14.7,
ytick style={color=black},
ytick={0,2,4,6,8,10,12,14,16},
yticklabels={
  \(\displaystyle {0}\),
  \(\displaystyle {2}\),
  \(\displaystyle {4}\),
  \(\displaystyle {6}\),
  \(\displaystyle {8}\),
  \(\displaystyle {10}\),
  \(\displaystyle {12}\),
  \(\displaystyle {14}\),
  \(\displaystyle {16}\)
}
]
\draw[draw=none,fill=color1] (axis cs:1.24768518518519,0) rectangle (axis cs:1.34442777777778,1);
\draw[draw=none,fill=color1] (axis cs:1.34442777777778,0) rectangle (axis cs:1.44117037037037,0);
\draw[draw=none,fill=color1] (axis cs:1.44117037037037,0) rectangle (axis cs:1.53791296296296,0);
\draw[draw=none,fill=color1] (axis cs:1.53791296296296,0) rectangle (axis cs:1.63465555555556,2);
\draw[draw=none,fill=color1] (axis cs:1.63465555555556,0) rectangle (axis cs:1.73139814814815,0);
\draw[draw=none,fill=color1] (axis cs:1.73139814814815,0) rectangle (axis cs:1.82814074074074,0);
\draw[draw=none,fill=color1] (axis cs:1.82814074074074,0) rectangle (axis cs:1.92488333333333,1);
\draw[draw=none,fill=color1] (axis cs:1.92488333333333,0) rectangle (axis cs:2.02162592592593,0);
\draw[draw=none,fill=color1] (axis cs:2.02162592592593,0) rectangle (axis cs:2.11836851851852,0);
\draw[draw=none,fill=color1] (axis cs:2.11836851851852,0) rectangle (axis cs:2.21511111111111,1);
\draw[draw=none,fill=color1] (axis cs:2.21511111111111,0) rectangle (axis cs:2.3118537037037,1);
\draw[draw=none,fill=color1] (axis cs:2.3118537037037,0) rectangle (axis cs:2.4085962962963,2);
\draw[draw=none,fill=color1] (axis cs:2.4085962962963,0) rectangle (axis cs:2.50533888888889,0);
\draw[draw=none,fill=color1] (axis cs:2.50533888888889,0) rectangle (axis cs:2.60208148148148,1);
\draw[draw=none,fill=color1] (axis cs:2.60208148148148,0) rectangle (axis cs:2.69882407407407,3);
\draw[draw=none,fill=color1] (axis cs:2.69882407407407,0) rectangle (axis cs:2.79556666666667,2);
\draw[draw=none,fill=color1] (axis cs:2.79556666666667,0) rectangle (axis cs:2.89230925925926,3);
\draw[draw=none,fill=color1] (axis cs:2.89230925925926,0) rectangle (axis cs:2.98905185185185,4);
\draw[draw=none,fill=color1] (axis cs:2.98905185185185,0) rectangle (axis cs:3.08579444444444,2);
\draw[draw=none,fill=color1] (axis cs:3.08579444444444,0) rectangle (axis cs:3.18253703703704,1);
\draw[draw=none,fill=color1] (axis cs:3.18253703703704,0) rectangle (axis cs:3.27927962962963,6);
\draw[draw=none,fill=color1] (axis cs:3.27927962962963,0) rectangle (axis cs:3.37602222222222,8);
\draw[draw=none,fill=color1] (axis cs:3.37602222222222,0) rectangle (axis cs:3.47276481481482,5);
\draw[draw=none,fill=color1] (axis cs:3.47276481481482,0) rectangle (axis cs:3.56950740740741,3);
\draw[draw=none,fill=color1] (axis cs:3.56950740740741,0) rectangle (axis cs:3.66625,8);
\draw[draw=none,fill=color1] (axis cs:3.66625,0) rectangle (axis cs:3.76299259259259,10);
\draw[draw=none,fill=color1] (axis cs:3.76299259259259,0) rectangle (axis cs:3.85973518518519,9);
\draw[draw=none,fill=color1] (axis cs:3.85973518518519,0) rectangle (axis cs:3.95647777777778,7);
\draw[draw=none,fill=color1] (axis cs:3.95647777777778,0) rectangle (axis cs:4.05322037037037,4);
\draw[draw=none,fill=color1] (axis cs:4.05322037037037,0) rectangle (axis cs:4.14996296296296,14);
\draw[draw=none,fill=color1] (axis cs:4.14996296296296,0) rectangle (axis cs:4.24670555555556,5);
\draw[draw=none,fill=color1] (axis cs:4.24670555555556,0) rectangle (axis cs:4.34344814814815,9);
\draw[draw=none,fill=color1] (axis cs:4.34344814814815,0) rectangle (axis cs:4.44019074074074,12);
\draw[draw=none,fill=color1] (axis cs:4.44019074074074,0) rectangle (axis cs:4.53693333333333,12);
\draw[draw=none,fill=color1] (axis cs:4.53693333333333,0) rectangle (axis cs:4.63367592592593,10);
\draw[draw=none,fill=color1] (axis cs:4.63367592592593,0) rectangle (axis cs:4.73041851851852,11);
\draw[draw=none,fill=color1] (axis cs:4.73041851851852,0) rectangle (axis cs:4.82716111111111,10);
\draw[draw=none,fill=color1] (axis cs:4.82716111111111,0) rectangle (axis cs:4.9239037037037,6);
\draw[draw=none,fill=color1] (axis cs:4.9239037037037,0) rectangle (axis cs:5.0206462962963,2);
\draw[draw=none,fill=color1] (axis cs:5.0206462962963,0) rectangle (axis cs:5.11738888888889,6);
\draw[draw=none,fill=color1] (axis cs:5.11738888888889,0) rectangle (axis cs:5.21413148148148,4);
\draw[draw=none,fill=color1] (axis cs:5.21413148148148,0) rectangle (axis cs:5.31087407407407,5);
\draw[draw=none,fill=color1] (axis cs:5.31087407407407,0) rectangle (axis cs:5.40761666666667,5);
\draw[draw=none,fill=color1] (axis cs:5.40761666666667,0) rectangle (axis cs:5.50435925925926,1);
\draw[draw=none,fill=color1] (axis cs:5.50435925925926,0) rectangle (axis cs:5.60110185185185,1);
\draw[draw=none,fill=color1] (axis cs:5.60110185185185,0) rectangle (axis cs:5.69784444444444,2);
\draw[draw=none,fill=color1] (axis cs:5.69784444444444,0) rectangle (axis cs:5.79458703703704,0);
\draw[draw=none,fill=color1] (axis cs:5.79458703703704,0) rectangle (axis cs:5.89132962962963,2);
\draw[draw=none,fill=color1] (axis cs:5.89132962962963,0) rectangle (axis cs:5.98807222222222,2);
\draw[draw=none,fill=color1] (axis cs:5.98807222222222,0) rectangle (axis cs:6.08481481481482,0);
\draw[draw=none,fill=color1] (axis cs:6.08481481481482,0) rectangle (axis cs:6.18155740740741,1);
\draw[draw=none,fill=color1] (axis cs:6.18155740740741,0) rectangle (axis cs:6.2783,1);
\draw[draw=none,fill=color1] (axis cs:6.2783,0) rectangle (axis cs:6.37504259259259,0);
\draw[draw=none,fill=color1] (axis cs:6.37504259259259,0) rectangle (axis cs:6.47178518518519,0);
\draw[draw=none,fill=color1] (axis cs:6.47178518518519,0) rectangle (axis cs:6.56852777777778,0);
\draw[draw=none,fill=color1] (axis cs:6.56852777777778,0) rectangle (axis cs:6.66527037037037,0);
\draw[draw=none,fill=color1] (axis cs:6.66527037037037,0) rectangle (axis cs:6.76201296296296,0);
\draw[draw=none,fill=color1] (axis cs:6.76201296296296,0) rectangle (axis cs:6.85875555555556,0);
\draw[draw=none,fill=color1] (axis cs:6.85875555555556,0) rectangle (axis cs:6.95549814814815,2);
\draw[draw=none,fill=color1] (axis cs:6.95549814814815,0) rectangle (axis cs:7.05224074074074,2);
\draw[draw=none,fill=color1] (axis cs:7.05224074074074,0) rectangle (axis cs:7.14898333333333,1);
\draw[draw=none,fill=color1] (axis cs:7.14898333333333,0) rectangle (axis cs:7.24572592592593,1);
\draw[draw=none,fill=color1] (axis cs:7.24572592592593,0) rectangle (axis cs:7.34246851851852,2);
\draw[draw=none,fill=color1] (axis cs:7.34246851851852,0) rectangle (axis cs:7.43921111111111,1);
\draw[draw=none,fill=color1] (axis cs:7.43921111111111,0) rectangle (axis cs:7.5359537037037,2);
\draw[draw=none,fill=color1] (axis cs:7.5359537037037,0) rectangle (axis cs:7.6326962962963,4);
\draw[draw=none,fill=color1] (axis cs:7.6326962962963,0) rectangle (axis cs:7.72943888888889,6);
\draw[draw=none,fill=color1] (axis cs:7.72943888888889,0) rectangle (axis cs:7.82618148148148,3);
\draw[draw=none,fill=color1] (axis cs:7.82618148148148,0) rectangle (axis cs:7.92292407407407,2);
\draw[draw=none,fill=color1] (axis cs:7.92292407407407,0) rectangle (axis cs:8.01966666666667,3);
\draw[draw=none,fill=color1] (axis cs:8.01966666666667,0) rectangle (axis cs:8.11640925925926,2);
\draw[draw=none,fill=color1] (axis cs:8.11640925925926,0) rectangle (axis cs:8.21315185185185,1);
\draw[draw=none,fill=color1] (axis cs:8.21315185185185,0) rectangle (axis cs:8.30989444444445,0);
\draw[draw=none,fill=color1] (axis cs:8.30989444444445,0) rectangle (axis cs:8.40663703703704,2);
\draw[draw=none,fill=color1] (axis cs:8.40663703703704,0) rectangle (axis cs:8.50337962962963,1);
\draw[draw=none,fill=color1] (axis cs:8.50337962962963,0) rectangle (axis cs:8.60012222222222,5);
\draw[draw=none,fill=color1] (axis cs:8.60012222222223,0) rectangle (axis cs:8.69686481481482,4);
\draw[draw=none,fill=color1] (axis cs:8.69686481481482,0) rectangle (axis cs:8.79360740740741,7);
\draw[draw=none,fill=color1] (axis cs:8.79360740740741,0) rectangle (axis cs:8.89035,6);
\draw[draw=none,fill=color1] (axis cs:8.89035,0) rectangle (axis cs:8.98709259259259,1);
\draw[draw=none,fill=color1] (axis cs:8.9870925925926,0) rectangle (axis cs:9.08383518518519,5);
\draw[draw=none,fill=color1] (axis cs:9.08383518518519,0) rectangle (axis cs:9.18057777777778,3);
\draw[draw=none,fill=color1] (axis cs:9.18057777777778,0) rectangle (axis cs:9.27732037037037,1);
\draw[draw=none,fill=color1] (axis cs:9.27732037037037,0) rectangle (axis cs:9.37406296296296,3);
\draw[draw=none,fill=color1] (axis cs:9.37406296296296,0) rectangle (axis cs:9.47080555555556,9);
\draw[draw=none,fill=color1] (axis cs:9.47080555555556,0) rectangle (axis cs:9.56754814814815,8);
\draw[draw=none,fill=color1] (axis cs:9.56754814814815,0) rectangle (axis cs:9.66429074074074,5);
\draw[draw=none,fill=color1] (axis cs:9.66429074074074,0) rectangle (axis cs:9.76103333333334,1);
\draw[draw=none,fill=color1] (axis cs:9.76103333333333,0) rectangle (axis cs:9.85777592592593,6);
\draw[draw=none,fill=color1] (axis cs:9.85777592592593,0) rectangle (axis cs:9.95451851851852,4);
\draw[draw=none,fill=color1] (axis cs:9.95451851851852,0) rectangle (axis cs:10.0512611111111,2);
\draw[draw=none,fill=color1] (axis cs:10.0512611111111,0) rectangle (axis cs:10.1480037037037,6);
\draw[draw=none,fill=color1] (axis cs:10.1480037037037,0) rectangle (axis cs:10.2447462962963,2);
\draw[draw=none,fill=color1] (axis cs:10.2447462962963,0) rectangle (axis cs:10.3414888888889,2);
\draw[draw=none,fill=color1] (axis cs:10.3414888888889,0) rectangle (axis cs:10.4382314814815,0);
\draw[draw=none,fill=color1] (axis cs:10.4382314814815,0) rectangle (axis cs:10.5349740740741,0);
\draw[draw=none,fill=color1] (axis cs:10.5349740740741,0) rectangle (axis cs:10.6317166666667,0);
\draw[draw=none,fill=color1] (axis cs:10.6317166666667,0) rectangle (axis cs:10.7284592592593,0);
\draw[draw=none,fill=color1] (axis cs:10.7284592592593,0) rectangle (axis cs:10.8252018518519,0);
\draw[draw=none,fill=color1] (axis cs:10.8252018518519,0) rectangle (axis cs:10.9219444444444,2);

\nextgroupplot[
tick align=outside,
tick pos=left,
title={O},
x grid style={white!69.0196078431373!black},
xmin=4.13990740740741, xmax=14.8883333333333,
xtick style={color=black},
xlabel={\emph{e-state Values}},
xtick={0,5,10,15},
xticklabels={
  \(\displaystyle {0}\),
  \(\displaystyle {5}\),
  \(\displaystyle {10}\),
  \(\displaystyle {15}\)
},
y grid style={white!69.0196078431373!black},
ymin=0, ymax=24.15,
ytick style={color=black},
ytick={0,5,10,15,20,25},
yticklabels={
  \(\displaystyle {0}\),
  \(\displaystyle {5}\),
  \(\displaystyle {10}\),
  \(\displaystyle {15}\),
  \(\displaystyle {20}\),
  \(\displaystyle {25}\)
}
]
\draw[draw=none,fill=color2] (axis cs:4.62847222222222,0) rectangle (axis cs:4.72618518518519,1);
\draw[draw=none,fill=color2] (axis cs:4.72618518518519,0) rectangle (axis cs:4.82389814814815,0);
\draw[draw=none,fill=color2] (axis cs:4.82389814814815,0) rectangle (axis cs:4.92161111111111,1);
\draw[draw=none,fill=color2] (axis cs:4.92161111111111,0) rectangle (axis cs:5.01932407407407,3);
\draw[draw=none,fill=color2] (axis cs:5.01932407407407,0) rectangle (axis cs:5.11703703703704,1);
\draw[draw=none,fill=color2] (axis cs:5.11703703703704,0) rectangle (axis cs:5.21475,1);
\draw[draw=none,fill=color2] (axis cs:5.21475,0) rectangle (axis cs:5.31246296296296,0);
\draw[draw=none,fill=color2] (axis cs:5.31246296296296,0) rectangle (axis cs:5.41017592592593,6);
\draw[draw=none,fill=color2] (axis cs:5.41017592592593,0) rectangle (axis cs:5.50788888888889,3);
\draw[draw=none,fill=color2] (axis cs:5.50788888888889,0) rectangle (axis cs:5.60560185185185,1);
\draw[draw=none,fill=color2] (axis cs:5.60560185185185,0) rectangle (axis cs:5.70331481481482,4);
\draw[draw=none,fill=color2] (axis cs:5.70331481481482,0) rectangle (axis cs:5.80102777777778,2);
\draw[draw=none,fill=color2] (axis cs:5.80102777777778,0) rectangle (axis cs:5.89874074074074,4);
\draw[draw=none,fill=color2] (axis cs:5.89874074074074,0) rectangle (axis cs:5.9964537037037,5);
\draw[draw=none,fill=color2] (axis cs:5.9964537037037,0) rectangle (axis cs:6.09416666666667,2);
\draw[draw=none,fill=color2] (axis cs:6.09416666666667,0) rectangle (axis cs:6.19187962962963,3);
\draw[draw=none,fill=color2] (axis cs:6.19187962962963,0) rectangle (axis cs:6.28959259259259,3);
\draw[draw=none,fill=color2] (axis cs:6.28959259259259,0) rectangle (axis cs:6.38730555555556,6);
\draw[draw=none,fill=color2] (axis cs:6.38730555555556,0) rectangle (axis cs:6.48501851851852,5);
\draw[draw=none,fill=color2] (axis cs:6.48501851851852,0) rectangle (axis cs:6.58273148148148,6);
\draw[draw=none,fill=color2] (axis cs:6.58273148148148,0) rectangle (axis cs:6.68044444444445,9);
\draw[draw=none,fill=color2] (axis cs:6.68044444444445,0) rectangle (axis cs:6.77815740740741,8);
\draw[draw=none,fill=color2] (axis cs:6.77815740740741,0) rectangle (axis cs:6.87587037037037,6);
\draw[draw=none,fill=color2] (axis cs:6.87587037037037,0) rectangle (axis cs:6.97358333333333,11);
\draw[draw=none,fill=color2] (axis cs:6.97358333333333,0) rectangle (axis cs:7.0712962962963,8);
\draw[draw=none,fill=color2] (axis cs:7.0712962962963,0) rectangle (axis cs:7.16900925925926,6);
\draw[draw=none,fill=color2] (axis cs:7.16900925925926,0) rectangle (axis cs:7.26672222222222,6);
\draw[draw=none,fill=color2] (axis cs:7.26672222222222,0) rectangle (axis cs:7.36443518518519,3);
\draw[draw=none,fill=color2] (axis cs:7.36443518518519,0) rectangle (axis cs:7.46214814814815,3);
\draw[draw=none,fill=color2] (axis cs:7.46214814814815,0) rectangle (axis cs:7.55986111111111,1);
\draw[draw=none,fill=color2] (axis cs:7.55986111111111,0) rectangle (axis cs:7.65757407407408,7);
\draw[draw=none,fill=color2] (axis cs:7.65757407407408,0) rectangle (axis cs:7.75528703703704,3);
\draw[draw=none,fill=color2] (axis cs:7.75528703703704,0) rectangle (axis cs:7.853,9);
\draw[draw=none,fill=color2] (axis cs:7.853,0) rectangle (axis cs:7.95071296296296,2);
\draw[draw=none,fill=color2] (axis cs:7.95071296296296,0) rectangle (axis cs:8.04842592592593,2);
\draw[draw=none,fill=color2] (axis cs:8.04842592592593,0) rectangle (axis cs:8.14613888888889,7);
\draw[draw=none,fill=color2] (axis cs:8.14613888888889,0) rectangle (axis cs:8.24385185185185,3);
\draw[draw=none,fill=color2] (axis cs:8.24385185185185,0) rectangle (axis cs:8.34156481481481,0);
\draw[draw=none,fill=color2] (axis cs:8.34156481481482,0) rectangle (axis cs:8.43927777777778,3);
\draw[draw=none,fill=color2] (axis cs:8.43927777777778,0) rectangle (axis cs:8.53699074074074,1);
\draw[draw=none,fill=color2] (axis cs:8.53699074074074,0) rectangle (axis cs:8.63470370370371,2);
\draw[draw=none,fill=color2] (axis cs:8.63470370370371,0) rectangle (axis cs:8.73241666666667,2);
\draw[draw=none,fill=color2] (axis cs:8.73241666666667,0) rectangle (axis cs:8.83012962962963,0);
\draw[draw=none,fill=color2] (axis cs:8.83012962962963,0) rectangle (axis cs:8.92784259259259,0);
\draw[draw=none,fill=color2] (axis cs:8.92784259259259,0) rectangle (axis cs:9.02555555555556,0);
\draw[draw=none,fill=color2] (axis cs:9.02555555555556,0) rectangle (axis cs:9.12326851851852,0);
\draw[draw=none,fill=color2] (axis cs:9.12326851851852,0) rectangle (axis cs:9.22098148148148,0);
\draw[draw=none,fill=color2] (axis cs:9.22098148148148,0) rectangle (axis cs:9.31869444444444,0);
\draw[draw=none,fill=color2] (axis cs:9.31869444444445,0) rectangle (axis cs:9.41640740740741,0);
\draw[draw=none,fill=color2] (axis cs:9.41640740740741,0) rectangle (axis cs:9.51412037037037,0);
\draw[draw=none,fill=color2] (axis cs:9.51412037037037,0) rectangle (axis cs:9.61183333333334,0);
\draw[draw=none,fill=color2] (axis cs:9.61183333333334,0) rectangle (axis cs:9.7095462962963,0);
\draw[draw=none,fill=color2] (axis cs:9.7095462962963,0) rectangle (axis cs:9.80725925925926,0);
\draw[draw=none,fill=color2] (axis cs:9.80725925925926,0) rectangle (axis cs:9.90497222222222,0);
\draw[draw=none,fill=color2] (axis cs:9.90497222222222,0) rectangle (axis cs:10.0026851851852,0);
\draw[draw=none,fill=color2] (axis cs:10.0026851851852,0) rectangle (axis cs:10.1003981481481,0);
\draw[draw=none,fill=color2] (axis cs:10.1003981481481,0) rectangle (axis cs:10.1981111111111,0);
\draw[draw=none,fill=color2] (axis cs:10.1981111111111,0) rectangle (axis cs:10.2958240740741,0);
\draw[draw=none,fill=color2] (axis cs:10.2958240740741,0) rectangle (axis cs:10.393537037037,4);
\draw[draw=none,fill=color2] (axis cs:10.393537037037,0) rectangle (axis cs:10.49125,2);
\draw[draw=none,fill=color2] (axis cs:10.49125,0) rectangle (axis cs:10.588962962963,2);
\draw[draw=none,fill=color2] (axis cs:10.588962962963,0) rectangle (axis cs:10.6866759259259,5);
\draw[draw=none,fill=color2] (axis cs:10.6866759259259,0) rectangle (axis cs:10.7843888888889,2);
\draw[draw=none,fill=color2] (axis cs:10.7843888888889,0) rectangle (axis cs:10.8821018518519,5);
\draw[draw=none,fill=color2] (axis cs:10.8821018518519,0) rectangle (axis cs:10.9798148148148,3);
\draw[draw=none,fill=color2] (axis cs:10.9798148148148,0) rectangle (axis cs:11.0775277777778,4);
\draw[draw=none,fill=color2] (axis cs:11.0775277777778,0) rectangle (axis cs:11.1752407407407,5);
\draw[draw=none,fill=color2] (axis cs:11.1752407407407,0) rectangle (axis cs:11.2729537037037,10);
\draw[draw=none,fill=color2] (axis cs:11.2729537037037,0) rectangle (axis cs:11.3706666666667,9);
\draw[draw=none,fill=color2] (axis cs:11.3706666666667,0) rectangle (axis cs:11.4683796296296,8);
\draw[draw=none,fill=color2] (axis cs:11.4683796296296,0) rectangle (axis cs:11.5660925925926,10);
\draw[draw=none,fill=color2] (axis cs:11.5660925925926,0) rectangle (axis cs:11.6638055555556,8);
\draw[draw=none,fill=color2] (axis cs:11.6638055555556,0) rectangle (axis cs:11.7615185185185,14);
\draw[draw=none,fill=color2] (axis cs:11.7615185185185,0) rectangle (axis cs:11.8592314814815,14);
\draw[draw=none,fill=color2] (axis cs:11.8592314814815,0) rectangle (axis cs:11.9569444444444,6);
\draw[draw=none,fill=color2] (axis cs:11.9569444444444,0) rectangle (axis cs:12.0546574074074,11);
\draw[draw=none,fill=color2] (axis cs:12.0546574074074,0) rectangle (axis cs:12.1523703703704,13);
\draw[draw=none,fill=color2] (axis cs:12.1523703703704,0) rectangle (axis cs:12.2500833333333,12);
\draw[draw=none,fill=color2] (axis cs:12.2500833333333,0) rectangle (axis cs:12.3477962962963,18);
\draw[draw=none,fill=color2] (axis cs:12.3477962962963,0) rectangle (axis cs:12.4455092592593,23);
\draw[draw=none,fill=color2] (axis cs:12.4455092592593,0) rectangle (axis cs:12.5432222222222,18);
\draw[draw=none,fill=color2] (axis cs:12.5432222222222,0) rectangle (axis cs:12.6409351851852,14);
\draw[draw=none,fill=color2] (axis cs:12.6409351851852,0) rectangle (axis cs:12.7386481481482,23);
\draw[draw=none,fill=color2] (axis cs:12.7386481481482,0) rectangle (axis cs:12.8363611111111,12);
\draw[draw=none,fill=color2] (axis cs:12.8363611111111,0) rectangle (axis cs:12.9340740740741,12);
\draw[draw=none,fill=color2] (axis cs:12.9340740740741,0) rectangle (axis cs:13.031787037037,13);
\draw[draw=none,fill=color2] (axis cs:13.031787037037,0) rectangle (axis cs:13.1295,14);
\draw[draw=none,fill=color2] (axis cs:13.1295,0) rectangle (axis cs:13.227212962963,8);
\draw[draw=none,fill=color2] (axis cs:13.227212962963,0) rectangle (axis cs:13.3249259259259,15);
\draw[draw=none,fill=color2] (axis cs:13.3249259259259,0) rectangle (axis cs:13.4226388888889,8);
\draw[draw=none,fill=color2] (axis cs:13.4226388888889,0) rectangle (axis cs:13.5203518518519,8);
\draw[draw=none,fill=color2] (axis cs:13.5203518518519,0) rectangle (axis cs:13.6180648148148,14);
\draw[draw=none,fill=color2] (axis cs:13.6180648148148,0) rectangle (axis cs:13.7157777777778,6);
\draw[draw=none,fill=color2] (axis cs:13.7157777777778,0) rectangle (axis cs:13.8134907407407,2);
\draw[draw=none,fill=color2] (axis cs:13.8134907407407,0) rectangle (axis cs:13.9112037037037,10);
\draw[draw=none,fill=color2] (axis cs:13.9112037037037,0) rectangle (axis cs:14.0089166666667,7);
\draw[draw=none,fill=color2] (axis cs:14.0089166666667,0) rectangle (axis cs:14.1066296296296,4);
\draw[draw=none,fill=color2] (axis cs:14.1066296296296,0) rectangle (axis cs:14.2043425925926,6);
\draw[draw=none,fill=color2] (axis cs:14.2043425925926,0) rectangle (axis cs:14.3020555555556,1);
\draw[draw=none,fill=color2] (axis cs:14.3020555555556,0) rectangle (axis cs:14.3997685185185,3);
\end{groupplot}

\end{tikzpicture}

%% file: images/learn_diag1.tex
\begin{tikzpicture}[,node distance=1.5cm]
\node (nstart) at (0,0) [rectangle,  rounded corners=2pt, draw=blue!50, fill=yellow!20,
                        ]{Molecule Codification(e.g. SMILES)};
\node (initial_step)[below of=nstart, rectangle,  rounded corners=2pt, align={left},
                draw=blue!50, fill=blue!20]
                {Construct connectivity};
\node (2n_step)[below of=initial_step, rectangle, rounded corners=2pt, align={left},
                draw=blue!50, fill=blue!20, ]
                {Splitting in blocks-class \\ 
                    based on connectivity};
\node (3t_step)[below of=2n_step, 
                diamond,
                aspect=2,
                xshift=3.9cm,
                yshift=-0.5cm,
                draw=blue!50, fill=green!20, ]
                {Last block-class? };
\node (4r_step)[below of=2n_step, rectangle, align={left},
                xshift=-0.7cm,
                yshift=-0.5cm,
                rounded corners=2pt, draw=blue!50, fill=blue!20, ]
                {Finding similar \\ 
                    block-class in $\tset$};
\node (5r_step)[below of=4r_step, rectangle, align={left},
                rounded corners=2pt, draw=blue!50, fill=blue!20, ]
                {Computing the scalar \\ 
                    product for $\tset$};
\node (6r_step)[below of=5r_step, rectangle, align={left},
                rounded corners=2pt, draw=blue!50, fill=blue!20, ]
                {Training the parameterized \\ 
                    KRR};
\node (7r_step)[below of=6r_step, rectangle, align={left},
                rounded corners=2pt, draw=blue!50, fill=blue!20, ]
                {Predicting scalar product \\ 
                    of the current block };
\node (8r_step)[below of=3t_step, rectangle, align={left},
                rounded corners=2pt,
                draw=blue!50, fill=yellow!20,
                xshift=-3cm,
                yshift=-4.5cm 
                ]
                {Constructing Cartesian coordinates \\ 
                    from the predicted scalar product};

\draw [->, >=stealth, style={thick}](nstart.south) -- (initial_step.north);
\draw [->, >=stealth, style={thick}] (initial_step.south) -- (2n_step.north);
\draw [->, >=stealth, style={thick}] (2n_step.east) -|(3t_step.north);
\draw [->, >=stealth, style={thick}] (5r_step.south) -- (6r_step.north);
\draw [->, >=stealth, style={thick}] (4r_step.south) -- (5r_step.north);
\draw [->, >=stealth, style={thick}] (6r_step.south) -- (7r_step.north);
\draw [->, >=stealth, style={thick}] (7r_step.east) -| (3t_step.south);
\draw [->, >=stealth, style={thick}]
            (3t_step.west) --  node[anchor=north] {No} (4r_step.east);
\draw [->, >=stealth, style={thick}]
             (3t_step.east)
            |- node[anchor=north] {Yes}
             (8r_step.east);
\end{tikzpicture}

%% file: images/error_block_hist_tikz.tex
\begin{tikzpicture}

\definecolor{color0}{rgb}{0.12156862745098,0.466666666666667,0.705882352941177}

\begin{axis}[
tick align=outside,
tick pos=left,
width=0.8\textwidth,
height=0.4\textwidth,
x grid style={white!69.0196078431373!black},
xlabel={Values of RMSD (\r A)},
xmin=-0.015087603380863, xmax=0.337790403312397,
xtick style={color=black},
xtick={-0.05,0,0.05,0.1,0.15,0.2,0.25,0.3,0.35},
xticklabels={
  \(\displaystyle {−0.05}\),
  \(\displaystyle {0.00}\),
  \(\displaystyle {0.05}\),
  \(\displaystyle {0.10}\),
  \(\displaystyle {0.15}\),
  \(\displaystyle {0.20}\),
  \(\displaystyle {0.25}\),
  \(\displaystyle {0.30}\),
  \(\displaystyle {0.35}\)
},
y grid style={white!69.0196078431373!black},
ylabel={Block Count},
ymin=0, ymax=18.9,
ytick style={color=black},
ytick={0,2.5,5,7.5,10,12.5,15,17.5,20},
yticklabels={0,2,5,7,10,12,15,17,20}
]
\draw[draw=none,fill=color0] (axis cs:0.000952306014285141,0) rectangle (axis cs:0.0169922154094333,18);
\draw[draw=none,fill=color0] (axis cs:0.0169922154094333,0) rectangle (axis cs:0.0330321248045815,7);
\draw[draw=none,fill=color0] (axis cs:0.0330321248045815,0) rectangle (axis cs:0.0490720341997297,2);
\draw[draw=none,fill=color0] (axis cs:0.0490720341997297,0) rectangle (axis cs:0.0651119435948779,2);
\draw[draw=none,fill=color0] (axis cs:0.0651119435948779,0) rectangle (axis cs:0.081151852990026,3);
\draw[draw=none,fill=color0] (axis cs:0.0811518529900261,0) rectangle (axis cs:0.0971917623851742,5);
\draw[draw=none,fill=color0] (axis cs:0.0971917623851742,0) rectangle (axis cs:0.113231671780322,2);
\draw[draw=none,fill=color0] (axis cs:0.113231671780322,0) rectangle (axis cs:0.129271581175471,4);
\draw[draw=none,fill=color0] (axis cs:0.129271581175471,0) rectangle (axis cs:0.145311490570619,2);
\draw[draw=none,fill=color0] (axis cs:0.145311490570619,0) rectangle (axis cs:0.161351399965767,1);
\draw[draw=none,fill=color0] (axis cs:0.161351399965767,0) rectangle (axis cs:0.177391309360915,2);
\draw[draw=none,fill=color0] (axis cs:0.177391309360915,0) rectangle (axis cs:0.193431218756063,0);
\draw[draw=none,fill=color0] (axis cs:0.193431218756063,0) rectangle (axis cs:0.209471128151211,0);
\draw[draw=none,fill=color0] (axis cs:0.209471128151211,0) rectangle (axis cs:0.22551103754636,0);
\draw[draw=none,fill=color0] (axis cs:0.22551103754636,0) rectangle (axis cs:0.241550946941508,1);
\draw[draw=none,fill=color0] (axis cs:0.241550946941508,0) rectangle (axis cs:0.257590856336656,1);
\draw[draw=none,fill=color0] (axis cs:0.257590856336656,0) rectangle (axis cs:0.273630765731804,0);
\draw[draw=none,fill=color0] (axis cs:0.273630765731804,0) rectangle (axis cs:0.289670675126952,0);
\draw[draw=none,fill=color0] (axis cs:0.289670675126952,0) rectangle (axis cs:0.3057105845221,0);
\draw[draw=none,fill=color0] (axis cs:0.305710584522101,0) rectangle (axis cs:0.321750493917249,3);
\end{axis}

\end{tikzpicture}

%% file: images/error_block_tikz.tex
\begin{tikzpicture}

\definecolor{color0}{rgb}{0.12156862745098,0.466666666666667,0.705882352941177}
\definecolor{color1}{rgb}{1,0.498039215686275,0.0549019607843137}
\definecolor{color2}{rgb}{0.172549019607843,0.627450980392157,0.172549019607843}
\definecolor{color3}{rgb}{0.83921568627451,0.152941176470588,0.156862745098039}
\definecolor{color4}{rgb}{0.580392156862745,0.403921568627451,0.741176470588235}
\definecolor{color5}{rgb}{0.549019607843137,0.337254901960784,0.294117647058824}
\definecolor{color6}{rgb}{0.890196078431372,0.466666666666667,0.76078431372549}
\definecolor{color7}{rgb}{0.737254901960784,0.741176470588235,0.133333333333333}
\definecolor{color8}{rgb}{0.0901960784313725,0.745098039215686,0.811764705882353}

\begin{axis}[
tick align=outside,
tick pos=left,
width=1\textwidth,
height=0.5\textwidth,
x grid style={white!69.0196078431373!black},
xlabel={\LARGE Blocks},
xmin=-1.6, xmax=55.6,
y grid style={white!69.0196078431373!black},
ylabel={RMSD (\r A)},
ymin=-0.0159691111234626, ymax=0.350346772582049,
ytick style={color=black},
ytick={-0.05,0,0.05,0.1,0.15,0.2,0.25,0.3,0.35,0.4},
yticklabels={
  \(\displaystyle {−0.05}\),
  \(\displaystyle {0.00}\),
  \(\displaystyle {0.05}\),
  \(\displaystyle {0.10}\),
  \(\displaystyle {0.15}\),
  \(\displaystyle {0.20}\),
  \(\displaystyle {0.25}\),
  \(\displaystyle {0.30}\),
  \(\displaystyle {0.35}\),
  \(\displaystyle {0.40}\)
}
]
\path [draw=color0, semithick]
(axis cs:1,0.00525254495387178)
--(axis cs:1,0.00551657317360135);

\path [draw=color1, semithick]
(axis cs:2,0.00877676201905509)
--(axis cs:2,0.0102028831452691);

\path [draw=color2, semithick]
(axis cs:3,0.0144234747825146)
--(axis cs:3,0.0160613448362837);

\path [draw=color3, semithick]
(axis cs:4,0.163013669510918)
--(axis cs:4,0.171021185161553);

\path [draw=color4, semithick]
(axis cs:5,0.0378822358733506)
--(axis cs:5,0.0459368436032339);

\path [draw=color5, semithick]
(axis cs:6,0.101490867730301)
--(axis cs:6,0.108376001791431);

\path [draw=color6, semithick]
(axis cs:7,0.00954393221439869)
--(axis cs:7,0.0105654363577977);

\path [draw=white!49.8039215686275!black, semithick]
(axis cs:8,0.00938263058821814)
--(axis cs:8,0.0102865812317831);

\path [draw=color7, semithick]
(axis cs:9,0.0551413425100731)
--(axis cs:9,0.0653655092983544);

\path [draw=color8, semithick]
(axis cs:10,0.0124638465056327)
--(axis cs:10,0.0132220642045602);

\path [draw=color0, semithick]
(axis cs:11,0.0797387505399886)
--(axis cs:11,0.100401668243664);

\path [draw=color1, semithick]
(axis cs:12,0.0176507625748713)
--(axis cs:12,0.0184623356709458);

\path [draw=color2, semithick]
(axis cs:13,0.0709828056522936)
--(axis cs:13,0.0919822040943442);

\path [draw=color3, semithick]
(axis cs:14,0.000681610863151599)
--(axis cs:14,0.00122300116541868);

\path [draw=color4, semithick]
(axis cs:15,0.0125370545753381)
--(axis cs:15,0.0139976933696787);

\path [draw=color5, semithick]
(axis cs:16,0.0111895182560726)
--(axis cs:16,0.012310777325196);

\path [draw=color6, semithick]
(axis cs:17,0.0212890633374949)
--(axis cs:17,0.0255113587255983);

\path [draw=white!49.8039215686275!black, semithick]
(axis cs:18,0.0154358381953301)
--(axis cs:18,0.0169009901337174);

\path [draw=color7, semithick]
(axis cs:19,0.0172584078884509)
--(axis cs:19,0.0186443232986019);

\path [draw=color8, semithick]
(axis cs:20,0.308160716282341)
--(axis cs:20,0.317846280508985);

\path [draw=color0, semithick]
(axis cs:21,0.0784077892427479)
--(axis cs:21,0.0808459112964211);

\path [draw=color1, semithick]
(axis cs:22,0.0660158140029417)
--(axis cs:22,0.0792629540119083);

\path [draw=color2, semithick]
(axis cs:23,0.0759561475413375)
--(axis cs:23,0.0783547579307813);

\path [draw=color3, semithick]
(axis cs:24,0.130303210609641)
--(axis cs:24,0.137708751530966);

\path [draw=color4, semithick]
(axis cs:25,0.140304915251069)
--(axis cs:25,0.160186891937437);

\path [draw=color5, semithick]
(axis cs:26,0.0128765261643402)
--(axis cs:26,0.0139685077657892);

\path [draw=color6, semithick]
(axis cs:27,0.0445731987327936)
--(axis cs:27,0.0485117688471025);

\path [draw=white!49.8039215686275!black, semithick]
(axis cs:28,0.00337716474283127)
--(axis cs:28,0.00612090289223867);

\path [draw=color7, semithick]
(axis cs:29,0.0568552859405748)
--(axis cs:29,0.0581205039558655);

\path [draw=color8, semithick]
(axis cs:30,0.0831413830986438)
--(axis cs:30,0.0844480749992771);

\path [draw=color0, semithick]
(axis cs:31,0.243206495649997)
--(axis cs:31,0.253879344724771);

\path [draw=color1, semithick]
(axis cs:32,0.0264785742278363)
--(axis cs:32,0.0302378009731478);

\path [draw=color2, semithick]
(axis cs:33,0.0126748837487962)
--(axis cs:33,0.0133467057473738);

\path [draw=color3, semithick]
(axis cs:34,0.0113837481457874)
--(axis cs:34,0.0130361320087297);

\path [draw=color4, semithick]
(axis cs:35,0.224001466051403)
--(axis cs:35,0.251990487511487);

\path [draw=color5, semithick]
(axis cs:36,0.115942007273305)
--(axis cs:36,0.120130378110757);

\path [draw=color6, semithick]
(axis cs:37,0.0834487960339566)
--(axis cs:37,0.08724883236206);

\path [draw=white!49.8039215686275!black, semithick]
(axis cs:38,0.300342042745185)
--(axis cs:38,0.314453414619613);

\path [draw=color7, semithick]
(axis cs:39,0.0177858707822375)
--(axis cs:39,0.0236864176803133);

\path [draw=color8, semithick]
(axis cs:40,0.0167786179327739)
--(axis cs:40,0.0189705945707227);

\path [draw=color0, semithick]
(axis cs:41,0.00224447560941512)
--(axis cs:41,0.00243534801989316);

\path [draw=color1, semithick]
(axis cs:42,0.0825659127616377)
--(axis cs:42,0.0937320797545665);

\path [draw=color2, semithick]
(axis cs:43,0.121805337432507)
--(axis cs:43,0.127820017941408);

\path [draw=color3, semithick]
(axis cs:44,0.00505371315081553)
--(axis cs:44,0.00581798451661808);

\path [draw=color4, semithick]
(axis cs:45,0.0188460387511878)
--(axis cs:45,0.0216073976607305);

\path [draw=color5, semithick]
(axis cs:46,0.0992534319970417)
--(axis cs:46,0.104515357259367);

\path [draw=color6, semithick]
(axis cs:47,0.309804937239063)
--(axis cs:47,0.333696050595435);

\path [draw=white!49.8039215686275!black, semithick]
(axis cs:48,0.118033871126328)
--(axis cs:48,0.138515048929779);

\path [draw=color7, semithick]
(axis cs:49,0.155758819274993)
--(axis cs:49,0.168939078764476);

\path [draw=color8, semithick]
(axis cs:50,0.0128236605788589)
--(axis cs:50,0.0174708421033888);

\path [draw=color0, semithick]
(axis cs:51,0.00818282090745734)
--(axis cs:51,0.00950898900005676);

\path [draw=color1, semithick]
(axis cs:52,0.129828234618645)
--(axis cs:52,0.147176690394348);

\path [draw=color2, semithick]
(axis cs:53,0.113706391286167)
--(axis cs:53,0.128836734981315);

\addplot [semithick, color0, mark=-, mark size=1.5, mark options={solid}, only marks]
table {%
1 0.00525254495387178
};
\addplot [semithick, color0, mark=-, mark size=1.5, mark options={solid}, only marks]
table {%
1 0.00551657317360135
};
\addplot [semithick, color1, mark=-, mark size=1.5, mark options={solid}, only marks]
table {%
2 0.00877676201905509
};
\addplot [semithick, color1, mark=-, mark size=1.5, mark options={solid}, only marks]
table {%
2 0.0102028831452691
};
\addplot [semithick, color2, mark=-, mark size=1.5, mark options={solid}, only marks]
table {%
3 0.0144234747825146
};
\addplot [semithick, color2, mark=-, mark size=1.5, mark options={solid}, only marks]
table {%
3 0.0160613448362837
};
\addplot [semithick, color3, mark=-, mark size=1.5, mark options={solid}, only marks]
table {%
4 0.163013669510918
};
\addplot [semithick, color3, mark=-, mark size=1.5, mark options={solid}, only marks]
table {%
4 0.171021185161553
};
\addplot [semithick, color4, mark=-, mark size=1.5, mark options={solid}, only marks]
table {%
5 0.0378822358733506
};
\addplot [semithick, color4, mark=-, mark size=1.5, mark options={solid}, only marks]
table {%
5 0.0459368436032339
};
\addplot [semithick, color5, mark=-, mark size=1.5, mark options={solid}, only marks]
table {%
6 0.101490867730301
};
\addplot [semithick, color5, mark=-, mark size=1.5, mark options={solid}, only marks]
table {%
6 0.108376001791431
};
\addplot [semithick, color6, mark=-, mark size=1.5, mark options={solid}, only marks]
table {%
7 0.00954393221439869
};
\addplot [semithick, color6, mark=-, mark size=1.5, mark options={solid}, only marks]
table {%
7 0.0105654363577977
};
\addplot [semithick, white!49.8039215686275!black, mark=-, mark size=1.5, mark options={solid}, only marks]
table {%
8 0.00938263058821814
};
\addplot [semithick, white!49.8039215686275!black, mark=-, mark size=1.5, mark options={solid}, only marks]
table {%
8 0.0102865812317831
};
\addplot [semithick, color7, mark=-, mark size=1.5, mark options={solid}, only marks]
table {%
9 0.0551413425100731
};
\addplot [semithick, color7, mark=-, mark size=1.5, mark options={solid}, only marks]
table {%
9 0.0653655092983544
};
\addplot [semithick, color8, mark=-, mark size=1.5, mark options={solid}, only marks]
table {%
10 0.0124638465056327
};
\addplot [semithick, color8, mark=-, mark size=1.5, mark options={solid}, only marks]
table {%
10 0.0132220642045602
};
\addplot [semithick, color0, mark=-, mark size=1.5, mark options={solid}, only marks]
table {%
11 0.0797387505399886
};
\addplot [semithick, color0, mark=-, mark size=1.5, mark options={solid}, only marks]
table {%
11 0.100401668243664
};
\addplot [semithick, color1, mark=-, mark size=1.5, mark options={solid}, only marks]
table {%
12 0.0176507625748713
};
\addplot [semithick, color1, mark=-, mark size=1.5, mark options={solid}, only marks]
table {%
12 0.0184623356709458
};
\addplot [semithick, color2, mark=-, mark size=1.5, mark options={solid}, only marks]
table {%
13 0.0709828056522936
};
\addplot [semithick, color2, mark=-, mark size=1.5, mark options={solid}, only marks]
table {%
13 0.0919822040943442
};
\addplot [semithick, color3, mark=-, mark size=1.5, mark options={solid}, only marks]
table {%
14 0.000681610863151599
};
\addplot [semithick, color3, mark=-, mark size=1.5, mark options={solid}, only marks]
table {%
14 0.00122300116541868
};
\addplot [semithick, color4, mark=-, mark size=1.5, mark options={solid}, only marks]
table {%
15 0.0125370545753381
};
\addplot [semithick, color4, mark=-, mark size=1.5, mark options={solid}, only marks]
table {%
15 0.0139976933696787
};
\addplot [semithick, color5, mark=-, mark size=1.5, mark options={solid}, only marks]
table {%
16 0.0111895182560726
};
\addplot [semithick, color5, mark=-, mark size=1.5, mark options={solid}, only marks]
table {%
16 0.012310777325196
};
\addplot [semithick, color6, mark=-, mark size=1.5, mark options={solid}, only marks]
table {%
17 0.0212890633374949
};
\addplot [semithick, color6, mark=-, mark size=1.5, mark options={solid}, only marks]
table {%
17 0.0255113587255983
};
\addplot [semithick, white!49.8039215686275!black, mark=-, mark size=1.5, mark options={solid}, only marks]
table {%
18 0.0154358381953301
};
\addplot [semithick, white!49.8039215686275!black, mark=-, mark size=1.5, mark options={solid}, only marks]
table {%
18 0.0169009901337174
};
\addplot [semithick, color7, mark=-, mark size=1.5, mark options={solid}, only marks]
table {%
19 0.0172584078884509
};
\addplot [semithick, color7, mark=-, mark size=1.5, mark options={solid}, only marks]
table {%
19 0.0186443232986019
};
\addplot [semithick, color8, mark=-, mark size=1.5, mark options={solid}, only marks]
table {%
20 0.308160716282341
};
\addplot [semithick, color8, mark=-, mark size=1.5, mark options={solid}, only marks]
table {%
20 0.317846280508985
};
\addplot [semithick, color0, mark=-, mark size=1.5, mark options={solid}, only marks]
table {%
21 0.0784077892427479
};
\addplot [semithick, color0, mark=-, mark size=1.5, mark options={solid}, only marks]
table {%
21 0.0808459112964211
};
\addplot [semithick, color1, mark=-, mark size=1.5, mark options={solid}, only marks]
table {%
22 0.0660158140029417
};
\addplot [semithick, color1, mark=-, mark size=1.5, mark options={solid}, only marks]
table {%
22 0.0792629540119083
};
\addplot [semithick, color2, mark=-, mark size=1.5, mark options={solid}, only marks]
table {%
23 0.0759561475413375
};
\addplot [semithick, color2, mark=-, mark size=1.5, mark options={solid}, only marks]
table {%
23 0.0783547579307813
};
\addplot [semithick, color3, mark=-, mark size=1.5, mark options={solid}, only marks]
table {%
24 0.130303210609641
};
\addplot [semithick, color3, mark=-, mark size=1.5, mark options={solid}, only marks]
table {%
24 0.137708751530966
};
\addplot [semithick, color4, mark=-, mark size=1.5, mark options={solid}, only marks]
table {%
25 0.140304915251069
};
\addplot [semithick, color4, mark=-, mark size=1.5, mark options={solid}, only marks]
table {%
25 0.160186891937437
};
\addplot [semithick, color5, mark=-, mark size=1.5, mark options={solid}, only marks]
table {%
26 0.0128765261643402
};
\addplot [semithick, color5, mark=-, mark size=1.5, mark options={solid}, only marks]
table {%
26 0.0139685077657892
};
\addplot [semithick, color6, mark=-, mark size=1.5, mark options={solid}, only marks]
table {%
27 0.0445731987327936
};
\addplot [semithick, color6, mark=-, mark size=1.5, mark options={solid}, only marks]
table {%
27 0.0485117688471025
};
\addplot [semithick, white!49.8039215686275!black, mark=-, mark size=1.5, mark options={solid}, only marks]
table {%
28 0.00337716474283127
};
\addplot [semithick, white!49.8039215686275!black, mark=-, mark size=1.5, mark options={solid}, only marks]
table {%
28 0.00612090289223867
};
\addplot [semithick, color7, mark=-, mark size=1.5, mark options={solid}, only marks]
table {%
29 0.0568552859405748
};
\addplot [semithick, color7, mark=-, mark size=1.5, mark options={solid}, only marks]
table {%
29 0.0581205039558655
};
\addplot [semithick, color8, mark=-, mark size=1.5, mark options={solid}, only marks]
table {%
30 0.0831413830986438
};
\addplot [semithick, color8, mark=-, mark size=1.5, mark options={solid}, only marks]
table {%
30 0.0844480749992771
};
\addplot [semithick, color0, mark=-, mark size=1.5, mark options={solid}, only marks]
table {%
31 0.243206495649997
};
\addplot [semithick, color0, mark=-, mark size=1.5, mark options={solid}, only marks]
table {%
31 0.253879344724771
};
\addplot [semithick, color1, mark=-, mark size=1.5, mark options={solid}, only marks]
table {%
32 0.0264785742278363
};
\addplot [semithick, color1, mark=-, mark size=1.5, mark options={solid}, only marks]
table {%
32 0.0302378009731478
};
\addplot [semithick, color2, mark=-, mark size=1.5, mark options={solid}, only marks]
table {%
33 0.0126748837487962
};
\addplot [semithick, color2, mark=-, mark size=1.5, mark options={solid}, only marks]
table {%
33 0.0133467057473738
};
\addplot [semithick, color3, mark=-, mark size=1.5, mark options={solid}, only marks]
table {%
34 0.0113837481457874
};
\addplot [semithick, color3, mark=-, mark size=1.5, mark options={solid}, only marks]
table {%
34 0.0130361320087297
};
\addplot [semithick, color4, mark=-, mark size=1.5, mark options={solid}, only marks]
table {%
35 0.224001466051403
};
\addplot [semithick, color4, mark=-, mark size=1.5, mark options={solid}, only marks]
table {%
35 0.251990487511487
};
\addplot [semithick, color5, mark=-, mark size=1.5, mark options={solid}, only marks]
table {%
36 0.115942007273305
};
\addplot [semithick, color5, mark=-, mark size=1.5, mark options={solid}, only marks]
table {%
36 0.120130378110757
};
\addplot [semithick, color6, mark=-, mark size=1.5, mark options={solid}, only marks]
table {%
37 0.0834487960339566
};
\addplot [semithick, color6, mark=-, mark size=1.5, mark options={solid}, only marks]
table {%
37 0.08724883236206
};
\addplot [semithick, white!49.8039215686275!black, mark=-, mark size=1.5, mark options={solid}, only marks]
table {%
38 0.300342042745185
};
\addplot [semithick, white!49.8039215686275!black, mark=-, mark size=1.5, mark options={solid}, only marks]
table {%
38 0.314453414619613
};
\addplot [semithick, color7, mark=-, mark size=1.5, mark options={solid}, only marks]
table {%
39 0.0177858707822375
};
\addplot [semithick, color7, mark=-, mark size=1.5, mark options={solid}, only marks]
table {%
39 0.0236864176803133
};
\addplot [semithick, color8, mark=-, mark size=1.5, mark options={solid}, only marks]
table {%
40 0.0167786179327739
};
\addplot [semithick, color8, mark=-, mark size=1.5, mark options={solid}, only marks]
table {%
40 0.0189705945707227
};
\addplot [semithick, color0, mark=-, mark size=1.5, mark options={solid}, only marks]
table {%
41 0.00224447560941512
};
\addplot [semithick, color0, mark=-, mark size=1.5, mark options={solid}, only marks]
table {%
41 0.00243534801989316
};
\addplot [semithick, color1, mark=-, mark size=1.5, mark options={solid}, only marks]
table {%
42 0.0825659127616377
};
\addplot [semithick, color1, mark=-, mark size=1.5, mark options={solid}, only marks]
table {%
42 0.0937320797545665
};
\addplot [semithick, color2, mark=-, mark size=1.5, mark options={solid}, only marks]
table {%
43 0.121805337432507
};
\addplot [semithick, color2, mark=-, mark size=1.5, mark options={solid}, only marks]
table {%
43 0.127820017941408
};
\addplot [semithick, color3, mark=-, mark size=1.5, mark options={solid}, only marks]
table {%
44 0.00505371315081553
};
\addplot [semithick, color3, mark=-, mark size=1.5, mark options={solid}, only marks]
table {%
44 0.00581798451661808
};
\addplot [semithick, color4, mark=-, mark size=1.5, mark options={solid}, only marks]
table {%
45 0.0188460387511878
};
\addplot [semithick, color4, mark=-, mark size=1.5, mark options={solid}, only marks]
table {%
45 0.0216073976607305
};
\addplot [semithick, color5, mark=-, mark size=1.5, mark options={solid}, only marks]
table {%
46 0.0992534319970417
};
\addplot [semithick, color5, mark=-, mark size=1.5, mark options={solid}, only marks]
table {%
46 0.104515357259367
};
\addplot [semithick, color6, mark=-, mark size=1.5, mark options={solid}, only marks]
table {%
47 0.309804937239063
};
\addplot [semithick, color6, mark=-, mark size=1.5, mark options={solid}, only marks]
table {%
47 0.333696050595435
};
\addplot [semithick, white!49.8039215686275!black, mark=-, mark size=1.5, mark options={solid}, only marks]
table {%
48 0.118033871126328
};
\addplot [semithick, white!49.8039215686275!black, mark=-, mark size=1.5, mark options={solid}, only marks]
table {%
48 0.138515048929779
};
\addplot [semithick, color7, mark=-, mark size=1.5, mark options={solid}, only marks]
table {%
49 0.155758819274993
};
\addplot [semithick, color7, mark=-, mark size=1.5, mark options={solid}, only marks]
table {%
49 0.168939078764476
};
\addplot [semithick, color8, mark=-, mark size=1.5, mark options={solid}, only marks]
table {%
50 0.0128236605788589
};
\addplot [semithick, color8, mark=-, mark size=1.5, mark options={solid}, only marks]
table {%
50 0.0174708421033888
};
\addplot [semithick, color0, mark=-, mark size=1.5, mark options={solid}, only marks]
table {%
51 0.00818282090745734
};
\addplot [semithick, color0, mark=-, mark size=1.5, mark options={solid}, only marks]
table {%
51 0.00950898900005676
};
\addplot [semithick, color1, mark=-, mark size=1.5, mark options={solid}, only marks]
table {%
52 0.129828234618645
};
\addplot [semithick, color1, mark=-, mark size=1.5, mark options={solid}, only marks]
table {%
52 0.147176690394348
};
\addplot [semithick, color2, mark=-, mark size=1.5, mark options={solid}, only marks]
table {%
53 0.113706391286167
};
\addplot [semithick, color2, mark=-, mark size=1.5, mark options={solid}, only marks]
table {%
53 0.128836734981315
};
\addplot [semithick, color0, dashed, mark=*, mark size=3, mark options={solid}]
table {%
1 0.00538455906373657
};
\addplot [semithick, color1, dashed, mark=*, mark size=3, mark options={solid}]
table {%
2 0.0094898225821621
};
\addplot [semithick, color2, dashed, mark=*, mark size=3, mark options={solid}]
table {%
3 0.0152424098093992
};
\addplot [semithick, color3, dashed, mark=*, mark size=3, mark options={solid}]
table {%
4 0.167017427336235
};
\addplot [semithick, color4, dashed, mark=*, mark size=3, mark options={solid}]
table {%
5 0.0419095397382922
};
\addplot [semithick, color5, dashed, mark=*, mark size=3, mark options={solid}]
table {%
6 0.104933434760866
};
\addplot [semithick, color6, dashed, mark=*, mark size=3, mark options={solid}]
table {%
7 0.0100546842860982
};
\addplot [semithick, white!49.8039215686275!black, dashed, mark=*, mark size=3, mark options={solid}]
table {%
8 0.0098346059100006
};
\addplot [semithick, color7, dashed, mark=*, mark size=3, mark options={solid}]
table {%
9 0.0602534259042138
};
\addplot [semithick, color8, dashed, mark=*, mark size=3, mark options={solid}]
table {%
10 0.0128429553550964
};
\addplot [semithick, color0, dashed, mark=*, mark size=3, mark options={solid}]
table {%
11 0.0900702093918262
};
\addplot [semithick, color1, dashed, mark=*, mark size=3, mark options={solid}]
table {%
12 0.0180565491229086
};
\addplot [semithick, color2, dashed, mark=*, mark size=3, mark options={solid}]
table {%
13 0.0814825048733189
};
\addplot [semithick, color3, dashed, mark=*, mark size=3, mark options={solid}]
table {%
14 0.000952306014285141
};
\addplot [semithick, color4, dashed, mark=*, mark size=3, mark options={solid}]
table {%
15 0.0132673739725084
};
\addplot [semithick, color5, dashed, mark=*, mark size=3, mark options={solid}]
table {%
16 0.0117501477906343
};
\addplot [semithick, color6, dashed, mark=*, mark size=3, mark options={solid}]
table {%
17 0.0234002110315466
};
\addplot [semithick, white!49.8039215686275!black, dashed, mark=*, mark size=3, mark options={solid}]
table {%
18 0.0161684141645238
};
\addplot [semithick, color7, dashed, mark=*, mark size=3, mark options={solid}]
table {%
19 0.0179513655935264
};
\addplot [semithick, color8, dashed, mark=*, mark size=3, mark options={solid}]
table {%
20 0.313003498395663
};
\addplot [semithick, color0, dashed, mark=*, mark size=3, mark options={solid}]
table {%
21 0.0796268502695845
};
\addplot [semithick, color1, dashed, mark=*, mark size=3, mark options={solid}]
table {%
22 0.072639384007425
};
\addplot [semithick, color2, dashed, mark=*, mark size=3, mark options={solid}]
table {%
23 0.0771554527360594
};
\addplot [semithick, color3, dashed, mark=*, mark size=3, mark options={solid}]
table {%
24 0.134005981070304
};
\addplot [semithick, color4, dashed, mark=*, mark size=3, mark options={solid}]
table {%
25 0.150245903594253
};
\addplot [semithick, color5, dashed, mark=*, mark size=3, mark options={solid}]
table {%
26 0.0134225169650647
};
\addplot [semithick, color6, dashed, mark=*, mark size=3, mark options={solid}]
table {%
27 0.046542483789948
};
\addplot [semithick, white!49.8039215686275!black, dashed, mark=*, mark size=3, mark options={solid}]
table {%
28 0.00474903381753497
};
\addplot [semithick, color7, dashed, mark=*, mark size=3, mark options={solid}]
table {%
29 0.0574878949482202
};
\addplot [semithick, color8, dashed, mark=*, mark size=3, mark options={solid}]
table {%
30 0.0837947290489604
};
\addplot [semithick, color0, dashed, mark=*, mark size=3, mark options={solid}]
table {%
31 0.248542920187384
};
\addplot [semithick, color1, dashed, mark=*, mark size=3, mark options={solid}]
table {%
32 0.028358187600492
};
\addplot [semithick, color2, dashed, mark=*, mark size=3, mark options={solid}]
table {%
33 0.013010794748085
};
\addplot [semithick, color3, dashed, mark=*, mark size=3, mark options={solid}]
table {%
34 0.0122099400772586
};
\addplot [semithick, color4, dashed, mark=*, mark size=3, mark options={solid}]
table {%
35 0.237995976781445
};
\addplot [semithick, color5, dashed, mark=*, mark size=3, mark options={solid}]
table {%
36 0.118036192692031
};
\addplot [semithick, color6, dashed, mark=*, mark size=3, mark options={solid}]
table {%
37 0.0853488141980083
};
\addplot [semithick, white!49.8039215686275!black, dashed, mark=*, mark size=3, mark options={solid}]
table {%
38 0.307397728682399
};
\addplot [semithick, color7, dashed, mark=*, mark size=3, mark options={solid}]
table {%
39 0.0207361442312754
};
\addplot [semithick, color8, dashed, mark=*, mark size=3, mark options={solid}]
table {%
40 0.0178746062517483
};
\addplot [semithick, color0, dashed, mark=*, mark size=3, mark options={solid}]
table {%
41 0.00233991181465414
};
\addplot [semithick, color1, dashed, mark=*, mark size=3, mark options={solid}]
table {%
42 0.0881489962581021
};
\addplot [semithick, color2, dashed, mark=*, mark size=3, mark options={solid}]
table {%
43 0.124812677686958
};
\addplot [semithick, color3, dashed, mark=*, mark size=3, mark options={solid}]
table {%
44 0.00543584883371681
};
\addplot [semithick, color4, dashed, mark=*, mark size=3, mark options={solid}]
table {%
45 0.0202267182059591
};
\addplot [semithick, color5, dashed, mark=*, mark size=3, mark options={solid}]
table {%
46 0.101884394628204
};
\addplot [semithick, color6, dashed, mark=*, mark size=3, mark options={solid}]
table {%
47 0.321750493917249
};
\addplot [semithick, white!49.8039215686275!black, dashed, mark=*, mark size=3, mark options={solid}]
table {%
48 0.128274460028053
};
\addplot [semithick, color7, dashed, mark=*, mark size=3, mark options={solid}]
table {%
49 0.162348949019734
};
\addplot [semithick, color8, dashed, mark=*, mark size=3, mark options={solid}]
table {%
50 0.0151472513411238
};
\addplot [semithick, color0, dashed, mark=*, mark size=3, mark options={solid}]
table {%
51 0.00884590495375705
};
\addplot [semithick, color1, dashed, mark=*, mark size=3, mark options={solid}]
table {%
52 0.138502462506497
};
\addplot [semithick, color2, dashed, mark=*, mark size=3, mark options={solid}]
table {%
53 0.121271563133741
};
\draw (axis cs:20.2,0.313003498395663) node[
  scale=0.5,
  anchor=base west,
  text=black,
  rotate=0.0
]{\LARGE N-C-C-C};
\draw (axis cs:31.2,0.258542920187384) node[
  scale=0.5,
  anchor=base west,
  text=black,
  rotate=0.0
]{\LARGE C-C-C-C-C};
\draw (axis cs:35.2,0.227995976781445) node[
  scale=0.5,
  anchor=base west,
  text=black,
  rotate=0.0
]{\LARGE N-C-H-H-H};
\draw (axis cs:38.2,0.307397728682399) node[
  scale=0.5,
  anchor=base west,
  text=black,
  rotate=0.0
]{\LARGE C-N-C-C-C};
\draw (axis cs:47.2,0.321750493917249) node[
  scale=0.5,
  anchor=base west,
  text=black,
  rotate=0.0
]{\LARGE N-N-N-H};
\end{axis}

\end{tikzpicture}

%% file: images/rmsd_error_conf.tex
\begin{tikzpicture}

\begin{groupplot}[group style={group size=2 by 1}]
\nextgroupplot[
tick align=outside,
tick pos=left,
title={(A)},
x grid style={white!69.0196078431373!black},
xlabel={Energy EMT/eV},
xmin=1.36459457094453, xmax=10.3956087333344,
xtick style={color=black},
xtick={0,2,4,6,8,10,12},
xticklabels={
  \(\displaystyle {0}\),
  \(\displaystyle {2}\),
  \(\displaystyle {4}\),
  \(\displaystyle {6}\),
  \(\displaystyle {8}\),
  \(\displaystyle {10}\),
  \(\displaystyle {12}\)
},
y grid style={white!69.0196078431373!black},
ylabel={Molecule Count},
ymin=0, ymax=14.7,
ytick style={color=black},
ytick={0,2,4,6,8,10,12,14,16},
yticklabels={
  \(\displaystyle {0}\),
  \(\displaystyle {2}\),
  \(\displaystyle {4}\),
  \(\displaystyle {6}\),
  \(\displaystyle {8}\),
  \(\displaystyle {10}\),
  \(\displaystyle {12}\),
  \(\displaystyle {14}\),
  \(\displaystyle {16}\)
}
]
\draw[draw=none,fill=red] (axis cs:1.77509521468953,0) rectangle (axis cs:1.85719534343853,1);

\draw[draw=none,fill=red] (axis cs:1.85719534343853,0) rectangle (axis cs:1.93929547218752,0);
\draw[draw=none,fill=red] (axis cs:1.93929547218753,0) rectangle (axis cs:2.02139560093652,0);
\draw[draw=none,fill=red] (axis cs:2.02139560093652,0) rectangle (axis cs:2.10349572968552,0);
\draw[draw=none,fill=red] (axis cs:2.10349572968552,0) rectangle (axis cs:2.18559585843452,0);
\draw[draw=none,fill=red] (axis cs:2.18559585843452,0) rectangle (axis cs:2.26769598718352,0);
\draw[draw=none,fill=red] (axis cs:2.26769598718352,0) rectangle (axis cs:2.34979611593252,0);
\draw[draw=none,fill=red] (axis cs:2.34979611593252,0) rectangle (axis cs:2.43189624468152,0);
\draw[draw=none,fill=red] (axis cs:2.43189624468152,0) rectangle (axis cs:2.51399637343052,0);
\draw[draw=none,fill=red] (axis cs:2.51399637343052,0) rectangle (axis cs:2.59609650217952,0);
\draw[draw=none,fill=red] (axis cs:2.59609650217952,0) rectangle (axis cs:2.67819663092851,0);
\draw[draw=none,fill=red] (axis cs:2.67819663092851,0) rectangle (axis cs:2.76029675967751,0);
\draw[draw=none,fill=red] (axis cs:2.76029675967751,0) rectangle (axis cs:2.84239688842651,0);
\draw[draw=none,fill=red] (axis cs:2.84239688842651,0) rectangle (axis cs:2.92449701717551,0);
\draw[draw=none,fill=red] (axis cs:2.92449701717551,0) rectangle (axis cs:3.00659714592451,0);
\draw[draw=none,fill=red] (axis cs:3.00659714592451,0) rectangle (axis cs:3.08869727467351,0);
\draw[draw=none,fill=red] (axis cs:3.08869727467351,0) rectangle (axis cs:3.17079740342251,2);
\draw[draw=none,fill=red] (axis cs:3.17079740342251,0) rectangle (axis cs:3.25289753217151,0);
\draw[draw=none,fill=red] (axis cs:3.25289753217151,0) rectangle (axis cs:3.33499766092051,0);
\draw[draw=none,fill=red] (axis cs:3.33499766092051,0) rectangle (axis cs:3.4170977896695,0);
\draw[draw=none,fill=red] (axis cs:3.4170977896695,0) rectangle (axis cs:3.4991979184185,2);
\draw[draw=none,fill=red] (axis cs:3.4991979184185,0) rectangle (axis cs:3.5812980471675,0);
\draw[draw=none,fill=red] (axis cs:3.5812980471675,0) rectangle (axis cs:3.6633981759165,2);
\draw[draw=none,fill=red] (axis cs:3.6633981759165,0) rectangle (axis cs:3.7454983046655,1);
\draw[draw=none,fill=red] (axis cs:3.7454983046655,0) rectangle (axis cs:3.8275984334145,0);
\draw[draw=none,fill=red] (axis cs:3.8275984334145,0) rectangle (axis cs:3.9096985621635,4);
\draw[draw=none,fill=red] (axis cs:3.9096985621635,0) rectangle (axis cs:3.9917986909125,1);
\draw[draw=none,fill=red] (axis cs:3.9917986909125,0) rectangle (axis cs:4.0738988196615,2);
\draw[draw=none,fill=red] (axis cs:4.0738988196615,0) rectangle (axis cs:4.15599894841049,1);
\draw[draw=none,fill=red] (axis cs:4.1559989484105,0) rectangle (axis cs:4.23809907715949,2);
\draw[draw=none,fill=red] (axis cs:4.23809907715949,0) rectangle (axis cs:4.32019920590849,3);
\draw[draw=none,fill=red] (axis cs:4.32019920590849,0) rectangle (axis cs:4.40229933465749,6);
\draw[draw=none,fill=red] (axis cs:4.40229933465749,0) rectangle (axis cs:4.48439946340649,7);
\draw[draw=none,fill=red] (axis cs:4.48439946340649,0) rectangle (axis cs:4.56649959215549,2);
\draw[draw=none,fill=red] (axis cs:4.56649959215549,0) rectangle (axis cs:4.64859972090449,2);
\draw[draw=none,fill=red] (axis cs:4.64859972090449,0) rectangle (axis cs:4.73069984965349,0);
\draw[draw=none,fill=red] (axis cs:4.73069984965349,0) rectangle (axis cs:4.81279997840249,4);
\draw[draw=none,fill=red] (axis cs:4.81279997840249,0) rectangle (axis cs:4.89490010715148,3);
\draw[draw=none,fill=red] (axis cs:4.89490010715148,0) rectangle (axis cs:4.97700023590048,3);
\draw[draw=none,fill=red] (axis cs:4.97700023590048,0) rectangle (axis cs:5.05910036464948,5);
\draw[draw=none,fill=red] (axis cs:5.05910036464948,0) rectangle (axis cs:5.14120049339848,4);
\draw[draw=none,fill=red] (axis cs:5.14120049339848,0) rectangle (axis cs:5.22330062214748,6);
\draw[draw=none,fill=red] (axis cs:5.22330062214748,0) rectangle (axis cs:5.30540075089648,4);
\draw[draw=none,fill=red] (axis cs:5.30540075089648,0) rectangle (axis cs:5.38750087964548,8);
\draw[draw=none,fill=red] (axis cs:5.38750087964548,0) rectangle (axis cs:5.46960100839448,6);
\draw[draw=none,fill=red] (axis cs:5.46960100839448,0) rectangle (axis cs:5.55170113714348,10);
\draw[draw=none,fill=red] (axis cs:5.55170113714347,0) rectangle (axis cs:5.63380126589247,8);
\draw[draw=none,fill=red] (axis cs:5.63380126589247,0) rectangle (axis cs:5.71590139464147,7);
\draw[draw=none,fill=red] (axis cs:5.71590139464147,0) rectangle (axis cs:5.79800152339047,6);
\draw[draw=none,fill=red] (axis cs:5.79800152339047,0) rectangle (axis cs:5.88010165213947,5);
\draw[draw=none,fill=red] (axis cs:5.88010165213947,0) rectangle (axis cs:5.96220178088847,11);
\draw[draw=none,fill=red] (axis cs:5.96220178088847,0) rectangle (axis cs:6.04430190963747,7);
\draw[draw=none,fill=red] (axis cs:6.04430190963747,0) rectangle (axis cs:6.12640203838647,9);
\draw[draw=none,fill=red] (axis cs:6.12640203838647,0) rectangle (axis cs:6.20850216713547,5);
\draw[draw=none,fill=red] (axis cs:6.20850216713547,0) rectangle (axis cs:6.29060229588447,8);
\draw[draw=none,fill=red] (axis cs:6.29060229588447,0) rectangle (axis cs:6.37270242463346,14);
\draw[draw=none,fill=red] (axis cs:6.37270242463346,0) rectangle (axis cs:6.45480255338246,10);
\draw[draw=none,fill=red] (axis cs:6.45480255338246,0) rectangle (axis cs:6.53690268213146,8);
\draw[draw=none,fill=red] (axis cs:6.53690268213146,0) rectangle (axis cs:6.61900281088046,8);
\draw[draw=none,fill=red] (axis cs:6.61900281088046,0) rectangle (axis cs:6.70110293962946,14);
\draw[draw=none,fill=red] (axis cs:6.70110293962946,0) rectangle (axis cs:6.78320306837846,13);
\draw[draw=none,fill=red] (axis cs:6.78320306837846,0) rectangle (axis cs:6.86530319712746,1);
\draw[draw=none,fill=red] (axis cs:6.86530319712746,0) rectangle (axis cs:6.94740332587646,7);
\draw[draw=none,fill=red] (axis cs:6.94740332587646,0) rectangle (axis cs:7.02950345462545,5);
\draw[draw=none,fill=red] (axis cs:7.02950345462546,0) rectangle (axis cs:7.11160358337445,12);
\draw[draw=none,fill=red] (axis cs:7.11160358337445,0) rectangle (axis cs:7.19370371212345,4);
\draw[draw=none,fill=red] (axis cs:7.19370371212345,0) rectangle (axis cs:7.27580384087245,5);
\draw[draw=none,fill=red] (axis cs:7.27580384087245,0) rectangle (axis cs:7.35790396962145,7);
\draw[draw=none,fill=red] (axis cs:7.35790396962145,0) rectangle (axis cs:7.44000409837045,5);
\draw[draw=none,fill=red] (axis cs:7.44000409837045,0) rectangle (axis cs:7.52210422711945,8);
\draw[draw=none,fill=red] (axis cs:7.52210422711945,0) rectangle (axis cs:7.60420435586845,7);
\draw[draw=none,fill=red] (axis cs:7.60420435586845,0) rectangle (axis cs:7.68630448461745,6);
\draw[draw=none,fill=red] (axis cs:7.68630448461745,0) rectangle (axis cs:7.76840461336645,6);
\draw[draw=none,fill=red] (axis cs:7.76840461336644,0) rectangle (axis cs:7.85050474211544,4);
\draw[draw=none,fill=red] (axis cs:7.85050474211544,0) rectangle (axis cs:7.93260487086444,8);
\draw[draw=none,fill=red] (axis cs:7.93260487086444,0) rectangle (axis cs:8.01470499961344,4);
\draw[draw=none,fill=red] (axis cs:8.01470499961344,0) rectangle (axis cs:8.09680512836244,6);
\draw[draw=none,fill=red] (axis cs:8.09680512836244,0) rectangle (axis cs:8.17890525711144,3);
\draw[draw=none,fill=red] (axis cs:8.17890525711144,0) rectangle (axis cs:8.26100538586044,7);
\draw[draw=none,fill=red] (axis cs:8.26100538586044,0) rectangle (axis cs:8.34310551460944,5);
\draw[draw=none,fill=red] (axis cs:8.34310551460944,0) rectangle (axis cs:8.42520564335844,2);
\draw[draw=none,fill=red] (axis cs:8.42520564335844,0) rectangle (axis cs:8.50730577210743,6);
\draw[draw=none,fill=red] (axis cs:8.50730577210744,0) rectangle (axis cs:8.58940590085644,6);
\draw[draw=none,fill=red] (axis cs:8.58940590085643,0) rectangle (axis cs:8.67150602960543,4);
\draw[draw=none,fill=red] (axis cs:8.67150602960543,0) rectangle (axis cs:8.75360615835443,3);
\draw[draw=none,fill=red] (axis cs:8.75360615835443,0) rectangle (axis cs:8.83570628710343,2);
\draw[draw=none,fill=red] (axis cs:8.83570628710343,0) rectangle (axis cs:8.91780641585243,2);
\draw[draw=none,fill=red] (axis cs:8.91780641585243,0) rectangle (axis cs:8.99990654460143,3);
\draw[draw=none,fill=red] (axis cs:8.99990654460143,0) rectangle (axis cs:9.08200667335043,1);
\draw[draw=none,fill=red] (axis cs:9.08200667335043,0) rectangle (axis cs:9.16410680209943,1);
\draw[draw=none,fill=red] (axis cs:9.16410680209943,0) rectangle (axis cs:9.24620693084842,0);
\draw[draw=none,fill=red] (axis cs:9.24620693084842,0) rectangle (axis cs:9.32830705959742,3);
\draw[draw=none,fill=red] (axis cs:9.32830705959742,0) rectangle (axis cs:9.41040718834642,2);
\draw[draw=none,fill=red] (axis cs:9.41040718834642,0) rectangle (axis cs:9.49250731709542,2);
\draw[draw=none,fill=red] (axis cs:9.49250731709542,0) rectangle (axis cs:9.57460744584442,1);
\draw[draw=none,fill=red] (axis cs:9.57460744584442,0) rectangle (axis cs:9.65670757459342,1);
\draw[draw=none,fill=red] (axis cs:9.65670757459342,0) rectangle (axis cs:9.73880770334242,1);
\draw[draw=none,fill=red] (axis cs:9.73880770334242,0) rectangle (axis cs:9.82090783209142,1);
\draw[draw=none,fill=red] (axis cs:9.82090783209142,0) rectangle (axis cs:9.90300796084042,0);
\draw[draw=none,fill=red] (axis cs:9.90300796084042,0) rectangle (axis cs:9.98510808958941,3);

\nextgroupplot[
tick align=outside,
tick pos=left,
title={(B)},
x grid style={white!69.0196078431373!black},
xlabel={RMSD/Å},
xmin=-0.013824924641471, xmax=0.879157963235338,
xtick style={color=black},
xtick={-0.2,0,0.2,0.4,0.6,0.8,1},
xticklabels={
  \(\displaystyle {−0.2}\),
  \(\displaystyle {0.0}\),
  \(\displaystyle {0.2}\),
  \(\displaystyle {0.4}\),
  \(\displaystyle {0.6}\),
  \(\displaystyle {0.8}\),
  \(\displaystyle {1.0}\)
},
y grid style={white!69.0196078431373!black},
ymin=0, ymax=12.6,
ytick style={color=black},
ytick={0,2,4,6,8,10,12,14},
yticklabels={
  \(\displaystyle {0}\),
  \(\displaystyle {2}\),
  \(\displaystyle {4}\),
  \(\displaystyle {6}\),
  \(\displaystyle {8}\),
  \(\displaystyle {10}\),
  \(\displaystyle {12}\),
  \(\displaystyle {14}\)
}
]
\draw[draw=none,fill=blue] (axis cs:0.0267652066256567,0) rectangle (axis cs:0.0348832328790822,1);
\draw[draw=none,fill=blue] (axis cs:0.0348832328790822,0) rectangle (axis cs:0.0430012591325078,0);
\draw[draw=none,fill=blue] (axis cs:0.0430012591325078,0) rectangle (axis cs:0.0511192853859333,3);
\draw[draw=none,fill=blue] (axis cs:0.0511192853859333,0) rectangle (axis cs:0.0592373116393588,0);
\draw[draw=none,fill=blue] (axis cs:0.0592373116393588,0) rectangle (axis cs:0.0673553378927844,4);
\draw[draw=none,fill=blue] (axis cs:0.0673553378927844,0) rectangle (axis cs:0.0754733641462099,0);
\draw[draw=none,fill=blue] (axis cs:0.0754733641462099,0) rectangle (axis cs:0.0835913903996355,2);
\draw[draw=none,fill=blue] (axis cs:0.0835913903996355,0) rectangle (axis cs:0.091709416653061,2);
\draw[draw=none,fill=blue] (axis cs:0.091709416653061,0) rectangle (axis cs:0.0998274429064865,0);
\draw[draw=none,fill=blue] (axis cs:0.0998274429064865,0) rectangle (axis cs:0.107945469159912,3);
\draw[draw=none,fill=blue] (axis cs:0.107945469159912,0) rectangle (axis cs:0.116063495413338,2);
\draw[draw=none,fill=blue] (axis cs:0.116063495413338,0) rectangle (axis cs:0.124181521666763,4);
\draw[draw=none,fill=blue] (axis cs:0.124181521666763,0) rectangle (axis cs:0.132299547920189,1);
\draw[draw=none,fill=blue] (axis cs:0.132299547920189,0) rectangle (axis cs:0.140417574173614,0);
\draw[draw=none,fill=blue] (axis cs:0.140417574173614,0) rectangle (axis cs:0.14853560042704,3);
\draw[draw=none,fill=blue] (axis cs:0.14853560042704,0) rectangle (axis cs:0.156653626680465,5);
\draw[draw=none,fill=blue] (axis cs:0.156653626680465,0) rectangle (axis cs:0.164771652933891,3);
\draw[draw=none,fill=blue] (axis cs:0.164771652933891,0) rectangle (axis cs:0.172889679187316,1);
\draw[draw=none,fill=blue] (axis cs:0.172889679187316,0) rectangle (axis cs:0.181007705440742,2);
\draw[draw=none,fill=blue] (axis cs:0.181007705440742,0) rectangle (axis cs:0.189125731694167,3);
\draw[draw=none,fill=blue] (axis cs:0.189125731694167,0) rectangle (axis cs:0.197243757947593,4);
\draw[draw=none,fill=blue] (axis cs:0.197243757947593,0) rectangle (axis cs:0.205361784201019,4);
\draw[draw=none,fill=blue] (axis cs:0.205361784201019,0) rectangle (axis cs:0.213479810454444,3);
\draw[draw=none,fill=blue] (axis cs:0.213479810454444,0) rectangle (axis cs:0.22159783670787,6);
\draw[draw=none,fill=blue] (axis cs:0.22159783670787,0) rectangle (axis cs:0.229715862961295,2);
\draw[draw=none,fill=blue] (axis cs:0.229715862961295,0) rectangle (axis cs:0.237833889214721,5);
\draw[draw=none,fill=blue] (axis cs:0.237833889214721,0) rectangle (axis cs:0.245951915468146,6);
\draw[draw=none,fill=blue] (axis cs:0.245951915468146,0) rectangle (axis cs:0.254069941721572,5);
\draw[draw=none,fill=blue] (axis cs:0.254069941721572,0) rectangle (axis cs:0.262187967974997,3);
\draw[draw=none,fill=blue] (axis cs:0.262187967974997,0) rectangle (axis cs:0.270305994228423,3);
\draw[draw=none,fill=blue] (axis cs:0.270305994228423,0) rectangle (axis cs:0.278424020481848,4);
\draw[draw=none,fill=blue] (axis cs:0.278424020481848,0) rectangle (axis cs:0.286542046735274,5);
\draw[draw=none,fill=blue] (axis cs:0.286542046735274,0) rectangle (axis cs:0.2946600729887,4);
\draw[draw=none,fill=blue] (axis cs:0.294660072988699,0) rectangle (axis cs:0.302778099242125,9);
\draw[draw=none,fill=blue] (axis cs:0.302778099242125,0) rectangle (axis cs:0.310896125495551,7);
\draw[draw=none,fill=blue] (axis cs:0.310896125495551,0) rectangle (axis cs:0.319014151748976,5);
\draw[draw=none,fill=blue] (axis cs:0.319014151748976,0) rectangle (axis cs:0.327132178002402,8);
\draw[draw=none,fill=blue] (axis cs:0.327132178002402,0) rectangle (axis cs:0.335250204255827,6);
\draw[draw=none,fill=blue] (axis cs:0.335250204255827,0) rectangle (axis cs:0.343368230509253,8);
\draw[draw=none,fill=blue] (axis cs:0.343368230509253,0) rectangle (axis cs:0.351486256762678,9);
\draw[draw=none,fill=blue] (axis cs:0.351486256762678,0) rectangle (axis cs:0.359604283016104,7);
\draw[draw=none,fill=blue] (axis cs:0.359604283016104,0) rectangle (axis cs:0.367722309269529,4);
\draw[draw=none,fill=blue] (axis cs:0.367722309269529,0) rectangle (axis cs:0.375840335522955,12);
\draw[draw=none,fill=blue] (axis cs:0.375840335522955,0) rectangle (axis cs:0.38395836177638,9);
\draw[draw=none,fill=blue] (axis cs:0.38395836177638,0) rectangle (axis cs:0.392076388029806,3);
\draw[draw=none,fill=blue] (axis cs:0.392076388029806,0) rectangle (axis cs:0.400194414283231,7);
\draw[draw=none,fill=blue] (axis cs:0.400194414283231,0) rectangle (axis cs:0.408312440536657,3);
\draw[draw=none,fill=blue] (axis cs:0.408312440536657,0) rectangle (axis cs:0.416430466790082,9);
\draw[draw=none,fill=blue] (axis cs:0.416430466790083,0) rectangle (axis cs:0.424548493043508,5);
\draw[draw=none,fill=blue] (axis cs:0.424548493043508,0) rectangle (axis cs:0.432666519296934,8);
\draw[draw=none,fill=blue] (axis cs:0.432666519296934,0) rectangle (axis cs:0.440784545550359,3);
\draw[draw=none,fill=blue] (axis cs:0.440784545550359,0) rectangle (axis cs:0.448902571803785,5);
\draw[draw=none,fill=blue] (axis cs:0.448902571803785,0) rectangle (axis cs:0.45702059805721,10);
\draw[draw=none,fill=blue] (axis cs:0.45702059805721,0) rectangle (axis cs:0.465138624310636,5);
\draw[draw=none,fill=blue] (axis cs:0.465138624310636,0) rectangle (axis cs:0.473256650564061,8);
\draw[draw=none,fill=blue] (axis cs:0.473256650564061,0) rectangle (axis cs:0.481374676817487,5);
\draw[draw=none,fill=blue] (axis cs:0.481374676817487,0) rectangle (axis cs:0.489492703070912,8);
\draw[draw=none,fill=blue] (axis cs:0.489492703070912,0) rectangle (axis cs:0.497610729324338,5);
\draw[draw=none,fill=blue] (axis cs:0.497610729324338,0) rectangle (axis cs:0.505728755577763,4);
\draw[draw=none,fill=blue] (axis cs:0.505728755577763,0) rectangle (axis cs:0.513846781831189,8);
\draw[draw=none,fill=blue] (axis cs:0.513846781831189,0) rectangle (axis cs:0.521964808084615,2);
\draw[draw=none,fill=blue] (axis cs:0.521964808084615,0) rectangle (axis cs:0.53008283433804,6);
\draw[draw=none,fill=blue] (axis cs:0.53008283433804,0) rectangle (axis cs:0.538200860591466,6);
\draw[draw=none,fill=blue] (axis cs:0.538200860591466,0) rectangle (axis cs:0.546318886844891,6);
\draw[draw=none,fill=blue] (axis cs:0.546318886844891,0) rectangle (axis cs:0.554436913098317,5);
\draw[draw=none,fill=blue] (axis cs:0.554436913098317,0) rectangle (axis cs:0.562554939351742,10);
\draw[draw=none,fill=blue] (axis cs:0.562554939351742,0) rectangle (axis cs:0.570672965605168,6);
\draw[draw=none,fill=blue] (axis cs:0.570672965605168,0) rectangle (axis cs:0.578790991858593,4);
\draw[draw=none,fill=blue] (axis cs:0.578790991858593,0) rectangle (axis cs:0.586909018112019,2);
\draw[draw=none,fill=blue] (axis cs:0.586909018112019,0) rectangle (axis cs:0.595027044365444,5);
\draw[draw=none,fill=blue] (axis cs:0.595027044365444,0) rectangle (axis cs:0.60314507061887,8);
\draw[draw=none,fill=blue] (axis cs:0.60314507061887,0) rectangle (axis cs:0.611263096872295,7);
\draw[draw=none,fill=blue] (axis cs:0.611263096872295,0) rectangle (axis cs:0.619381123125721,6);
\draw[draw=none,fill=blue] (axis cs:0.619381123125721,0) rectangle (axis cs:0.627499149379147,3);
\draw[draw=none,fill=blue] (axis cs:0.627499149379147,0) rectangle (axis cs:0.635617175632572,4);
\draw[draw=none,fill=blue] (axis cs:0.635617175632572,0) rectangle (axis cs:0.643735201885998,2);
\draw[draw=none,fill=blue] (axis cs:0.643735201885998,0) rectangle (axis cs:0.651853228139423,4);
\draw[draw=none,fill=blue] (axis cs:0.651853228139423,0) rectangle (axis cs:0.659971254392849,1);
\draw[draw=none,fill=blue] (axis cs:0.659971254392849,0) rectangle (axis cs:0.668089280646274,3);
\draw[draw=none,fill=blue] (axis cs:0.668089280646274,0) rectangle (axis cs:0.6762073068997,2);
\draw[draw=none,fill=blue] (axis cs:0.6762073068997,0) rectangle (axis cs:0.684325333153125,2);
\draw[draw=none,fill=blue] (axis cs:0.684325333153125,0) rectangle (axis cs:0.692443359406551,0);
\draw[draw=none,fill=blue] (axis cs:0.692443359406551,0) rectangle (axis cs:0.700561385659976,2);
\draw[draw=none,fill=blue] (axis cs:0.700561385659976,0) rectangle (axis cs:0.708679411913402,2);
\draw[draw=none,fill=blue] (axis cs:0.708679411913402,0) rectangle (axis cs:0.716797438166828,1);
\draw[draw=none,fill=blue] (axis cs:0.716797438166828,0) rectangle (axis cs:0.724915464420253,1);
\draw[draw=none,fill=blue] (axis cs:0.724915464420253,0) rectangle (axis cs:0.733033490673679,0);
\draw[draw=none,fill=blue] (axis cs:0.733033490673678,0) rectangle (axis cs:0.741151516927104,0);
\draw[draw=none,fill=blue] (axis cs:0.741151516927104,0) rectangle (axis cs:0.74926954318053,0);
\draw[draw=none,fill=blue] (axis cs:0.74926954318053,0) rectangle (axis cs:0.757387569433955,0);
\draw[draw=none,fill=blue] (axis cs:0.757387569433955,0) rectangle (axis cs:0.765505595687381,0);
\draw[draw=none,fill=blue] (axis cs:0.765505595687381,0) rectangle (axis cs:0.773623621940806,0);
\draw[draw=none,fill=blue] (axis cs:0.773623621940806,0) rectangle (axis cs:0.781741648194232,2);
\draw[draw=none,fill=blue] (axis cs:0.781741648194232,0) rectangle (axis cs:0.789859674447657,0);
\draw[draw=none,fill=blue] (axis cs:0.789859674447657,0) rectangle (axis cs:0.797977700701083,0);
\draw[draw=none,fill=blue] (axis cs:0.797977700701083,0) rectangle (axis cs:0.806095726954508,0);
\draw[draw=none,fill=blue] (axis cs:0.806095726954508,0) rectangle (axis cs:0.814213753207934,0);
\draw[draw=none,fill=blue] (axis cs:0.814213753207934,0) rectangle (axis cs:0.822331779461359,1);
\draw[draw=none,fill=blue] (axis cs:0.822331779461359,0) rectangle (axis cs:0.830449805714785,1);
\draw[draw=none,fill=blue] (axis cs:0.830449805714785,0) rectangle (axis cs:0.83856783196821,1);
\end{groupplot}

\end{tikzpicture}

%% file: images/rmsd_error_mol_set.tex
\begin{tikzpicture}

\begin{groupplot}[group style={group size=2 by 1}]
\nextgroupplot[
tick align=outside,
tick pos=left,
title={(C)},
x grid style={white!69.0196078431373!black},
xlabel={Energy EMT/eV},
xmin=1.5376925131935, xmax=9.97462468074528,
xtick style={color=black},
xtick={0,2,4,6,8,10},
xticklabels={
  \(\displaystyle {0}\),
  \(\displaystyle {2}\),
  \(\displaystyle {4}\),
  \(\displaystyle {6}\),
  \(\displaystyle {8}\),
  \(\displaystyle {10}\)
},
y grid style={white!69.0196078431373!black},
ylabel={Molecule Count},
ymin=0, ymax=13.65,
ytick style={color=black},
ytick={0,2,4,6,8,10,12,14},
yticklabels={
  \(\displaystyle {0}\),
  \(\displaystyle {2}\),
  \(\displaystyle {4}\),
  \(\displaystyle {6}\),
  \(\displaystyle {8}\),
  \(\displaystyle {10}\),
  \(\displaystyle {12}\),
  \(\displaystyle {14}\)
}
]
\draw[draw=none,fill=red] (axis cs:1.9211894299004,0) rectangle (axis cs:1.99788881324178,1);

\draw[draw=none,fill=red] (axis cs:1.99788881324178,0) rectangle (axis cs:2.07458819658316,0);
\draw[draw=none,fill=red] (axis cs:2.07458819658316,0) rectangle (axis cs:2.15128757992454,0);
\draw[draw=none,fill=red] (axis cs:2.15128757992454,0) rectangle (axis cs:2.22798696326592,0);
\draw[draw=none,fill=red] (axis cs:2.22798696326592,0) rectangle (axis cs:2.3046863466073,0);
\draw[draw=none,fill=red] (axis cs:2.3046863466073,0) rectangle (axis cs:2.38138572994868,0);
\draw[draw=none,fill=red] (axis cs:2.38138572994868,0) rectangle (axis cs:2.45808511329006,0);
\draw[draw=none,fill=red] (axis cs:2.45808511329006,0) rectangle (axis cs:2.53478449663144,0);
\draw[draw=none,fill=red] (axis cs:2.53478449663144,0) rectangle (axis cs:2.61148387997282,0);
\draw[draw=none,fill=red] (axis cs:2.61148387997282,0) rectangle (axis cs:2.6881832633142,0);
\draw[draw=none,fill=red] (axis cs:2.6881832633142,0) rectangle (axis cs:2.76488264665558,0);
\draw[draw=none,fill=red] (axis cs:2.76488264665558,0) rectangle (axis cs:2.84158202999696,0);
\draw[draw=none,fill=red] (axis cs:2.84158202999696,0) rectangle (axis cs:2.91828141333834,0);
\draw[draw=none,fill=red] (axis cs:2.91828141333834,0) rectangle (axis cs:2.99498079667972,1);
\draw[draw=none,fill=red] (axis cs:2.99498079667972,0) rectangle (axis cs:3.0716801800211,0);
\draw[draw=none,fill=red] (axis cs:3.0716801800211,0) rectangle (axis cs:3.14837956336248,0);
\draw[draw=none,fill=red] (axis cs:3.14837956336248,0) rectangle (axis cs:3.22507894670386,2);
\draw[draw=none,fill=red] (axis cs:3.22507894670386,0) rectangle (axis cs:3.30177833004524,0);
\draw[draw=none,fill=red] (axis cs:3.30177833004524,0) rectangle (axis cs:3.37847771338662,0);
\draw[draw=none,fill=red] (axis cs:3.37847771338662,0) rectangle (axis cs:3.455177096728,2);
\draw[draw=none,fill=red] (axis cs:3.455177096728,0) rectangle (axis cs:3.53187648006938,2);
\draw[draw=none,fill=red] (axis cs:3.53187648006938,0) rectangle (axis cs:3.60857586341076,1);
\draw[draw=none,fill=red] (axis cs:3.60857586341076,0) rectangle (axis cs:3.68527524675214,0);
\draw[draw=none,fill=red] (axis cs:3.68527524675214,0) rectangle (axis cs:3.76197463009352,2);
\draw[draw=none,fill=red] (axis cs:3.76197463009352,0) rectangle (axis cs:3.8386740134349,2);
\draw[draw=none,fill=red] (axis cs:3.8386740134349,0) rectangle (axis cs:3.91537339677628,2);
\draw[draw=none,fill=red] (axis cs:3.91537339677628,0) rectangle (axis cs:3.99207278011766,3);
\draw[draw=none,fill=red] (axis cs:3.99207278011766,0) rectangle (axis cs:4.06877216345904,2);
\draw[draw=none,fill=red] (axis cs:4.06877216345904,0) rectangle (axis cs:4.14547154680042,2);
\draw[draw=none,fill=red] (axis cs:4.14547154680042,0) rectangle (axis cs:4.22217093014179,1);
\draw[draw=none,fill=red] (axis cs:4.22217093014179,0) rectangle (axis cs:4.29887031348318,3);
\draw[draw=none,fill=red] (axis cs:4.29887031348318,0) rectangle (axis cs:4.37556969682455,3);
\draw[draw=none,fill=red] (axis cs:4.37556969682455,0) rectangle (axis cs:4.45226908016593,4);
\draw[draw=none,fill=red] (axis cs:4.45226908016593,0) rectangle (axis cs:4.52896846350731,6);
\draw[draw=none,fill=red] (axis cs:4.52896846350732,0) rectangle (axis cs:4.60566784684869,5);
\draw[draw=none,fill=red] (axis cs:4.60566784684869,0) rectangle (axis cs:4.68236723019007,7);
\draw[draw=none,fill=red] (axis cs:4.68236723019007,0) rectangle (axis cs:4.75906661353145,3);
\draw[draw=none,fill=red] (axis cs:4.75906661353145,0) rectangle (axis cs:4.83576599687283,6);
\draw[draw=none,fill=red] (axis cs:4.83576599687283,0) rectangle (axis cs:4.91246538021421,1);
\draw[draw=none,fill=red] (axis cs:4.91246538021421,0) rectangle (axis cs:4.98916476355559,5);
\draw[draw=none,fill=red] (axis cs:4.98916476355559,0) rectangle (axis cs:5.06586414689697,3);
\draw[draw=none,fill=red] (axis cs:5.06586414689697,0) rectangle (axis cs:5.14256353023835,6);
\draw[draw=none,fill=red] (axis cs:5.14256353023835,0) rectangle (axis cs:5.21926291357973,7);
\draw[draw=none,fill=red] (axis cs:5.21926291357973,0) rectangle (axis cs:5.29596229692111,6);
\draw[draw=none,fill=red] (axis cs:5.29596229692111,0) rectangle (axis cs:5.37266168026249,8);
\draw[draw=none,fill=red] (axis cs:5.37266168026249,0) rectangle (axis cs:5.44936106360387,7);
\draw[draw=none,fill=red] (axis cs:5.44936106360387,0) rectangle (axis cs:5.52606044694525,7);
\draw[draw=none,fill=red] (axis cs:5.52606044694525,0) rectangle (axis cs:5.60275983028663,7);
\draw[draw=none,fill=red] (axis cs:5.60275983028663,0) rectangle (axis cs:5.67945921362801,5);
\draw[draw=none,fill=red] (axis cs:5.67945921362801,0) rectangle (axis cs:5.75615859696939,5);
\draw[draw=none,fill=red] (axis cs:5.75615859696939,0) rectangle (axis cs:5.83285798031077,8);
\draw[draw=none,fill=red] (axis cs:5.83285798031077,0) rectangle (axis cs:5.90955736365215,8);
\draw[draw=none,fill=red] (axis cs:5.90955736365215,0) rectangle (axis cs:5.98625674699353,5);
\draw[draw=none,fill=red] (axis cs:5.98625674699353,0) rectangle (axis cs:6.06295613033491,11);
\draw[draw=none,fill=red] (axis cs:6.06295613033491,0) rectangle (axis cs:6.13965551367629,10);
\draw[draw=none,fill=red] (axis cs:6.13965551367629,0) rectangle (axis cs:6.21635489701767,13);
\draw[draw=none,fill=red] (axis cs:6.21635489701767,0) rectangle (axis cs:6.29305428035905,9);
\draw[draw=none,fill=red] (axis cs:6.29305428035905,0) rectangle (axis cs:6.36975366370043,10);
\draw[draw=none,fill=red] (axis cs:6.36975366370043,0) rectangle (axis cs:6.44645304704181,7);
\draw[draw=none,fill=red] (axis cs:6.44645304704181,0) rectangle (axis cs:6.52315243038319,13);
\draw[draw=none,fill=red] (axis cs:6.52315243038319,0) rectangle (axis cs:6.59985181372457,6);
\draw[draw=none,fill=red] (axis cs:6.59985181372457,0) rectangle (axis cs:6.67655119706595,8);
\draw[draw=none,fill=red] (axis cs:6.67655119706595,0) rectangle (axis cs:6.75325058040733,5);
\draw[draw=none,fill=red] (axis cs:6.75325058040733,0) rectangle (axis cs:6.82994996374871,7);
\draw[draw=none,fill=red] (axis cs:6.82994996374871,0) rectangle (axis cs:6.90664934709009,9);
\draw[draw=none,fill=red] (axis cs:6.90664934709009,0) rectangle (axis cs:6.98334873043147,11);
\draw[draw=none,fill=red] (axis cs:6.98334873043147,0) rectangle (axis cs:7.06004811377285,6);
\draw[draw=none,fill=red] (axis cs:7.06004811377285,0) rectangle (axis cs:7.13674749711423,13);
\draw[draw=none,fill=red] (axis cs:7.13674749711423,0) rectangle (axis cs:7.21344688045561,4);
\draw[draw=none,fill=red] (axis cs:7.21344688045561,0) rectangle (axis cs:7.29014626379699,2);
\draw[draw=none,fill=red] (axis cs:7.29014626379699,0) rectangle (axis cs:7.36684564713837,6);
\draw[draw=none,fill=red] (axis cs:7.36684564713837,0) rectangle (axis cs:7.44354503047975,6);
\draw[draw=none,fill=red] (axis cs:7.44354503047975,0) rectangle (axis cs:7.52024441382113,1);
\draw[draw=none,fill=red] (axis cs:7.52024441382113,0) rectangle (axis cs:7.59694379716251,2);
\draw[draw=none,fill=red] (axis cs:7.59694379716251,0) rectangle (axis cs:7.67364318050388,4);
\draw[draw=none,fill=red] (axis cs:7.67364318050388,0) rectangle (axis cs:7.75034256384526,7);
\draw[draw=none,fill=red] (axis cs:7.75034256384526,0) rectangle (axis cs:7.82704194718664,11);
\draw[draw=none,fill=red] (axis cs:7.82704194718664,0) rectangle (axis cs:7.90374133052802,11);
\draw[draw=none,fill=red] (axis cs:7.90374133052802,0) rectangle (axis cs:7.9804407138694,2);
\draw[draw=none,fill=red] (axis cs:7.98044071386941,0) rectangle (axis cs:8.05714009721078,4);
\draw[draw=none,fill=red] (axis cs:8.05714009721078,0) rectangle (axis cs:8.13383948055216,1);
\draw[draw=none,fill=red] (axis cs:8.13383948055217,0) rectangle (axis cs:8.21053886389354,4);
\draw[draw=none,fill=red] (axis cs:8.21053886389354,0) rectangle (axis cs:8.28723824723492,2);
\draw[draw=none,fill=red] (axis cs:8.28723824723492,0) rectangle (axis cs:8.3639376305763,0);
\draw[draw=none,fill=red] (axis cs:8.3639376305763,0) rectangle (axis cs:8.44063701391768,1);
\draw[draw=none,fill=red] (axis cs:8.44063701391768,0) rectangle (axis cs:8.51733639725906,3);
\draw[draw=none,fill=red] (axis cs:8.51733639725906,0) rectangle (axis cs:8.59403578060044,5);
\draw[draw=none,fill=red] (axis cs:8.59403578060044,0) rectangle (axis cs:8.67073516394182,1);
\draw[draw=none,fill=red] (axis cs:8.67073516394182,0) rectangle (axis cs:8.7474345472832,4);
\draw[draw=none,fill=red] (axis cs:8.7474345472832,0) rectangle (axis cs:8.82413393062458,4);
\draw[draw=none,fill=red] (axis cs:8.82413393062458,0) rectangle (axis cs:8.90083331396596,0);
\draw[draw=none,fill=red] (axis cs:8.90083331396596,0) rectangle (axis cs:8.97753269730734,0);
\draw[draw=none,fill=red] (axis cs:8.97753269730734,0) rectangle (axis cs:9.05423208064872,0);
\draw[draw=none,fill=red] (axis cs:9.05423208064872,0) rectangle (axis cs:9.1309314639901,0);
\draw[draw=none,fill=red] (axis cs:9.1309314639901,0) rectangle (axis cs:9.20763084733148,0);
\draw[draw=none,fill=red] (axis cs:9.20763084733148,0) rectangle (axis cs:9.28433023067286,1);
\draw[draw=none,fill=red] (axis cs:9.28433023067286,0) rectangle (axis cs:9.36102961401424,1);
\draw[draw=none,fill=red] (axis cs:9.36102961401424,0) rectangle (axis cs:9.43772899735562,0);
\draw[draw=none,fill=red] (axis cs:9.43772899735562,0) rectangle (axis cs:9.514428380697,0);
\draw[draw=none,fill=red] (axis cs:9.514428380697,0) rectangle (axis cs:9.59112776403838,5);

\nextgroupplot[
tick align=outside,
tick pos=left,
title={(D)},
x grid style={white!69.0196078431373!black},
xlabel={RMSD/Å},
xmin=-0.0988628474825842, xmax=2.73114210790299,
xtick style={color=black},
xtick={-1,0,1,2,3},
xticklabels={
  \(\displaystyle {−1}\),
  \(\displaystyle {0}\),
  \(\displaystyle {1}\),
  \(\displaystyle {2}\),
  \(\displaystyle {3}\)
},
y grid style={white!69.0196078431373!black},
ymin=0, ymax=13.65,
ytick style={color=black},
ytick={0,2,4,6,8,10,12,14},
yticklabels={
  \(\displaystyle {0}\),
  \(\displaystyle {2}\),
  \(\displaystyle {4}\),
  \(\displaystyle {6}\),
  \(\displaystyle {8}\),
  \(\displaystyle {10}\),
  \(\displaystyle {12}\),
  \(\displaystyle {14}\)
}
]
\draw[draw=none,fill=blue] (axis cs:0.0297737413985782,0) rectangle (axis cs:0.0555010591748106,1);
\draw[draw=none,fill=blue] (axis cs:0.0555010591748106,0) rectangle (axis cs:0.0812283769510431,0);
\draw[draw=none,fill=blue] (axis cs:0.0812283769510431,0) rectangle (axis cs:0.106955694727276,2);
\draw[draw=none,fill=blue] (axis cs:0.106955694727276,0) rectangle (axis cs:0.132683012503508,5);
\draw[draw=none,fill=blue] (axis cs:0.132683012503508,0) rectangle (axis cs:0.15841033027974,4);
\draw[draw=none,fill=blue] (axis cs:0.15841033027974,0) rectangle (axis cs:0.184137648055973,0);
\draw[draw=none,fill=blue] (axis cs:0.184137648055973,0) rectangle (axis cs:0.209864965832205,4);
\draw[draw=none,fill=blue] (axis cs:0.209864965832205,0) rectangle (axis cs:0.235592283608438,0);
\draw[draw=none,fill=blue] (axis cs:0.235592283608438,0) rectangle (axis cs:0.26131960138467,4);
\draw[draw=none,fill=blue] (axis cs:0.26131960138467,0) rectangle (axis cs:0.287046919160903,5);
\draw[draw=none,fill=blue] (axis cs:0.287046919160903,0) rectangle (axis cs:0.312774236937135,5);
\draw[draw=none,fill=blue] (axis cs:0.312774236937135,0) rectangle (axis cs:0.338501554713368,3);
\draw[draw=none,fill=blue] (axis cs:0.338501554713368,0) rectangle (axis cs:0.3642288724896,4);
\draw[draw=none,fill=blue] (axis cs:0.3642288724896,0) rectangle (axis cs:0.389956190265833,8);
\draw[draw=none,fill=blue] (axis cs:0.389956190265833,0) rectangle (axis cs:0.415683508042065,10);
\draw[draw=none,fill=blue] (axis cs:0.415683508042065,0) rectangle (axis cs:0.441410825818298,9);
\draw[draw=none,fill=blue] (axis cs:0.441410825818298,0) rectangle (axis cs:0.46713814359453,12);
\draw[draw=none,fill=blue] (axis cs:0.46713814359453,0) rectangle (axis cs:0.492865461370763,4);
\draw[draw=none,fill=blue] (axis cs:0.492865461370762,0) rectangle (axis cs:0.518592779146995,13);
\draw[draw=none,fill=blue] (axis cs:0.518592779146995,0) rectangle (axis cs:0.544320096923227,10);
\draw[draw=none,fill=blue] (axis cs:0.544320096923227,0) rectangle (axis cs:0.57004741469946,11);
\draw[draw=none,fill=blue] (axis cs:0.57004741469946,0) rectangle (axis cs:0.595774732475692,13);
\draw[draw=none,fill=blue] (axis cs:0.595774732475692,0) rectangle (axis cs:0.621502050251925,10);
\draw[draw=none,fill=blue] (axis cs:0.621502050251925,0) rectangle (axis cs:0.647229368028157,11);
\draw[draw=none,fill=blue] (axis cs:0.647229368028157,0) rectangle (axis cs:0.67295668580439,13);
\draw[draw=none,fill=blue] (axis cs:0.67295668580439,0) rectangle (axis cs:0.698684003580622,11);
\draw[draw=none,fill=blue] (axis cs:0.698684003580622,0) rectangle (axis cs:0.724411321356855,6);
\draw[draw=none,fill=blue] (axis cs:0.724411321356855,0) rectangle (axis cs:0.750138639133087,8);
\draw[draw=none,fill=blue] (axis cs:0.750138639133087,0) rectangle (axis cs:0.77586595690932,6);
\draw[draw=none,fill=blue] (axis cs:0.77586595690932,0) rectangle (axis cs:0.801593274685552,8);
\draw[draw=none,fill=blue] (axis cs:0.801593274685552,0) rectangle (axis cs:0.827320592461784,4);
\draw[draw=none,fill=blue] (axis cs:0.827320592461784,0) rectangle (axis cs:0.853047910238017,10);
\draw[draw=none,fill=blue] (axis cs:0.853047910238017,0) rectangle (axis cs:0.878775228014249,5);
\draw[draw=none,fill=blue] (axis cs:0.87877522801425,0) rectangle (axis cs:0.904502545790482,5);
\draw[draw=none,fill=blue] (axis cs:0.904502545790482,0) rectangle (axis cs:0.930229863566714,9);
\draw[draw=none,fill=blue] (axis cs:0.930229863566714,0) rectangle (axis cs:0.955957181342947,3);
\draw[draw=none,fill=blue] (axis cs:0.955957181342947,0) rectangle (axis cs:0.981684499119179,7);
\draw[draw=none,fill=blue] (axis cs:0.981684499119179,0) rectangle (axis cs:1.00741181689541,10);
\draw[draw=none,fill=blue] (axis cs:1.00741181689541,0) rectangle (axis cs:1.03313913467164,4);
\draw[draw=none,fill=blue] (axis cs:1.03313913467164,0) rectangle (axis cs:1.05886645244788,3);
\draw[draw=none,fill=blue] (axis cs:1.05886645244788,0) rectangle (axis cs:1.08459377022411,1);
\draw[draw=none,fill=blue] (axis cs:1.08459377022411,0) rectangle (axis cs:1.11032108800034,2);
\draw[draw=none,fill=blue] (axis cs:1.11032108800034,0) rectangle (axis cs:1.13604840577657,4);
\draw[draw=none,fill=blue] (axis cs:1.13604840577657,0) rectangle (axis cs:1.16177572355281,7);
\draw[draw=none,fill=blue] (axis cs:1.16177572355281,0) rectangle (axis cs:1.18750304132904,6);
\draw[draw=none,fill=blue] (axis cs:1.18750304132904,0) rectangle (axis cs:1.21323035910527,8);
\draw[draw=none,fill=blue] (axis cs:1.21323035910527,0) rectangle (axis cs:1.2389576768815,2);
\draw[draw=none,fill=blue] (axis cs:1.2389576768815,0) rectangle (axis cs:1.26468499465774,4);
\draw[draw=none,fill=blue] (axis cs:1.26468499465774,0) rectangle (axis cs:1.29041231243397,10);
\draw[draw=none,fill=blue] (axis cs:1.29041231243397,0) rectangle (axis cs:1.3161396302102,1);
\draw[draw=none,fill=blue] (axis cs:1.3161396302102,0) rectangle (axis cs:1.34186694798643,4);
\draw[draw=none,fill=blue] (axis cs:1.34186694798643,0) rectangle (axis cs:1.36759426576267,2);
\draw[draw=none,fill=blue] (axis cs:1.36759426576267,0) rectangle (axis cs:1.3933215835389,2);
\draw[draw=none,fill=blue] (axis cs:1.3933215835389,0) rectangle (axis cs:1.41904890131513,2);
\draw[draw=none,fill=blue] (axis cs:1.41904890131513,0) rectangle (axis cs:1.44477621909136,6);
\draw[draw=none,fill=blue] (axis cs:1.44477621909136,0) rectangle (axis cs:1.4705035368676,4);
\draw[draw=none,fill=blue] (axis cs:1.4705035368676,0) rectangle (axis cs:1.49623085464383,3);
\draw[draw=none,fill=blue] (axis cs:1.49623085464383,0) rectangle (axis cs:1.52195817242006,3);
\draw[draw=none,fill=blue] (axis cs:1.52195817242006,0) rectangle (axis cs:1.54768549019629,2);
\draw[draw=none,fill=blue] (axis cs:1.54768549019629,0) rectangle (axis cs:1.57341280797253,5);
\draw[draw=none,fill=blue] (axis cs:1.57341280797253,0) rectangle (axis cs:1.59914012574876,2);
\draw[draw=none,fill=blue] (axis cs:1.59914012574876,0) rectangle (axis cs:1.62486744352499,1);
\draw[draw=none,fill=blue] (axis cs:1.62486744352499,0) rectangle (axis cs:1.65059476130122,6);
\draw[draw=none,fill=blue] (axis cs:1.65059476130122,0) rectangle (axis cs:1.67632207907746,2);
\draw[draw=none,fill=blue] (axis cs:1.67632207907746,0) rectangle (axis cs:1.70204939685369,2);
\draw[draw=none,fill=blue] (axis cs:1.70204939685369,0) rectangle (axis cs:1.72777671462992,2);
\draw[draw=none,fill=blue] (axis cs:1.72777671462992,0) rectangle (axis cs:1.75350403240615,2);
\draw[draw=none,fill=blue] (axis cs:1.75350403240615,0) rectangle (axis cs:1.77923135018239,2);
\draw[draw=none,fill=blue] (axis cs:1.77923135018239,0) rectangle (axis cs:1.80495866795862,1);
\draw[draw=none,fill=blue] (axis cs:1.80495866795862,0) rectangle (axis cs:1.83068598573485,2);
\draw[draw=none,fill=blue] (axis cs:1.83068598573485,0) rectangle (axis cs:1.85641330351108,2);
\draw[draw=none,fill=blue] (axis cs:1.85641330351108,0) rectangle (axis cs:1.88214062128732,1);
\draw[draw=none,fill=blue] (axis cs:1.88214062128732,0) rectangle (axis cs:1.90786793906355,1);
\draw[draw=none,fill=blue] (axis cs:1.90786793906355,0) rectangle (axis cs:1.93359525683978,2);
\draw[draw=none,fill=blue] (axis cs:1.93359525683978,0) rectangle (axis cs:1.95932257461601,0);
\draw[draw=none,fill=blue] (axis cs:1.95932257461601,0) rectangle (axis cs:1.98504989239225,1);
\draw[draw=none,fill=blue] (axis cs:1.98504989239225,0) rectangle (axis cs:2.01077721016848,1);
\draw[draw=none,fill=blue] (axis cs:2.01077721016848,0) rectangle (axis cs:2.03650452794471,0);
\draw[draw=none,fill=blue] (axis cs:2.03650452794471,0) rectangle (axis cs:2.06223184572094,0);
\draw[draw=none,fill=blue] (axis cs:2.06223184572094,0) rectangle (axis cs:2.08795916349718,2);
\draw[draw=none,fill=blue] (axis cs:2.08795916349718,0) rectangle (axis cs:2.11368648127341,0);
\draw[draw=none,fill=blue] (axis cs:2.11368648127341,0) rectangle (axis cs:2.13941379904964,1);
\draw[draw=none,fill=blue] (axis cs:2.13941379904964,0) rectangle (axis cs:2.16514111682587,2);
\draw[draw=none,fill=blue] (axis cs:2.16514111682587,0) rectangle (axis cs:2.19086843460211,0);
\draw[draw=none,fill=blue] (axis cs:2.19086843460211,0) rectangle (axis cs:2.21659575237834,0);
\draw[draw=none,fill=blue] (axis cs:2.21659575237834,0) rectangle (axis cs:2.24232307015457,0);
\draw[draw=none,fill=blue] (axis cs:2.24232307015457,0) rectangle (axis cs:2.2680503879308,0);
\draw[draw=none,fill=blue] (axis cs:2.2680503879308,0) rectangle (axis cs:2.29377770570703,1);
\draw[draw=none,fill=blue] (axis cs:2.29377770570704,0) rectangle (axis cs:2.31950502348327,0);
\draw[draw=none,fill=blue] (axis cs:2.31950502348327,0) rectangle (axis cs:2.3452323412595,0);
\draw[draw=none,fill=blue] (axis cs:2.3452323412595,0) rectangle (axis cs:2.37095965903573,0);
\draw[draw=none,fill=blue] (axis cs:2.37095965903573,0) rectangle (axis cs:2.39668697681196,1);
\draw[draw=none,fill=blue] (axis cs:2.39668697681197,0) rectangle (axis cs:2.4224142945882,0);
\draw[draw=none,fill=blue] (axis cs:2.4224142945882,0) rectangle (axis cs:2.44814161236443,0);
\draw[draw=none,fill=blue] (axis cs:2.44814161236443,0) rectangle (axis cs:2.47386893014066,0);
\draw[draw=none,fill=blue] (axis cs:2.47386893014066,0) rectangle (axis cs:2.49959624791689,1);
\draw[draw=none,fill=blue] (axis cs:2.49959624791689,0) rectangle (axis cs:2.52532356569313,0);
\draw[draw=none,fill=blue] (axis cs:2.52532356569313,0) rectangle (axis cs:2.55105088346936,0);
\draw[draw=none,fill=blue] (axis cs:2.55105088346936,0) rectangle (axis cs:2.57677820124559,0);
\draw[draw=none,fill=blue] (axis cs:2.57677820124559,0) rectangle (axis cs:2.60250551902182,2);
\end{groupplot}

\end{tikzpicture}

%% file: ml_geometries.bbl
\providecommand{\latin}[1]{#1}
\makeatletter
\providecommand{\doi}
  {\begingroup\let\do\@makeother\dospecials
  \catcode`\{=1 \catcode`\}=2 \doi@aux}
\providecommand{\doi@aux}[1]{\endgroup\texttt{#1}}
\makeatother
\providecommand*\mcitethebibliography{\thebibliography}
\csname @ifundefined\endcsname{endmcitethebibliography}
  {\let\endmcitethebibliography\endthebibliography}{}
\begin{mcitethebibliography}{56}
\providecommand*\natexlab[1]{#1}
\providecommand*\mciteSetBstSublistMode[1]{}
\providecommand*\mciteSetBstMaxWidthForm[2]{}
\providecommand*\mciteBstWouldAddEndPuncttrue
  {\def\EndOfBibitem{\unskip.}}
\providecommand*\mciteBstWouldAddEndPunctfalse
  {\let\EndOfBibitem\relax}
\providecommand*\mciteSetBstMidEndSepPunct[3]{}
\providecommand*\mciteSetBstSublistLabelBeginEnd[3]{}
\providecommand*\EndOfBibitem{}
\mciteSetBstSublistMode{f}
\mciteSetBstMaxWidthForm{subitem}{(\alph{mcitesubitemcount})}
\mciteSetBstSublistLabelBeginEnd
  {\mcitemaxwidthsubitemform\space}
  {\relax}
  {\relax}

\bibitem[Halgren(1996)]{ml_85}
Halgren,~T.~A. Merck molecular force field. I. Basis, form, scope,
  parameterization, and performance of MMFF94. \emph{Journal of Computational
  Chemistry} \textbf{1996}, \emph{17}, 490--519\relax
\mciteBstWouldAddEndPuncttrue
\mciteSetBstMidEndSepPunct{\mcitedefaultmidpunct}
{\mcitedefaultendpunct}{\mcitedefaultseppunct}\relax
\EndOfBibitem
\bibitem[Rappe \latin{et~al.}(1992)Rappe, Casewit, Colwell, Goddard, and
  Skiff]{ml_86}
Rappe,~A.~K.; Casewit,~C.~J.; Colwell,~K.~S.; Goddard,~W.~A.; Skiff,~W.~M.
  {UFF}, a full periodic table force field for molecular mechanics and
  molecular dynamics simulations. \emph{J . Am. Chem. SOC} \textbf{1992},
  \emph{114}, 10024--10035\relax
\mciteBstWouldAddEndPuncttrue
\mciteSetBstMidEndSepPunct{\mcitedefaultmidpunct}
{\mcitedefaultendpunct}{\mcitedefaultseppunct}\relax
\EndOfBibitem
\bibitem[Thiel(2014)]{ml_127}
Thiel,~W. Semiempirical quantum–chemical methods. \emph{WIREs Computational
  Molecular Science} \textbf{2014}, \emph{4}, 145--157\relax
\mciteBstWouldAddEndPuncttrue
\mciteSetBstMidEndSepPunct{\mcitedefaultmidpunct}
{\mcitedefaultendpunct}{\mcitedefaultseppunct}\relax
\EndOfBibitem
\bibitem[Vuckovic and Burke(2020)Vuckovic, and Burke]{ml_102}
Vuckovic,~S.; Burke,~K. Quantifying and understanding errors in molecular
  geometries. \emph{arXiv:2007.15076} \textbf{2020}, \relax
\mciteBstWouldAddEndPunctfalse
\mciteSetBstMidEndSepPunct{\mcitedefaultmidpunct}
{}{\mcitedefaultseppunct}\relax
\EndOfBibitem
\bibitem[Snyman and Wilke(2018)Snyman, and Wilke]{ml_59}
Snyman,~J.~A.; Wilke,~D.~N. \emph{Practical Mathematical Optimization Basic
  Optimization Theory and Gradient-Based Algorithms}, 2nd ed.; Springer,
  2018\relax
\mciteBstWouldAddEndPuncttrue
\mciteSetBstMidEndSepPunct{\mcitedefaultmidpunct}
{\mcitedefaultendpunct}{\mcitedefaultseppunct}\relax
\EndOfBibitem
\bibitem[Ouyang \latin{et~al.}(2018)Ouyang, Curtarolo, Ahmetcik, Scheffler, and
  Ghiringhelli]{ml_3}
Ouyang,~R.; Curtarolo,~S.; Ahmetcik,~E.; Scheffler,~M.; Ghiringhelli,~L. SISSO:
  a compressed-sensing method for systematically identifying efficient physical
  models of materials properties. \emph{Phys. Rev. Materials} \textbf{2018},
  \emph{2}, 083802\relax
\mciteBstWouldAddEndPuncttrue
\mciteSetBstMidEndSepPunct{\mcitedefaultmidpunct}
{\mcitedefaultendpunct}{\mcitedefaultseppunct}\relax
\EndOfBibitem
\bibitem[De \latin{et~al.}(2016)De, Bart{\'o}k, Cs{\'a}nyic, and
  Ceriotti]{ml_11}
De,~S.; Bart{\'o}k,~A.~P.; Cs{\'a}nyic,~G.; Ceriotti,~M. Comparing molecules
  and solids across structural and alchemical space. \emph{Phisycal Chemistry
  Chemical Physics} \textbf{2016}, \emph{18}, 13754--13769\relax
\mciteBstWouldAddEndPuncttrue
\mciteSetBstMidEndSepPunct{\mcitedefaultmidpunct}
{\mcitedefaultendpunct}{\mcitedefaultseppunct}\relax
\EndOfBibitem
\bibitem[Bart{\'o}k \latin{et~al.}(2013)Bart{\'o}k, Kondor, and
  Cs{\'a}nyi]{ml_12}
Bart{\'o}k,~A.~P.; Kondor,~R.; Cs{\'a}nyi,~G. On representing chemical
  environments. \emph{Physical Review B} \textbf{2013}, \emph{87}, 184115\relax
\mciteBstWouldAddEndPuncttrue
\mciteSetBstMidEndSepPunct{\mcitedefaultmidpunct}
{\mcitedefaultendpunct}{\mcitedefaultseppunct}\relax
\EndOfBibitem
\bibitem[Bart{\'o}k and Cs{\'a}nyi(2015)Bart{\'o}k, and Cs{\'a}nyi]{ml_13}
Bart{\'o}k,~A.~P.; Cs{\'a}nyi,~G. Gaussian Approximation Potentials: A Brief
  Tutorial Introduction. \emph{International Journal of Quantum Chemistry}
  \textbf{2015}, 1051--1057\relax
\mciteBstWouldAddEndPuncttrue
\mciteSetBstMidEndSepPunct{\mcitedefaultmidpunct}
{\mcitedefaultendpunct}{\mcitedefaultseppunct}\relax
\EndOfBibitem
\bibitem[Mar\'{\i} \latin{et~al.}(2008)Mar\'{\i}, Aguirre, and Daza]{ml_10}
Mar\'{\i},~R.~M.; Aguirre,~N.~F.; Daza,~E.~E. Graph Theoretical Similarity
  Approach To Compare Molecular Electrostatic Potentials. \emph{Journal of
  Chemical Information and Modeling} \textbf{2008}, \emph{48}, 109--118\relax
\mciteBstWouldAddEndPuncttrue
\mciteSetBstMidEndSepPunct{\mcitedefaultmidpunct}
{\mcitedefaultendpunct}{\mcitedefaultseppunct}\relax
\EndOfBibitem
\bibitem[Guido \latin{et~al.}(2013)Guido, Cortona, Mennucci, and Adamo]{ml_34}
Guido,~C.~A.; Cortona,~P.; Mennucci,~B.; Adamo,~C. On the Metric of Charge
  Transfer Molecular Excitations: A Simple Chemical Descriptor. \emph{Journal
  of Chemical Theory and Computation} \textbf{2013}, \emph{9}, 3118--3126,
  PMID: 26583991\relax
\mciteBstWouldAddEndPuncttrue
\mciteSetBstMidEndSepPunct{\mcitedefaultmidpunct}
{\mcitedefaultendpunct}{\mcitedefaultseppunct}\relax
\EndOfBibitem
\bibitem[Sadeghi \latin{et~al.}(2013)Sadeghi, Ghasemi, Schaefer, Mohr, Lill,
  and Goedecker]{ml_56}
Sadeghi,~A.; Ghasemi,~S.~A.; Schaefer,~B.; Mohr,~S.; Lill,~M.~A.; Goedecker,~S.
  Metrics for measuring distances in configuration spaces. \emph{The Journal of
  Chemical Physics} \textbf{2013}, \emph{139}, 184118\relax
\mciteBstWouldAddEndPuncttrue
\mciteSetBstMidEndSepPunct{\mcitedefaultmidpunct}
{\mcitedefaultendpunct}{\mcitedefaultseppunct}\relax
\EndOfBibitem
\bibitem[Dong \latin{et~al.}(2015)Dong, DS, HY, Liu, BC, YH, NN, AP, WB, and
  AF]{ml_79}
Dong,~J.; DS,~C.; HY,~M.; Liu,~S.; BC,~D.; YH,~Y.; NN,~W.; AP,~L.; WB,~Z.;
  AF,~C. ChemDes: An integrated web-based platform for molecular descriptor and
  fingerprint computation. \emph{J Cheminform} \textbf{2015}, \emph{7},
  60\relax
\mciteBstWouldAddEndPuncttrue
\mciteSetBstMidEndSepPunct{\mcitedefaultmidpunct}
{\mcitedefaultendpunct}{\mcitedefaultseppunct}\relax
\EndOfBibitem
\bibitem[Huang and von Lilienfeld(2016)Huang, and von Lilienfeld]{ml_92}
Huang,~B.; von Lilienfeld,~O.~A. Communication: Understanding molecular
  representations in machine learning: The role of uniqueness and target
  similarity. \emph{The Journal of Chemical Physics} \textbf{2016}, \emph{145},
  161102\relax
\mciteBstWouldAddEndPuncttrue
\mciteSetBstMidEndSepPunct{\mcitedefaultmidpunct}
{\mcitedefaultendpunct}{\mcitedefaultseppunct}\relax
\EndOfBibitem
\bibitem[Collins \latin{et~al.}(2018)Collins, Gordon, von Lilienfeld, and
  Yaron]{ml_93}
Collins,~C.~R.; Gordon,~G.~J.; von Lilienfeld,~O.~A.; Yaron,~D.~J. Constant
  size descriptors for accurate machine learning models of molecular
  properties. \emph{The Journal of Chemical Physics} \textbf{2018}, \emph{148},
  241718\relax
\mciteBstWouldAddEndPuncttrue
\mciteSetBstMidEndSepPunct{\mcitedefaultmidpunct}
{\mcitedefaultendpunct}{\mcitedefaultseppunct}\relax
\EndOfBibitem
\bibitem[Sifain \latin{et~al.}(2018)Sifain, Lubbers, Nebgen, Smith, Lokhov,
  Isayev, Roitberg, Barros, and Tretiak]{ml_94}
Sifain,~A.~E.; Lubbers,~N.; Nebgen,~B.~T.; Smith,~J.~S.; Lokhov,~A.~Y.;
  Isayev,~O.; Roitberg,~A.~E.; Barros,~K.; Tretiak,~S. Discovering a
  Transferable Charge Assignment Model Using Machine Learning. \emph{The
  Journal of Physical Chemistry Letters} \textbf{2018}, \emph{9},
  4495--4501\relax
\mciteBstWouldAddEndPuncttrue
\mciteSetBstMidEndSepPunct{\mcitedefaultmidpunct}
{\mcitedefaultendpunct}{\mcitedefaultseppunct}\relax
\EndOfBibitem
\bibitem[Himanen \latin{et~al.}(2020)Himanen, J{\"a}ger, {Eiaki V. Morooka},
  Canova, Ranawat, Gao, Rinke, and Foster]{ml_95}
Himanen,~L.; J{\"a}ger,~M. O.~J.; {Eiaki V. Morooka},; Canova,~F.~F.;
  Ranawat,~Y.~S.; Gao,~D.~Z.; Rinke,~P.; Foster,~A.~S. DScribe: Library of
  Descriptors for Machine Learning in Materials Science. \emph{Computer Physics
  Communications} \textbf{2020}, \emph{247}, 106949\relax
\mciteBstWouldAddEndPuncttrue
\mciteSetBstMidEndSepPunct{\mcitedefaultmidpunct}
{\mcitedefaultendpunct}{\mcitedefaultseppunct}\relax
\EndOfBibitem
\bibitem[Hall \latin{et~al.}(1991)Hall, Mohney, and Kier]{ml_9}
Hall,~L.~H.; Mohney,~B.; Kier,~L.~B. The Electrotopological State: Structure
  Information at the Atomic Level for Molecular Graphs. \emph{Journal of
  Chemical Information and Computer Sciences} \textbf{1991}, \emph{31},
  76--82\relax
\mciteBstWouldAddEndPuncttrue
\mciteSetBstMidEndSepPunct{\mcitedefaultmidpunct}
{\mcitedefaultendpunct}{\mcitedefaultseppunct}\relax
\EndOfBibitem
\bibitem[Grisafi \latin{et~al.}(2018)Grisafi, Wilkins, Cs{\'a}nyi, and
  Ceriotti]{ml_97}
Grisafi,~A.; Wilkins,~D.~M.; Cs{\'a}nyi,~G.; Ceriotti,~M. Symmetry-Adapted
  Machine Learning for Tensorial Properties of Atomistic Systems.
  \emph{Physical Review Letters} \textbf{2018}, \emph{120}\relax
\mciteBstWouldAddEndPuncttrue
\mciteSetBstMidEndSepPunct{\mcitedefaultmidpunct}
{\mcitedefaultendpunct}{\mcitedefaultseppunct}\relax
\EndOfBibitem
\bibitem[Rupp \latin{et~al.}(2015)Rupp, Ramakrishnan, and von
  Lilienfeld]{ml_98}
Rupp,~M.; Ramakrishnan,~R.; von Lilienfeld,~O.~A. Machine Learning for Quantum
  Mechanical Properties of Atoms in Molecules. \emph{The Journal of Physical
  Chemistry Letters} \textbf{2015}, \emph{6}, 3309--3313\relax
\mciteBstWouldAddEndPuncttrue
\mciteSetBstMidEndSepPunct{\mcitedefaultmidpunct}
{\mcitedefaultendpunct}{\mcitedefaultseppunct}\relax
\EndOfBibitem
\bibitem[Huang \latin{et~al.}(2018)Huang, Symonds, and von Lilienfeld]{ml_99}
Huang,~B.; Symonds,~N.~O.; von Lilienfeld,~O.~A. The fundamentals of quantum
  machine learning. \emph{arXiv:1807.04259} \textbf{2018}, \relax
\mciteBstWouldAddEndPunctfalse
\mciteSetBstMidEndSepPunct{\mcitedefaultmidpunct}
{}{\mcitedefaultseppunct}\relax
\EndOfBibitem
\bibitem[von Lilienfeld(2018)]{ml_130}
von Lilienfeld,~O.~A. Quantum Machine Learning in Chemical Compound Space.
  \emph{Computational Chemistry} \textbf{2018}, \emph{57}, 4164--4169\relax
\mciteBstWouldAddEndPuncttrue
\mciteSetBstMidEndSepPunct{\mcitedefaultmidpunct}
{\mcitedefaultendpunct}{\mcitedefaultseppunct}\relax
\EndOfBibitem
\bibitem[Artrith \latin{et~al.}(2011)Artrith, Morawietz, and Behler]{ml_131}
Artrith,~N.; Morawietz,~T.; Behler,~J. High-dimensional neural-network
  potentials for multicomponent systems: Applications to zinc oxide.
  \emph{Physical Review B} \textbf{2011}, \emph{83}, 153101\relax
\mciteBstWouldAddEndPuncttrue
\mciteSetBstMidEndSepPunct{\mcitedefaultmidpunct}
{\mcitedefaultendpunct}{\mcitedefaultseppunct}\relax
\EndOfBibitem
\bibitem[Artrith \latin{et~al.}(2017)Artrith, Urban, and Ceder]{ml_132}
Artrith,~N.; Urban,~A.; Ceder,~G. Efficient and accurate machine-learning
  interpolation of atomic energies in compositions with many species.
  \emph{Physical Review B} \textbf{2017}, \emph{96}, 014114\relax
\mciteBstWouldAddEndPuncttrue
\mciteSetBstMidEndSepPunct{\mcitedefaultmidpunct}
{\mcitedefaultendpunct}{\mcitedefaultseppunct}\relax
\EndOfBibitem
\bibitem[Bartók \latin{et~al.}(2017)Bartók, De, Poelking, Bernstein, Kermode,
  Csányi, and Ceriotti]{ml_133}
Bartók,~A.~P.; De,~S.; Poelking,~C.; Bernstein,~N.; Kermode,~J.~R.;
  Csányi,~G.; Ceriotti,~M. Machine learning unifies the modeling of materials
  and molecules. \emph{Scientific Advance} \textbf{2017}, \emph{3}\relax
\mciteBstWouldAddEndPuncttrue
\mciteSetBstMidEndSepPunct{\mcitedefaultmidpunct}
{\mcitedefaultendpunct}{\mcitedefaultseppunct}\relax
\EndOfBibitem
\bibitem[Behle(2011)]{ml_134}
Behle,~J. Atom-centered symmetry functions for constructing high-dimensional
  neural network potentials. \emph{JOURNAL OF CHEMICAL PHYSICS} \textbf{2011},
  \emph{134}, 074106\relax
\mciteBstWouldAddEndPuncttrue
\mciteSetBstMidEndSepPunct{\mcitedefaultmidpunct}
{\mcitedefaultendpunct}{\mcitedefaultseppunct}\relax
\EndOfBibitem
\bibitem[Behler and Parrinello(2007)Behler, and Parrinello]{ml_135}
Behler,~J.; Parrinello,~M. Generalized Neural-Network Representation of
  High-Dimensional Potential-Energy Surfaces. \emph{PHYSICAL REVIEW LETTERS}
  \textbf{2007}, \emph{98}, 146401\relax
\mciteBstWouldAddEndPuncttrue
\mciteSetBstMidEndSepPunct{\mcitedefaultmidpunct}
{\mcitedefaultendpunct}{\mcitedefaultseppunct}\relax
\EndOfBibitem
\bibitem[Bianucci \latin{et~al.}(2000)Bianucci, Alessio~Micheli, and
  Starita]{ml_136}
Bianucci,~A.~M.; Alessio~Micheli,~A.~S.; Starita,~A. Machine Learning Force
  Fields: Construction, Validation, and Outlook. \emph{Applied Intelligence}
  \textbf{2000}, \emph{12}, 117–146\relax
\mciteBstWouldAddEndPuncttrue
\mciteSetBstMidEndSepPunct{\mcitedefaultmidpunct}
{\mcitedefaultendpunct}{\mcitedefaultseppunct}\relax
\EndOfBibitem
\bibitem[Chandrasekaran \latin{et~al.}(2019)Chandrasekaran, Kamal, Batra, Kim,
  Chen, and Ramprasad]{ml_137}
Chandrasekaran,~A.; Kamal,~D.; Batra,~R.; Kim,~C.; Chen,~L.; Ramprasad,~R.
  Solving the electronic structure problem with machine learning.
  \emph{Computational Materials} \textbf{2019}, \emph{22}\relax
\mciteBstWouldAddEndPuncttrue
\mciteSetBstMidEndSepPunct{\mcitedefaultmidpunct}
{\mcitedefaultendpunct}{\mcitedefaultseppunct}\relax
\EndOfBibitem
\bibitem[Hughes \latin{et~al.}(2019)Hughes, Thacker, Wilson, and
  Popelier]{ml_138}
Hughes,~Z.~E.; Thacker,~J. C.~R.; Wilson,~A.~L.; Popelier,~P. L.~A. Description
  of Potential Energy Surfaces of Molecules Using FFLUX Machine Learning
  Models. \emph{Journal of Chemical Theory and Computation} \textbf{2019},
  \emph{15}, 116--126\relax
\mciteBstWouldAddEndPuncttrue
\mciteSetBstMidEndSepPunct{\mcitedefaultmidpunct}
{\mcitedefaultendpunct}{\mcitedefaultseppunct}\relax
\EndOfBibitem
\bibitem[Gastegger \latin{et~al.}(2017)Gastegger, Behler, and
  Marquetand]{ml_139}
Gastegger,~M.; Behler,~J.; Marquetand,~P. Machine Learning Molecular Dynamics
  for the Simulation of Infrared Spectra. \emph{arXiv:1705.05907v1}
  \textbf{2017}, \relax
\mciteBstWouldAddEndPunctfalse
\mciteSetBstMidEndSepPunct{\mcitedefaultmidpunct}
{}{\mcitedefaultseppunct}\relax
\EndOfBibitem
\bibitem[Huang \latin{et~al.}(2018)Huang, Symonds, and von Lilienfeld]{ml_140}
Huang,~B.; Symonds,~N.~O.; von Lilienfeld,~O.~A. \emph{Quantum Machine Learning
  in Chemistry and Materials}; Springer International Publishing, 2018\relax
\mciteBstWouldAddEndPuncttrue
\mciteSetBstMidEndSepPunct{\mcitedefaultmidpunct}
{\mcitedefaultendpunct}{\mcitedefaultseppunct}\relax
\EndOfBibitem
\bibitem[Hu \latin{et~al.}(2018)Hu, Xie, Li, Li, and Lan]{ml_141}
Hu,~D.; Xie,~Y.; Li,~X.; Li,~L.; Lan,~Z. The Inclusion of Machine Learning
  Kernel Ridge Regression Potential Energy Surfaces in On-the-Fly Nonadiabatic
  Molecular Dynamics Simulation. \emph{Journal of Physical Chemistry Letters}
  \textbf{2018}, \relax
\mciteBstWouldAddEndPunctfalse
\mciteSetBstMidEndSepPunct{\mcitedefaultmidpunct}
{}{\mcitedefaultseppunct}\relax
\EndOfBibitem
\bibitem[Ramakrishnan \latin{et~al.}(2014)Ramakrishnan, Dral, Rupp, and von
  Lilienfeld]{ml_142}
Ramakrishnan,~R.; Dral,~P.~O.; Rupp,~M.; von Lilienfeld,~O.~A. Quantum
  chemistry structures and properties of 134 kilo molecules. \emph{Scientific
  Data} \textbf{2014}, \emph{1}, 140022\relax
\mciteBstWouldAddEndPuncttrue
\mciteSetBstMidEndSepPunct{\mcitedefaultmidpunct}
{\mcitedefaultendpunct}{\mcitedefaultseppunct}\relax
\EndOfBibitem
\bibitem[Huo and von Matthias~Rupp(2018)Huo, and von Matthias~Rupp]{ml_143}
Huo,~H.; von Matthias~Rupp, Unified Representation of Molecules and Crystals
  for Machine Learning. \emph{arXiv:1704.06439v3} \textbf{2018}, \relax
\mciteBstWouldAddEndPunctfalse
\mciteSetBstMidEndSepPunct{\mcitedefaultmidpunct}
{}{\mcitedefaultseppunct}\relax
\EndOfBibitem
\bibitem[Imbalzano \latin{et~al.}(2018)Imbalzano, Anelli, Giofr{\'e}, Klees,
  Behler, and Ceriotti]{ml_144}
Imbalzano,~G.; Anelli,~A.; Giofr{\'e},~D.; Klees,~S.; Behler,~J.; Ceriotti,~M.
  Automatic selection of atomic fingerprints and reference configurations for
  machine-learning potentials. \emph{JOURNAL OF CHEMICAL PHYSICS}
  \textbf{2018}, \emph{148}, 241730\relax
\mciteBstWouldAddEndPuncttrue
\mciteSetBstMidEndSepPunct{\mcitedefaultmidpunct}
{\mcitedefaultendpunct}{\mcitedefaultseppunct}\relax
\EndOfBibitem
\bibitem[Jaeger \latin{et~al.}(2018)Jaeger, Fulle, and Turk]{ml_145}
Jaeger,~S.; Fulle,~S.; Turk,~S. Mol2vec: Unsupervised Machine Learning Approach
  with Chemical Intuition. \emph{J. Chem. Inf. Model} \textbf{2018}, \emph{58},
  27--35\relax
\mciteBstWouldAddEndPuncttrue
\mciteSetBstMidEndSepPunct{\mcitedefaultmidpunct}
{\mcitedefaultendpunct}{\mcitedefaultseppunct}\relax
\EndOfBibitem
\bibitem[Unke and Meuwly(2019)Unke, and Meuwly]{ml_72}
Unke,~O.~T.; Meuwly,~M. PhysNet: A Neural Network for Predicting Energies,
  Forces, Dipole Moments, and Partial Charges. \emph{Journal of Chemical Theory
  and Computation} \textbf{2019}, \emph{15}, 3678--3693\relax
\mciteBstWouldAddEndPuncttrue
\mciteSetBstMidEndSepPunct{\mcitedefaultmidpunct}
{\mcitedefaultendpunct}{\mcitedefaultseppunct}\relax
\EndOfBibitem
\bibitem[Mansimov \latin{et~al.}(2019)Mansimov, Kang, Mahmood, and Cho]{ml_81}
Mansimov,~E.; Kang,~S.; Mahmood,~O.; Cho,~K. Molecular Geometry Prediction
  using a Deep Generative Graph Neural Network. \emph{Sci Rep} \textbf{2019},
  \emph{9}, 20381\relax
\mciteBstWouldAddEndPuncttrue
\mciteSetBstMidEndSepPunct{\mcitedefaultmidpunct}
{\mcitedefaultendpunct}{\mcitedefaultseppunct}\relax
\EndOfBibitem
\bibitem[Ramakrishnan \latin{et~al.}(2015)Ramakrishnan, Hartmann, Tapavicza,
  and von Lilienfeld]{ml_6}
Ramakrishnan,~R.; Hartmann,~M.; Tapavicza,~E.; von Lilienfeld,~O.~A. Electronic
  Spectra from TDDFT and Machine Learning in Chemical Space. \emph{The Journal
  of Chemical Physics} \textbf{2015}, \emph{143}, 084111\relax
\mciteBstWouldAddEndPuncttrue
\mciteSetBstMidEndSepPunct{\mcitedefaultmidpunct}
{\mcitedefaultendpunct}{\mcitedefaultseppunct}\relax
\EndOfBibitem
\bibitem[Hall and Kier(1995)Hall, and Kier]{ml_8}
Hall,~L.~H.; Kier,~L.~B. Electrotopological State Indices for Atom Types: A
  Novel Combination of Electronic, Topological, and Valence State Information.
  \emph{Journal of Chemical Information and Computer Sciences} \textbf{1995},
  \emph{35}, 1039--1045\relax
\mciteBstWouldAddEndPuncttrue
\mciteSetBstMidEndSepPunct{\mcitedefaultmidpunct}
{\mcitedefaultendpunct}{\mcitedefaultseppunct}\relax
\EndOfBibitem
\bibitem[Scharfer and Schulz-Gasch(2013)Scharfer, and Schulz-Gasch]{ml_30}
Scharfer,~C.; Schulz-Gasch,~T. Torsion Angle Preferences in Druglike Chemical
  Space: A Comprehensive Guide. \emph{Journal of Medicinal Chemistry}
  \textbf{2013}, \emph{56}, 2016--2028\relax
\mciteBstWouldAddEndPuncttrue
\mciteSetBstMidEndSepPunct{\mcitedefaultmidpunct}
{\mcitedefaultendpunct}{\mcitedefaultseppunct}\relax
\EndOfBibitem
\bibitem[Landrum(2006)]{ml_31}
Landrum,~G. RDKit: Open-source cheminformatics. 2006;
  \url{http://www.rdkit.org}\relax
\mciteBstWouldAddEndPuncttrue
\mciteSetBstMidEndSepPunct{\mcitedefaultmidpunct}
{\mcitedefaultendpunct}{\mcitedefaultseppunct}\relax
\EndOfBibitem
\bibitem[Riniker and Landrum(2015)Riniker, and Landrum]{ml_84}
Riniker,~S.; Landrum,~G.~A. Better Informed Distance Geometry: Using What We
  Know To Improve Conformation Generation. \emph{Chem. Inf. Model.}
  \textbf{2015}, \emph{55}, 2562--2574\relax
\mciteBstWouldAddEndPuncttrue
\mciteSetBstMidEndSepPunct{\mcitedefaultmidpunct}
{\mcitedefaultendpunct}{\mcitedefaultseppunct}\relax
\EndOfBibitem
\bibitem[Landrum(2016)]{ml_146}
Landrum,~G. RDKit: Open-Source Cheminformatics Software. \textbf{2016}, \relax
\mciteBstWouldAddEndPunctfalse
\mciteSetBstMidEndSepPunct{\mcitedefaultmidpunct}
{}{\mcitedefaultseppunct}\relax
\EndOfBibitem
\bibitem[Ruddigkeit \latin{et~al.}(2012)Ruddigkeit, van Deursen, Blum, and
  Reymond]{ml_125}
Ruddigkeit,~L.; van Deursen,~R.; Blum,~L.~C.; Reymond,~J.-L. Enumeration of 166
  Billion Organic Small Molecules in the Chemical Universe Database GDB-17.
  \emph{Journal of Chemical Information and Modeling} \textbf{2012}, \emph{52},
  2864–2875\relax
\mciteBstWouldAddEndPuncttrue
\mciteSetBstMidEndSepPunct{\mcitedefaultmidpunct}
{\mcitedefaultendpunct}{\mcitedefaultseppunct}\relax
\EndOfBibitem
\bibitem[Ramakrishnan \latin{et~al.}(2014)Ramakrishnan, Dral, Rupp, and von
  Lilienfeld]{ml_126}
Ramakrishnan,~R.; Dral,~P.~O.; Rupp,~M.; von Lilienfeld,~O.~A. Quantum
  chemistry structures and properties of 134 kilo molecules. \emph{Scientific
  Data} \textbf{2014}, \emph{1}, 140022\relax
\mciteBstWouldAddEndPuncttrue
\mciteSetBstMidEndSepPunct{\mcitedefaultmidpunct}
{\mcitedefaultendpunct}{\mcitedefaultseppunct}\relax
\EndOfBibitem
\bibitem[Stephens \latin{et~al.}(1994)Stephens, Devlin, Chabalowski, and
  Frisch]{ml_43}
Stephens,~P.~J.; Devlin,~F.~J.; Chabalowski,~C.~F.; Frisch,~M.~J. Ab Initio
  Calculation of Vibrational Absorption and Circular Dichroism Spectra Using
  Density Functional Force Fields. \emph{ACS Publications} \textbf{1994},
  \emph{98}, 11623--11627\relax
\mciteBstWouldAddEndPuncttrue
\mciteSetBstMidEndSepPunct{\mcitedefaultmidpunct}
{\mcitedefaultendpunct}{\mcitedefaultseppunct}\relax
\EndOfBibitem
\bibitem[Ghiringhelli \latin{et~al.}(2015)Ghiringhelli, Vybiral, {V.
  Levchenko}, Draxl, and Scheffler]{ml_4}
Ghiringhelli,~L.; Vybiral,~J.; {V. Levchenko},~S.; Draxl,~C.; Scheffler,~M. Big
  Data of Materials Science - Critical Role of the Descriptor. \emph{Physical
  Review Letters} \textbf{2015}, \emph{114}, 105503\relax
\mciteBstWouldAddEndPuncttrue
\mciteSetBstMidEndSepPunct{\mcitedefaultmidpunct}
{\mcitedefaultendpunct}{\mcitedefaultseppunct}\relax
\EndOfBibitem
\bibitem[Raghunathan and von Lilienfeld(2015)Raghunathan, and von
  Lilienfeld]{ml_128}
Raghunathan,~R.; von Lilienfeld,~O.~A. Many Molecular Properties from One
  Kernel in Chemical Space. \emph{CHIMIA International Journal for Chemistry}
  \textbf{2015}, \emph{69}, 182--186\relax
\mciteBstWouldAddEndPuncttrue
\mciteSetBstMidEndSepPunct{\mcitedefaultmidpunct}
{\mcitedefaultendpunct}{\mcitedefaultseppunct}\relax
\EndOfBibitem
\bibitem[Weininger(1988)]{ml_129}
Weininger,~D. SMILES, a chemical language and information system. 1.
  Introduction to methodology and encoding rules. \emph{Journal of Chemical
  Information and Computer Sciences} \textbf{1988}, \emph{28}, 31--36\relax
\mciteBstWouldAddEndPuncttrue
\mciteSetBstMidEndSepPunct{\mcitedefaultmidpunct}
{\mcitedefaultendpunct}{\mcitedefaultseppunct}\relax
\EndOfBibitem
\bibitem[Crippen and Havel(1988)Crippen, and Havel]{ml_121}
Crippen,~G.; Havel,~T. \emph{Distance Geometry and Molecular Conformation};
  1988\relax
\mciteBstWouldAddEndPuncttrue
\mciteSetBstMidEndSepPunct{\mcitedefaultmidpunct}
{\mcitedefaultendpunct}{\mcitedefaultseppunct}\relax
\EndOfBibitem
\bibitem[Blaney and Dixon(1994)Blaney, and Dixon]{ml_122}
Blaney,~J.~M.; Dixon,~J.~S. \emph{Distance Geometry in Molecular Modeling};
  1994\relax
\mciteBstWouldAddEndPuncttrue
\mciteSetBstMidEndSepPunct{\mcitedefaultmidpunct}
{\mcitedefaultendpunct}{\mcitedefaultseppunct}\relax
\EndOfBibitem
\bibitem[Jacobsen \latin{et~al.}(1996)Jacobsen, Stoltze, and Norskov]{ml_100}
Jacobsen,~K.; Stoltze,~P.; Norskov,~J. A semi-empirical effective medium theory
  for metals and alloys. \emph{ELSEVIER Surface Science} \textbf{1996},
  \emph{366}, 394--402\relax
\mciteBstWouldAddEndPuncttrue
\mciteSetBstMidEndSepPunct{\mcitedefaultmidpunct}
{\mcitedefaultendpunct}{\mcitedefaultseppunct}\relax
\EndOfBibitem
\bibitem[Larsen \latin{et~al.}(2017)Larsen, Mortensen, Blomqvist, Castelli,
  Christensen, Du{\l}ak, Friis, Groves, Hammer, Hargus, Hermes, Jennings,
  Jensen, Kermode, Kitchin, Kolsbjerg, Kubal, Kaasbjerg, Lysgaard, Maronsson,
  Maxson, Olsen, Pastewka, Peterson, Rostgaard, Schi{\o}tz, Sch{\"u}tt,
  Strange, Thygesen, Vegge, Vilhelmsen, Walter, Zeng, and Jacobsen]{ase_paper}
Larsen,~A.~H. \latin{et~al.}  The atomic simulation environment--a Python
  library for working with atoms. \emph{Journal of Physics: Condensed Matter}
  \textbf{2017}, \emph{29}, 273002\relax
\mciteBstWouldAddEndPuncttrue
\mciteSetBstMidEndSepPunct{\mcitedefaultmidpunct}
{\mcitedefaultendpunct}{\mcitedefaultseppunct}\relax
\EndOfBibitem
\end{mcitethebibliography}
